\documentclass[12pt]{article}

\usepackage[letterpaper, margin=1in, headheight=14pt]{geometry}
\usepackage{setspace}
\usepackage{tikz}
\usepackage{tabularx}

\usepackage[T1]{fontenc}
\usepackage[utf8]{inputenc}
\usepackage{lmodern}
\usepackage{amsmath, amssymb, amsthm, mathtools}
\usepackage{dsfont}

\usepackage{graphicx}
\usepackage{subcaption}
\usepackage{xcolor}
\usepackage{booktabs}
\usepackage{multirow}
\usepackage{enumitem}

\setlist{noitemsep}
\setlist{nosep}

\graphicspath{{Figures/}{./Figures/}}

\usepackage[numbers, square, comma, sort&compress]{natbib}
\usepackage{algorithm,algorithmic}

\theoremstyle{definition}

\newtheorem*{assumption*}{\assumptionnumber}
\providecommand{\assumptionnumber}{}


\DeclareMathOperator*{\argmin}{argmin}

\newcommand{\given}{\, \vert \,}
\renewcommand{\hat}{\widehat}
\newcommand{\US}{U.S.}

\def\nth{^{\textnormal{th}}}

\makeatletter
\newcommand*\rel@kern[1]{\kern#1\dimexpr\macc@kerna}
\newcommand*\widebar[1]{%
  \begingroup
  \def\mathaccent##1##2{%
    \rel@kern{0.8}%
    \overline{\rel@kern{-0.8}\macc@nucleus\rel@kern{0.2}}%
    \rel@kern{-0.2}%
  }%
  \macc@depth\@ne
  \let\math@bgroup\@empty \let\math@egroup\macc@set@skewchar
  \mathsurround\z@ \frozen@everymath{\mathgroup\macc@group\relax}%
  \macc@set@skewchar\relax
  \let\mathaccentV\macc@nested@a
  \macc@nested@a\relax111{#1}%
  \endgroup
}
\makeatother
\mathtoolsset{showonlyrefs}

\usepackage{hyperref}
\hypersetup{colorlinks=true, linkcolor=black, citecolor=blue, urlcolor=blue}
\usepackage[capitalise, noabbrev]{cleveref}
\usepackage{bookmark}

\usepackage{mdframed}

\newmdenv[
  topline=false,
  bottomline=false,
  rightline=false,
  leftline=true,
  linewidth=2pt,
  linecolor=black,
  leftmargin=40pt,
  rightmargin=40pt,
  innerleftmargin=15pt,
  innerrightmargin=0pt,
  innertopmargin=5pt,
  innerbottommargin=5pt,
  skipabove=\baselineskip,
  skipbelow=\baselineskip
]{leftbar}

\title{Fast, Frequentist Estimation of Epidemic Reproduction Numbers}

\usepackage{authblk}

\author[1]{Jeremy Goldwasser}
\author[1]{Ryan J.\ Tibshirani}
\author[2]{Alyssa Bilinski}
\affil[1]{Department of Statistics, University of California, Berkeley}
\affil[2]{Departments of Health Policy and Biostatistics, Brown University}
\date{}

\begin{document}

\maketitle

\begin{abstract}
The effective reproduction number $R_t$ is one of the most important indicators of epidemic dynamics.
Estimating $R_t$, typically from case reports or hospitalization counts, poses a challenging inverse problem. 
One key issue is lag: $R_t$ acts at the moment of transmission, while the data it generates surface days later.
To handle this delay and infer recent infections in real time,
popular methods take a Bayesian approach, which can be slow and sensitive to prior specification.
As an alternative, we propose ConvRt, a frequentist method for retrospective and real-time estimation.
ConvRt deconvolves latent infections and then estimates $R_t$ with successive penalized-likelihood steps, using a spline basis to model smooth curves. 
Across both stylized and data-driven simulations,
we demonstrate favorable performance in point estimation, uncertainty quantification, and runtime. 
Moreover, by untangling smoothness from future projections---assumptions conventionally bundled in priors---ConvRt enables researchers to assess which qualitative narratives about $R_t$ the data support.




\end{abstract}

\section{Introduction}\label{sec:intro}

The effective reproduction number $R_t$ is a central metric in infectious disease epidemiology. It summarizes the transmissibility of a pathogen as the average number of secondary infections generated by a single infected individual at time $t$.
Reliable, timely estimates of $R_t$
are critical for public health decision-making, enabling policymakers to adopt and adjust intervention measures and assess their effectiveness.


Most methods estimate $R_t$ using a Bayesian framework: priors simultaneously encode beliefs about $R_t$ and regularize predictions over time.
One strategy, exemplified by EpiEstim \citep{cori2013new}, substitutes observed case reports for daily transmissions, effectively estimating $R_t$ at a lag.
An alternative, exemplified by EpiNow2 \citep{abbott2020estimating}, explicitly models the latent infection process and the observation delay in a Bayesian generative framework. 
This yields more principled estimates, but fitting can take an hour or more, and convergence is not guaranteed. 
The estimates are also sensitive to chosen priors, which are subjective and may be poorly specified.
A natural question is whether one can retain the mechanistic structure underpinning methods like EpiNow2, while avoiding these drawbacks.

We propose ConvRt, a frequentist estimator that uses spline-based smoothing to incorporate complex deconvolutions.
Unlike similar frequentist approaches \citep{rtestim}, ConvRt models the same generative structure as EpiNow2, and thus avoids bias that arises from substituting cases as infections.
It recovers latent infections by deconvolving the observed cases with penalized maximum likelihood;
infection counts are then plugged into a second deconvolution step, this time for $R_t$.
Benchmarking across six synthetic datasets, we show that ConvRt matches or outperforms existing methods in both accuracy and calibration.
On real COVID-19 and influenza data, it makes similar predictions to leading methods at a fraction of the computational cost.

Another practical limitation of Bayesian $R_t$ models is that a single prior conflates two distinct modeling choices: temporal smoothness and tail behavior. ConvRt decouples these, enabling practitioners to tune each independently---a flexibility we exploit in scenario analyses in \cref{sec:scenario}. These analyses are further aided by ConvRt's quick runtime, while methods like EpiNow2 are orders of magnitude slower.

The rest of this paper proceeds as follows. \cref{sec:background} reviews definitions of $R_t$. \cref{sec:rt-methods} develops ConvRt, including retrospective and real-time versions, and situates it among existing estimators.
\cref{sec:experiments} evaluates ConvRt on several real and synthetic datasets, benchmarking against standard methods. \cref{sec:scenario} demonstrates how ConvRt can analyze possible scenarios in real time. \cref{sec:discussion} discusses further opportunities to overcome remaining limitations.

\subsection*{Notation}\label{sec:notation}

\cref{tab:notation} summarizes notation used throughout this paper.

\begin{table}[!ht]
\centering
\caption{Summary of notation.}
\label{tab:notation}
\renewcommand{\arraystretch}{1.15}
\begin{tabularx}{\linewidth}{@{}l >{\raggedright\arraybackslash}X@{}}
\toprule
\textbf{Symbol} & \textbf{Meaning} \\
\midrule
\multicolumn{2}{@{}l}{\textsc{Reproduction numbers}} \\
\addlinespace[2pt]
$R_0$ & Basic reproduction number \\
$R_t$ & Effective reproduction number (instantaneous) \\
\midrule
\multicolumn{2}{@{}l}{\textsc{Infection process}} \\
\addlinespace[2pt]
$x_t$ & Infections at $t$ \\
$x_{<t}$ & Infection history $(x_0, \dots, x_{t-1})$ \\
$g$ & Generation interval distribution \\
\midrule
\multicolumn{2}{@{}l}{\textsc{Observation model}} \\
\addlinespace[2pt]
$y_t$ & Observed counts (e.g.\ cases, hospitalizations) at $t$ \\
$\pi$ & Infection-to-report delay distribution \\
$\Lambda_t$ & Unscaled conditional mean of $y_t \given x_{<t}$ \\
$\mu_t$ & Conditional mean of $y_t \given x_{<t}$ \\
$\rho_t$ & Case ascertainment rate (proportion of infections reported at $t$) \\
$\omega_d$ & Day-of-week multiplicative effect, $d \in \{0,\ldots,6\}$ \\
$\varphi$ & Quasi-Poisson dispersion parameter \\
\midrule
\multicolumn{2}{@{}l}{\textsc{Estimation}} \\
\addlinespace[2pt]
$\theta, \omega$ & Spline coefficients and day-of-week effects \\
$S$ & Spline basis matrix; $R_t(\theta) = S(t)^\top \theta$ \\
$Z_t$ & Covariate row at $t$ from \eqref{eq:linear-mean-cases} \\
$\lambda$ & Curvature-penalty hyperparameter \\
$\gamma$ & Tail-smoothness hyperparameter (real-time setting) \\
$\Omega, \Psi$ & Curvature and tapered-smoothness penalty matrices \\
$D^{(m)}$ & Finite difference operator of order $m$ \\
\bottomrule
\end{tabularx}
\end{table}

\section{Reproduction Numbers in Epidemiology}\label{sec:background}
\label{sec:rt-definitions}

Reproduction numbers have been studied extensively in epidemiology
\citep{dublin1925true, kermack1927contribution, macdonald1952equilibrium, diekmann1990definition, anderson1991infectious, heesterbeek2002brief}.
The basic reproduction number, often referred to as $R_0$, is the expected number of secondary infections produced by one infectious individual in a fully susceptible population.
First defined in 1925, it delineates a useful threshold: in expectation, epidemics grow when $R_0 > 1$ and die out when $R_0 < 1$.

Since the assumption of full susceptibility rarely holds beyond an outbreak's earliest moments, the effective reproduction number $R_t$ is often a more practical measure.
Informally, $R_t$ is the expected secondary infections from an individual infected at time $t$ given the current population state.
\citet{fraser2007} proposes two formalizations amenable to estimation.
While his ideas were originally presented in continuous time, we work in discrete time, following common practice since data are observed discretely (e.g., daily).

\paragraph{Preliminaries.}
Let $x_t$ be the number of individuals infected at $t$.
This is indexed by the date of transmission, not the date the individual becomes infectious.
Further define $w(t,k)$ as the rate of secondary transmissions for a primary infection acquired $k$ timesteps before time $t$:
$$w(t,k)=\text{average \# transmissions at $t$ per individual infected at $t-k$}.$$
The number of new infections at $t$ is then
\begin{equation}\label{eq:fraser_alt}
    x_t = \sum_{k> 0} x_{t-k} \cdot w(t,k).
\end{equation}

We treat infections $x_t$ as deterministic, a modeling convenience rather than a claim that infections are noiseless. 
This setup is shared by the $R_t$ formalization in \citet{fraser2007} and estimation methods including EpiNow2 \citep{abbott2020estimating}. 
Other works instead model $x_t$ as random, such as \citet{cori2013new}.
Doing so would not change our $R_t$ estimator described in \cref{sec:rt-methods}.
The inferential target would become the conditional mean curve $\mathbb{E}[x_t \mid x_{<t}]$, which satisfies the same renewal process in \cref{eq:fraser_alt};
our estimator would then use the inferred mean as a plug-in approximation for $x_t$.
In addition, infection-level noise is not separately identifiable from observation noise. 
Modeling it does not generally aid inference, a point we revisit in \cref{sec:discussion}.

Before presenting $R_t$ definitions, we introduce the generation interval distribution $g(t)$.
The generation interval is the time between a primary individual's infection and their transmission to a secondary individual.
This may be expressed as a probability distribution, aggregating secondary transmissions at time $t$.
\citet{fraser2007} formalizes $g_k(t)$ as the $k$-step transmission rate at time $t$ normalized by its total mass:
\begin{align}\label{eq:gi_basic}
    g_k(t) &= \frac{w(t,k)}{\sum_j w(t,j)}\\
    &= \mathbb{P}(\text{transmission occurs $k$ steps after infection} \given 
    \text{transmission occurs at $t$}).
\end{align}
In general, $g_k(t)$ may vary over time---for instance, due to behavioral changes or 
the emergence of new variants. 
For exposition, we assume the generation interval distribution is stationary over time ($g_k(t) = g_k$ for all $t$), 
though our method does not require it.
This convention is shared by most papers and software packages, including EpiEstim and EpiNow2 \citep{cori2013new,abbott2020estimating}.

\paragraph{Instantaneous $R_t$.}
The instantaneous reproduction number is defined as the sum of all age-specific transmission rates at time $t$ \citep{fraser2007}:
\begin{align}\label{eq:inst}
    R_t &= \sum_k w(t,k)
\end{align}
We can understand this as the average number of secondary infections per primary infection, under conditions prevailing at time $t$.
Instantaneous $R_t$ quantifies real-time transmission, which makes it well-suited for understanding present conditions, such as the effect of interventions.

Invoking the stationarity assumption introduced above, 
plugging \eqref{eq:inst} into the definition of generation interval \eqref{eq:gi_basic} yields \mbox{$g_k=w(t,k)/R_t$}.
Substituting this relation into the transmission model \eqref{eq:fraser_alt} produces the renewal equation:
\begin{equation}\label{eq:renewal}
    x_t = R_t \sum_{k> 0} x_{t-k} \cdot g_k.
\end{equation}

The renewal equation is the basis for most modern $R_t$ estimation.
Other definitions of $R_t$ exist in the literature. The \emph{case} (or cohort) reproduction number \citep{fraser2007} is forward-looking, summing transmission rates from $t$ into the future rather than from the past. Standard compartmental models such as SIR and SEIR yield a \emph{mechanistic} $R_t$ based on the contact rate and infectious duration \citep{anderson1991infectious}; under homogeneous mixing, mechanistic $R_t$ coincides with the instantaneous definition \citep{gostic2020practical,goldwasser2026equivalenceinstantaneousmechanisticreproduction}. We focus on instantaneous $R_t$ throughout, in line with the mainstream estimation literature.

%
%
%

\section{Estimating \texorpdfstring{$R_t$}{Rt} via Penalized Deconvolution}\label{sec:rt-methods}


Our aim is to estimate $R_t$ as a smooth yet flexible curve over time.
As defined in \cref{sec:background}, $R_t$ is a function of daily infection counts $x_t$, but these are unobserved in practice.
Instead, reported cases appear at a lag of several days, or even weeks.
Methods that substitute these cases for infections effectively estimate $R_t$ at a lag, which can obscure rapid changes in transmission.

We propose a two-stage deconvolution procedure that addresses this directly. 
The first stage recovers latent infections from observed reports, and the second estimates $R_t$ from the recovered infections.
Each stage parameterizes its time series as a spline, fitting its coefficients with penalized maximum-likelihood.
We call the resulting procedure ConvRt, for convolutional $R_t$.
This approach is fast to fit and admits standard frequentist confidence intervals.

\cref{sec:rt-seir} introduces the observation model for reported cases and the likelihood used throughout.
\cref{sec:inf-deconv} deconvolves latent infections from this likelihood.
\cref{sec:rt-retro} proposes a retrospective $R_t$
estimator, which \cref{sec:rt-rt} extends for the real-time setting.
\cref{sec:rt-uncertainty} introduces our approaches for uncertainty quantification.
\cref{sec:method-contrasts} surveys existing
$R_t$ estimators and situates our approach among them.

\subsection{Observation Model}\label{sec:rt-seir}

We observe epidemic incidence $y_t$, such as positive cases or hospitalizations.
To characterize the distribution of $y_t$ conditional on the infection path $x_{<t}$, we impose a Poisson noise model with rate $\mu_t$.
This is a standard model in practice (e.g., \citet{cori2013new}), though some works allow for overdispersion (e.g., \citet{abbott2020estimating}).
Assuming reports are independent 
from time $t_0$ to $t_1$, the joint distribution is
\begin{align}\label{eq:joint-likelihood}
    \mathbb{P}(y_{t_0:t_1} \given x_{<t_1}) &= \prod_{t=t_0}^{t_1} \text{Poisson}(y_t;\, \mu_t), \\
    \log\mathbb{P}(y_{t_0:t_1}\given x_{<t_1})&= \sum_{t=t_0}^{t_1} \left( y_t \log \mu_t - \mu_t \right) + \text{const}.
\end{align}

The Poisson model can be relaxed to quasi-Poisson noise, characterized by a dispersion term $\varphi$.
This does not affect parameter estimation since quasi-Poisson point estimates coincide with Poisson MLE, so we maintain Poisson distributions in the following expressions.
Quasi-Poisson inference simply inflates the Poisson variance, estimating $\varphi$ with the Pearson chi-squared statistic.

We now turn to identifying $\mu_t$, the mean number of reports $y_t$ conditional on infections $x_{<t}$.
Each infection is recorded several days after it actually occurs.
For example, COVID-19 cases were reported around 5 days after symptom onset \citep{abbott2020estimating}, which itself came 5 days after exposure \citep{kucharski2020early}.
Define $\pi$ as the infection-to-report delay distribution:
$$\pi_k = \mathbb{P}(\text{Infection reported after $k$ timesteps}\given\text{Infection reported}),\qquad\forall\ k>0.$$

Throughout this section, we treat $\pi$ as known.
In practice, it is typically estimated from line-list data or household contact tracing studies.
For ease of notation, we also treat $\pi$ as stationary.

If all infections were reported, expected counts would be the convolution of the delay kernel $\pi$ against infection counts $x_{<t}$.
However, cases rarely comprise all infections.
For example, \citet{reese2021estimated} estimated a case ascertainment rate of 13\% for COVID-19 through September 2020;
the CDC uses hospitalized cases to track $R_t$ for influenza and RSV \citep{cdccfa2026rt}.
Let $\rho_t$ be the proportion of infections observed at time $t$.
In addition, optionally include multiplicative day-of-week (DoW) effects $\omega_{(t \bmod 7)}$. 
The conditional mean is thus
\begin{equation}\label{eq:mean-cases}
\mu_t = \omega_{(t \bmod 7)} \rho_t\Lambda_t,\quad\text{where}\quad
\Lambda_t = \sum_{s<t} x_s \pi_{t-s}.
\end{equation}

The likelihood \eqref{eq:joint-likelihood} depends on incident infections $x_{<t_1}$, which are unobserved.
It is also not directly parameterized by $R_t$, though the two are linked through the renewal equation \eqref{eq:renewal}.
We can make this link explicit.
To express mean cases in terms of $R_t$, we replace each $x_s$ with the renewal equation, \mbox{$x_s =R_s \sum_{u<s} x_u\, g_{s-u}$}.
The convolution becomes
\begin{equation}\label{eq:exp-cases-rt}
    \Lambda_t = \sum_{s<t} R_s \left(\sum_{u<s} x_u\, g_{s-u}\right) \pi_{t-s}.
\end{equation}

Substituting \eqref{eq:exp-cases-rt} into \eqref{eq:mean-cases} and then into \eqref{eq:joint-likelihood} gives the full log-likelihood in terms of $R_t$, infections, and DoW effects:
\begin{align}\label{eq:full-likelihood2}
\ell(R, x,\omega \mid y) = \sum_{t=t_0}^{t_1} \Bigl[ y_t \log\bigl(&\omega_{(t \bmod 7)}\,\rho_t \sum_{s<t} R_s \, \pi_{t-s}\sum_{u<s} x_u\, g_{s-u}\bigr) \notag \\
-\ &\omega_{(t \bmod 7)}\,\rho_t \sum_{s<t} R_s\, \pi_{t-s}  \sum_{u<s} x_u\, g_{s-u}\Bigr] + \text{const}.
\end{align}

Both $x_t$ and $R_t$ are unobserved, and must be estimated from the data.
However, joint estimation is infeasible in the formulation \eqref{eq:full-likelihood2}, since the products of $R_s$ and $x_u$ render the problem nonconvex.
The approach we pursue estimates them one at a time: first $x_t$, then $R_t$.
We deconvolve infections in \cref{sec:inf-deconv}, then plug the estimates $\hat x_{<t_1}$ into \eqref{eq:exp-cases-rt} and solve for $R_t$ in \cref{sec:rt-retro}.

\subsection{Deconvolving Latent Infections}\label{sec:inf-deconv}




To obtain a plug-in estimate of infections $x_t$, we deconvolve counts with penalized maximum likelihood.
This takes the place of sampling infections during MCMC in standard Bayesian methods.
Expanding the Poisson negative log likelihood \eqref{eq:joint-likelihood} with the formula for $\mu_t$ \eqref{eq:mean-cases},
\begin{equation}\label{eq:infection-likelihood}
-\ell(x,\omega \mid y) = \sum_{t=t_0}^{t_1} \Bigl[\omega_{(t \bmod 7)}\,\rho_t \sum_{s<t} \pi_{t-s}\, x_s - y_t \log\bigl(\omega_{(t \bmod 7)}\,\rho_t \sum_{s<t} \pi_{t-s}\, x_s\bigr)\Bigr] + \text{const}.
\end{equation}

We can expect the curve of daily infections to be reasonably smooth over time.
Therefore, we parameterize $x_t$ as a spline---a piecewise polynomial curve whose curvature changes at a set of knots.
Defining $S$ as the spline basis matrix with basis vectors $S(t)$, and $\theta^{(x)}$ the coefficients, 
we model \smash{$x_t(\theta) = S(t)^\top \theta^{(x)}$}. 
While this framework accommodates all parameterizations (e.g., B-splines, P-splines), our experiments use a natural cubic spline basis.

To recover infections, we minimize the negative log likelihood \eqref{eq:infection-likelihood} subject to constraints and a smoothness penalty. $\theta^{(x)}$ is constrained to produce non-negative $x_t$ values, and $\omega^{(x)}$ to have geometric mean one (normalizing the multiplicative DoW effects). 
Smoothness is imposed through the quadratic regularization term $\theta^{(x)\top} \Omega^{(x)}\, \theta^{(x)}$, weighted by a hyperparameter $\lambda^{(x)}$. 
We describe the penalty matrix $\Omega^{(x)}$ and strategies to tune $\lambda^{(x)}$ in \cref{apx:deconvolution}.

The spline coefficients solve a penalized Poisson generalized linear model (GLM),
\begin{align}
\hat\theta^{(x)}, \hat\omega^{(x)} &= 
\argmin_{\bar{\omega}_{GM}^{(x)}=1,\;  S\theta^{(x)} \succeq 0}\;
\sum_t \biggl[\bigl(\omega_{(t \bmod 7)}^{(x)}\,\rho_t\, \theta^{(x)\top}\sum_{s<t} \pi_{t-s}\, S(s)  \bigr)\\
&\hspace{4cm}-
y_t \log \bigl(\omega_{(t \bmod 7)}^{(x)}\,\rho_t\, \theta^{(x)\top}\sum_{s<t} \pi_{t-s}\, S(s)  \bigr)\biggr]
+ \lambda^{(x)}\, \theta^{(x)\top} \Omega^{(x)}\, \theta^{(x)}.
\end{align}
Finally, we take $\hat x_t = S(t)^\top \theta^{(x)}$.
The GLM can be fit quickly with off-the-shelf IRLS solvers.
Details on optimization are in \cref{apx:optim}; the strategy is shared with $R_t$, introduced in the following section.

\subsection{\texorpdfstring{$R_t$}{Rt} Deconvolution in Retrospect}\label{sec:rt-retro}

Having estimated infections, we now turn our focus to $R_t$, using a similar method as above.
We drop the $(x)$ superscripts, reusing $S$, $\theta$, $\Omega$, $\lambda$ for the $R_t$ stage.
We first study $R_t$ estimation in retrospect, where analysts are more interested in historical levels of $R_t$ than its recent trends.
Our method estimates $R_t$ at all dates until the final observation date, $t_1$.
However, tail predictions are highly unstable, as we discuss in \cref{sec:rt-rt}.

As with infections $x_t$, we parameterize $R_t$ as a spline.
Defining the (natural cubic) spline basis $S$ and coefficients $\theta$, we learn $R_t(\theta) = S(t)^\top \theta$.
Substituting deconvolved infections $\hat x$ into \eqref{eq:exp-cases-rt} makes the expected counts linear in $\theta$:
\begin{equation}\label{eq:linear-mean-cases}
    \Lambda_t(\theta) = \sum_{s<t} R_s(\theta) \left(\sum_{u<s} \hat x_u\, g_{s-u}\right) \pi_{t-s} =
             \theta^\top \underbrace{\sum_{s<t} S(s) \left(\sum_{u<s} \hat x_u\, g_{s-u}\right) \pi_{t-s}}_{Z_t} = \theta^\top Z_t.
\end{equation}
We again fit a penalized Poisson GLM, with a similar smoothness penalty. 
By default, we let $\Omega$ be the integrated second-derivative penalty,
\mbox{$\theta^\top \Omega\, \theta = \int_{t_0}^{t_1}\!\bigl[R''(u)\bigr]^2\,du$,}
familiar from the smoothing spline literature.
Applying \eqref{eq:linear-mean-cases} to the likelihood \eqref{eq:full-likelihood2}, we solve for $\theta$ and $\omega$:
\begin{align}\label{eq:glm-retro}
\hat{\theta}, \hat\omega &=
\argmin_{\substack{\bar{\omega}_{GM}=1 \\ S\theta \succeq 0}}\
             \sum_{t=t_0}^{t_1}\!\bigl[ \mu_t(\omega, \theta) - y_t \log \mu_t(\omega, \theta) \bigr]
             \;+\; \lambda\,\theta^\top \Omega\, \theta,\\
             &\qquad\text{where}\quad \mu_t(\omega, \theta) = \omega_{(\text{$t$ mod 7})}\,\rho_t\, \Lambda_t(\theta)
             \quad\text{and}\quad \Lambda_t(\theta) = Z_t^\top\theta.
\end{align}

We tune the smoothness hyperparameter $\lambda$ with $K$-fold cross-validation, holding out every $K\nth$ timestep from the loss.
The cross-validated error curve evaluates predictions on held-out timesteps.
Per \citet{rtestim}, we use Poisson deviance, a goodness-of-fit metric based on the likelihood ratio.
We tune $\lambda$ via the 1se rule, selecting the largest $\lambda$ whose cross-validated error is within one standard error of the minimum.

Consider the special case in which ascertainment rate is a constant $\rho$.
Infections $\hat x_t$ are then scaled by a factor of $\rho^{-1}$, per \eqref{eq:mean-cases}.
This cancels out with the factor in the $R_t$ deconvolution.
Consequently, the rate itself is irrelevant and does not need to be specified.
That stationarity assumption is plausible in certain settings.
Otherwise, it is necessary to plug in ascertainment rates estimated \textit{a priori}.
Jointly solving for ascertainment rates is underspecified in this setup, but $\rho_t$ may be estimated separately, for example from seroprevalence data \citep{reese2021estimated, Chitwood2022}.

\subsection{Real-Time Methods}\label{sec:rt-rt}

Recall that in the retrospective setting, $R_t$ is estimated through time $t_1$, but the final predictions before $t_1$ are ignored.
In contrast, these $R_t$ estimates are of utmost importance in the real-time setting, where $t_1$ denotes the present.
Real-time $R_t$ estimation is central to forecasting and policy evaluation, yet is a more challenging task.
On the days leading up to $t_1$, a growing number of infections $x_t$ will not be reported as cases until \textit{after} $t_1$, through the delay kernel $\pi$.
Therefore, the $R_t$ signal that corresponds to these infections does not appear in the available data, so without extra regularization, predictions become highly unstable.

We address this in two ways, borrowing from \citet{Jahja2022}.
First, parametrizing $R_t$ as a smoothing spline constrains the tail to be linear past the boundary knots.
This adds a degree of stability without
any explicit constraint on $\theta$.
(As an alternative, \cref{apx:optim-constraints} shows how to enforce a constant fit.)

Second, we add a tapered smoothing penalty to the loss.
This progressively encourages tail predictions towards stationarity as the cutoff date approaches.
Squared first-order differences in $R_t$ are penalized with a weight
inversely proportional to the CDF of the exposure-to-observation delay distribution.
Here, the final timesteps correspond to early timesteps of the delay distribution, at which little mass has accumulated; therefore the reciprocal of the CDF is high, so nonstationarity is heavily penalized.
Aggregating root weights into diagonal matrix $W$, define \mbox{$\Psi = D^{(1)\top} W^\top W D^{(1)}$} such that \mbox{$\theta^\top \Psi \theta = \|W D^{(1)}\theta\|_2^2$}.
Together, these real-time penalties modify \eqref{eq:glm-retro} to

\begin{align}\label{eq:glm-real-time}
\hat{\theta}, \hat\omega &=
\argmin_{\substack{\bar{\omega}_{GM}=1 \\ S\theta \succeq 0}}\
     \sum_{t=t_0}^{t_1}\!\bigl[\mu_t(\omega, \theta)-y_t \log \mu_t(\omega, \theta) \bigr]
     \;+\; \lambda\,\theta^\top \Omega\, \theta
     \;+\; \gamma \,\theta^\top \Psi \, \theta,\\
     &\qquad\text{where}\quad \mu_t(\omega, \theta) = \omega_{(\text{$t$ mod 7})}\,\rho_t\, \Lambda_t(\theta)
     \quad\text{and}\quad \Lambda_t(\theta) = Z_t^\top\theta.
\end{align}

The tail smoothness hyperparameter, $\gamma$, is tuned via forward-validation.
For each timestep in the recent past (e.g.\ 7 days),
we fit our estimator, then convolve the $R_t$ estimate to predict observations one day ahead.
Aggregating the absolute error over the validation window, we select $\gamma$
via the min- or 1se rule.

Real-time surveillance counts are typically incomplete at the most recent timesteps due to reporting delays, and are revised upward as backfill accumulates. We discuss this limitation further in \cref{sec:discussion}.

\subsection{Inference}\label{sec:rt-uncertainty}


Uncertainty quantification is important for $R_t$, particularly as small differences can have large epidemiological consequences.
Fortunately, our spline parameterization can naturally produce asymptotic confidence intervals, at least in the retrospective case.
We survey this first, then introduce our approach for real-time uncertainty quantification.
Both methods are explored in depth in \cref{apx:rt-uncertainty}.

Suppose there is some vector $\theta^*$ for which the spline is well-specified, meaning $R_t = S(t)^\top\theta^*\ \forall t$.
This condition is shared with essentially all nonparametric inference, not specific to our approach.
Further assume our procedure yields an unbiased estimate for $\theta^*$, such that $R_t = S(t)^\top \mathbb{E}[\hat\theta]=S(t)^\top \theta^*$.
Unbiasedness is a stronger condition, since regularization almost always adds bias. 
Nevertheless, this assumption is still palatable in the retrospective case, while the real-time method is more heavily regularized.

Under these assumptions, $\hat R_t$ is unbiased, so its error can be attributed only to variance.
Because $R_t(\theta)=S(t)^\top\theta$ is linear in $\theta$, estimating $\mathrm{Cov}(\hat\theta)$ suffices to produce confidence intervals for $R_t$ via \smash{$\widehat{\mathrm{Var}}(\hat R_t) =
   S(t)^\top\,\widehat{\mathrm{Cov}}(\hat\theta)\,S(t)$}.
In \cref{apx:rt-uncertainty}, we derive $\widehat{\mathrm{Cov}}(\hat\theta)$ as a sandwich estimator for the penalized Poisson score equations of \eqref{eq:glm-retro}.
DoW effects, quasi-Poisson overdispersion, and plug-in asymptotics also preserve this structure with minor modifications.
Invoking the central limit theorem, we thus produce pointwise Wald confidence intervals,
\begin{equation}\label{eq:wald}
    \text{CI}(t,\alpha)=\hat R_t \;\pm\; z_{1-\alpha/2}\,\mathrm{SE}(\hat R_t).
\end{equation}
Simultaneous coverage bands, which guarantee all timesteps are covered with high probability, may also be attained.
To do so, $z_{1-\alpha/2}$ is replaced with a simulation-based critical value that accounts for the correlation among adjacent $\hat R_t$.


The real-time method imposes strong tail regularization on $R_t$ in order to stabilize the final predictions.
As a result, the variance-only intervals \eqref{eq:wald} are often too narrow near the boundary.
To address this, we borrow techniques from conformal inference to obtain real-time confidence intervals $\text{CI}(t,\alpha)$.
\cref{apx:rt-realtime-ci} details our approach, as well as less effective alternatives.
These uncertainty bands are not necessarily calibrated, since their assumption of exchangeable data does not hold for time series, and they rely on a surrogate ground truth.
Even so, empirical results demonstrate their practical utility, especially relative to the Wald baseline.

\subsection{Existing Methods and Comparison}\label{sec:method-contrasts}

%
%
%

We benchmark ConvRt against six widely used $R_t$ estimators, all with open-source \texttt{R} implementations.
They span the major structural choices in the contemporary literature, summarized in \cref{tab:method-contrast}.
The central axis of variation is how each method handles the gap between latent infections and observed reports---the gap that motivates our two-stage deconvolution.

Most baselines do not model latent infections at all, instead applying the renewal equation directly to observed reports.
\textbf{EpiEstim} \citep{cori2013new}, the most widely deployed $R_t$ tool in practice \citep{nash2023real}, slides a gamma--Poisson conjugate update along the renewal equation, treating reports as infections.
\textbf{EpiLPS} \citep{gressani2022epilps} parameterizes log-$R_t$ as a penalized B-spline and fits the renewal-equation likelihood by Laplace approximation.
\textbf{Rtestim} \citep{rtestim} applies trend filtering to the same renewal-equation likelihood, with confidence bands from a quadratic relaxation.
Both rtestim and our method are frequentist penalized-likelihood estimators producing piecewise-polynomial $R_t$.
However, it does not model latent infections, and its $\ell_1$ penalty allows locally sharper changes than our globally smooth fit.

Three baselines do model latent infections.
\textbf{EpiNow2} \citep{abbott2020estimating}, the basis for the CDC's published real-time influenza $R_t$ estimates \citep{cdccfa2026rt}, is a fully Bayesian generative model.
It places a Gaussian-process prior on log-$R_t$, links $R_t$ to latent infections through the renewal equation, and maps infections to observed counts via a reporting-delay convolution---the same structure as our approach, but fit with Hamiltonian Monte Carlo in Stan.
\textbf{EstimateR} \citep{scire2023estimateR}, developed for Swiss COVID-19 surveillance, takes a lighter approach: it smooths observed counts with LOESS, deconvolves infections with the Richardson--Lucy algorithm, and passes them to EpiEstim.

The third, \textbf{CovidEstim} \citep{Chitwood2022}, is a Bayesian generative model specific to COVID-19.
It models log-$R_t$ as a random walk and fits jointly to case and death series by Hamiltonian Monte Carlo, learning time-varying ascertainment rates from seroprevalence data.
We include it only as a reference on real COVID-19 data (\cref{sec:rt-experiments-real}), using its publicly released outputs rather than refitting it.

\begin{table}[!h]
\centering
\small
\caption{Structural comparison of $R_t$ estimators benchmarked in this
paper. ``Latent infections'' indicates
how the method recovers latent infections from observed reports, if at
all. ``$R_t$ form'' is the functional class for $R_t$ (or log-$R_t$,
where applicable). All methods are run on both retrospective and weekly
real-time vintages except CovidEstim, for which we use the project's
publicly released outputs; configurations are detailed in
\cref{apx:methods-config}.}
\label{tab:method-contrast}
\begin{tabular}{lccc}
\toprule
Method      & Latent infections & $R_t$ form        & Inference   \\
\midrule
\textbf{ConvRt (ours)} & Deconvolution     & Spline             & Frequentist \\
EpiNow2     & MCMC              & Gaussian process   & Bayesian    \\
EpiEstim    & --                & Sliding window     & Bayesian    \\
estimateR   & Deconvolution     & Sliding window     & Bayesian    \\
EpiLPS      & --                & Spline             & Bayesian    \\
rtestim     & --                & Trend filter       & Frequentist \\
CovidEstim  & MCMC              & Random walk        & Bayesian    \\
\bottomrule
\end{tabular}
\end{table}

\section{Experimental Results}\label{sec:experiments}

We evaluate ConvRt against the estimators described in \cref{sec:method-contrasts} on synthetic benchmarks with known ground-truth $R_t$ (\cref{sec:sim-study}) and on real surveillance data (\cref{sec:rt-experiments-real}).
Throughout, we ran ConvRt with knots spaced evenly every 5 days,
tuned $\lambda$ via $K=5$-fold cross-validation with the min rule,
and, in the real-time setting, tuned $\gamma$ with the min rule on 7-day forward-validation.

\subsection{Synthetic-Data Experiments}\label{sec:sim-study}

We benchmark ConvRt against the five methods detailed in \cref{sec:method-contrasts}---EpiEstim, estimateR, EpiNow2, EpiLPS, and rtestim---on six synthetic datasets for which the true $R_t$ is known.
CovidEstim was omitted due to its reliance on seroprevalence data.

\subsubsection*{Data}

Four of our synthetic $R_t$ benchmarks were introduced in \citet{rtestim}.
They pose unique estimation challenges and mimic some real-world phenomena.
The first three define piecewise $R_t$ curves, where the jump discontinuities
could represent sweeping policy changes like school closures or lockdowns.
The fourth is sinusoidal, modeling multiple smooth yet rapid $R_t$ shifts.
Our experiments on these benchmarks only evaluated methods' retrospective predictions, matching rtestim.

The other two simulated datasets are based on real seasonal influenza data (see \cref{fig:flu-sim-both} in \cref{apx:flu-sim}).
We simulated hospitalizations using the delay distributions described in \cref{sec:rt-experiments-real} and a 1.5\% severity rate \citep{reed2015estimating}.
For the first dataset's ground truth, we used $R_t$ predictions on the 2022/23 season, made by the method from \citet{wallinga_teunis}.
For the second dataset, we heavily smoothed the $R_t$ curve with LOESS.
We also added a higher degree of noise to mimic real-world data.

\subsubsection*{Evaluation}

We compare the predictions of each method over a relevant evaluation window.
For retrospective experiments, this window excludes a burn-in and burn-out period at the boundaries, where methods have insufficient data to estimate $R_t$ reliably.
For real-time experiments, it covers the week leading up to the current vintage date.

We evaluate point estimates with mean absolute error (MAE). 
Given a true curve $R_t$ and estimates $\hat R_t$ over a time window $\mathcal{T}$, 
\[
\mathrm{MAE} \;=\; \frac{1}{|\mathcal{T}|}\sum_{t\in\mathcal{T}} \bigl|R_t - \hat R_t\bigr|.
\]

To evaluate uncertainty bands, we consider their proximity to the targeted coverage level.
For a nominal central level $\alpha \in (0,1)$, let $\hat c(\alpha)$ denote
the empirical coverage of the corresponding two-sided prediction interval:
\[
\hat c(\alpha)=\frac{1}{|\mathcal{T}|}\sum_{t\in\mathcal{T}}1\{R_t\in \text{CI}(t,\alpha)\}.
\]
We define calibration error
\[
\mathrm{CE} \;=\; \frac{1}{|\mathcal{A}|}\sum_{\alpha \in \mathcal{A}} \bigl|\hat c(\alpha) - \alpha\bigr|,
\qquad \mathcal{A} = \{0.5,\,0.6,\,0.7,\,0.8,\,0.9,\,0.95\},
\]
which averages the absolute gap between empirical and nominal coverage over a grid of central levels.
While frequentist confidence intervals and Bayesian credible intervals have different interpretations, we score them with the same coverage criterion.

\subsubsection*{Retrospective Results}

\Cref{tab:combined-retro} reports retrospective performance across all six benchmarks.
ConvRt is highly accurate: on the three datasets without jump discontinuities, ConvRt's MAE is extremely low, below 0.01. 
Its worst MAE, on the piecewise linear dataset, is still below 0.04.

\begin{table}[ht]
\centering
\caption{Retrospective performance across simulation settings. Mean absolute error (MAE)
and $\ell_1$ calibration error (CE) reported in units of $10^{-2}$; runtime is for a single
fit on the full input series. Influenza benchmarks run from October 1 to January 31.
The vertical line separates methods that model latent infections (left) from those that do not (right).}
\label{tab:combined-retro}
\setlength{\tabcolsep}{5pt}
\begin{tabular}{l l ccc|ccc}
\toprule
& & ConvRt & EpiNow2 & estimateR & EpiEstim & EpiLPS & rtestim \\
\midrule
\multicolumn{8}{l}{\textbf{\textit{rtestim benchmarks}}} \\
\hline
\multirow{3}{2.2cm}{Piecewise Constant}    & MAE     & 1.58    & 2.64     & 1.93    & 4.86    & 3.96    & 3.37    \\
                                           & CE      & 9.75    & 6.89     & 19.69   & 11.84   & 15.08   & 20.03   \\
                                           & Runtime & 1.25\,s & 117.8\,m & 0.18\,s & 0.03\,s & 0.07\,s & 22.9\,s \\
\multirow{3}{2.2cm}{Piecewise Exponential} & MAE     & 1.77    & 2.14     & 2.42    & 11.78   & 8.45    & 8.60    \\
                                           & CE      & 3.35    & 3.44     & 12.29   & 73.89   & 21.34   & 22.73   \\
                                           & Runtime & 1.21\,s & 45.1\,m  & 0.18\,s & 0.03\,s & 0.07\,s & 22.9\,s \\
\multirow{3}{2.2cm}{Piecewise Linear}      & MAE     & 3.79    & 4.45     & 5.20    & 17.25   & 12.54   & 12.10   \\
                                           & CE      & 11.43   & 5.61     & 10.37   & 74.17   & 9.29    & 11.19   \\
                                           & Runtime & 1.18\,s & 130.5\,m & 0.18\,s & 0.03\,s & 0.07\,s & 22.9\,s \\
\multirow{3}{2.2cm}{Periodic}              & MAE     & 0.62    & 1.95     & 3.20    & 29.91   & 21.27   & 21.43   \\
                                           & CE      & 3.86    & 2.86     & 18.59   & 71.89   & 58.67   & 24.51   \\
                                           & Runtime & 1.26\,s & 38.1\,m  & 0.18\,s & 0.03\,s & 0.07\,s & 22.9\,s \\
\hline
\multicolumn{8}{l}{\textbf{\textit{Influenza benchmarks}}} \\
\hline
\multirow{3}{2.2cm}{Wiggly $R_t$} & MAE     & 0.84    & 0.83    & 1.74    & 7.40    & 4.64    & 4.46 \\
                                  & CE      & 7.36    & 11.47   & 25.66   & 52.49   & 66.71   & 22.58 \\
                                  & Runtime & 3.25\,s & 35.6\,m & 0.06\,s & 0.04\,s & 0.20\,s & 0.69\,s \\
\multirow{3}{2.2cm}{Smooth $R_t$} & MAE     & 0.62    & 0.82    & 1.18    & 3.45    & 2.01    & 2.18 \\
                                  & CE      & 8.22    & 19.83& 3.04    & 32.57   & 51.13   & 17.84 \\
                                  & Runtime & 2.38\,s & 46.1\,m & 0.07\,s & 0.04\,s & 0.17\,s & 0.61\,s \\
\bottomrule
\end{tabular}
\end{table}

\begin{figure}[ht]
    \centering
    \includegraphics[width=.95\linewidth]{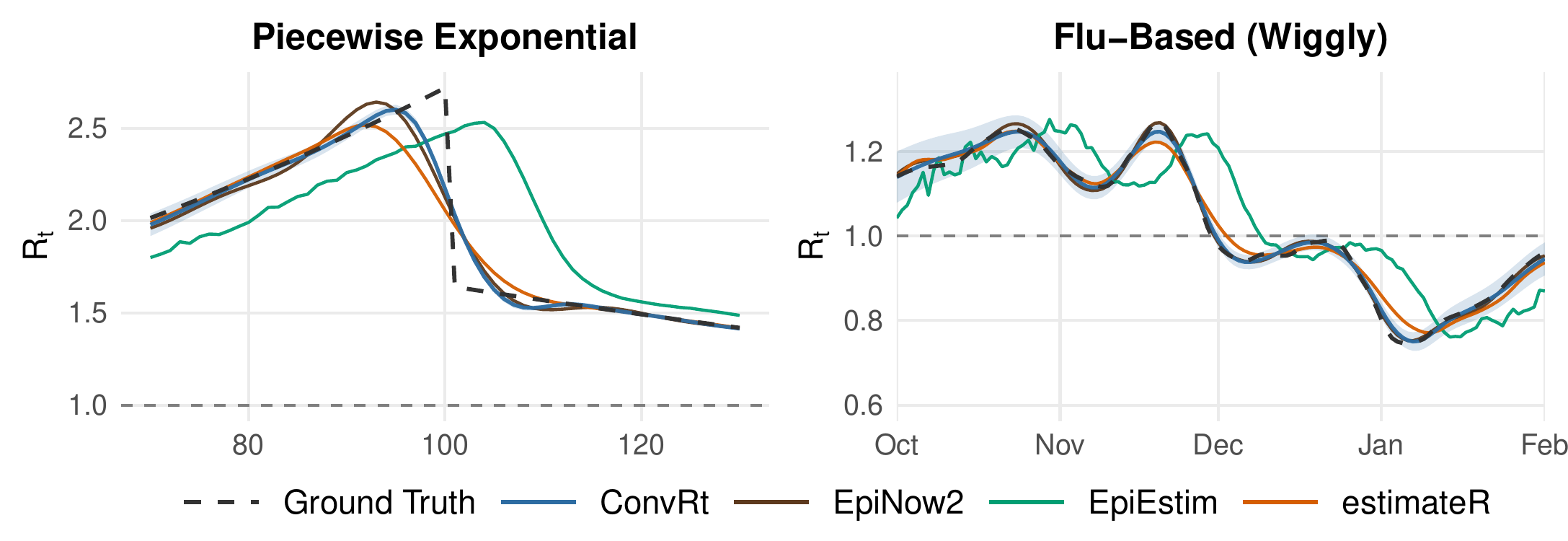}
    \caption{Comparison of retrospective $R_t$ estimates on benchmark datasets.}
    \label{fig:retro-comparison}
  \end{figure}

We organize our comparison around three axes: point accuracy (MAE), uncertainty quantification (CE), and runtime.
We begin with EpiNow2, ConvRt's closest competitor, before turning to the others.
Relative to EpiNow2, ConvRt reduces MAE by 20\% or more on 5 of 6 datasets, and is effectively tied on the last.
In the best case---the periodic rtestim benchmark---ConvRt has nearly 70\% lower MAE than EpiNow2.
\Cref{fig:retro-comparison} compares predictions on selected rtestim and influenza datasets.
Both methods track the true $R_t$ curves closely.

ConvRt and EpiNow2 have similarly strong uncertainty quantification.
Each has lower CE in exactly half of the benchmarks.
Notably, ConvRt outperforms EpiNow2 in both influenza benchmarks, which are designed to be realistic.
In the third benchmark with no jumps (Periodic), EpiNow2 has lower CE, but ConvRt is still very strong.
Its low CE, under 0.04, indicates confidence intervals are well-calibrated. 
Examining more precisely, we see ConvRt tends to produce wider, more conservative intervals, while EpiNow2's are sometimes too narrow.

ConvRt is also several orders of magnitude faster than EpiNow2, running in seconds versus up to 2.5 hours.
Since a single fit is rarely the end of the analysis, this gap compounds: refitting under many hyperparameter settings multiplies EpiNow2's cost while leaving ConvRt's negligible.
Doing so may be necessary for EpiNow2, which frequently failed to converge on these benchmarks. 

Among methods that run in seconds, ConvRt outperforms the baselines in terms of both MAE and CE.
The methods that do not model latent infections track $R_t$ at a delay, and thus have much higher MAE and CE. \cref{fig:retro-comparison} omits EpiLPS and rtestim, which have similar bias to EpiEstim but are much less widely used.
In contrast, estimateR fares better because it recovers latent infections.
However, it uses imprecise statistical models for infections and $R_t$. 
Consequently, ConvRt has over 20\% lower MAE than estimateR on all datasets, and over 50\% on half.

ConvRt's main weakness is at the jump discontinuity, where its point estimates are too smooth and its confidence intervals fail to cover (\cref{fig:retro-comparison}).
Such sharp gaps are unlikely in practice, as interventions may roll out gradually rather than instantaneously.
Moreover, this has an easy fix.
To handle jumps in $R_t$, one option is to augment its basis in ConvRt
with a step function that switches on at the intervention day.
Performance then improves dramatically, with MAE falling well below 0.01.
(\cref{apx:rtestim}).

\subsubsection*{Real-time Results}

\Cref{tab:combined-realtime} reports real-time performance on the flu benchmark.
We refit weekly, with vintage end dates from October through January, and evaluate the last seven days of each.
ConvRt's confidence intervals use the conformalized approach of \cref{apx:rt-realtime-ci}.
Predictions are shown in \cref{apx:flu-sim}.

  \begin{table}[ht]
  \centering
  \caption{Real-time performance on the influenza simulations: last-7-day estimates
  across 18 weekly vintages (week-ending dates through January~31). MAE and $\ell_1$
  calibration error (CE) in units of $10^{-2}$; runtime is mean wall-clock time per
  vintage. The vertical line differentiates whether methods model latent infections.}
  \label{tab:combined-realtime}
  \setlength{\tabcolsep}{5pt}
  \begin{tabular}{ll ccc!{\vrule}ccc}
  \toprule
  & & ConvRt & EpiNow2 & estimateR & EpiEstim & EpiLPS & rtestim \\
  \midrule
  \multirow{3}{*}{Flu (wiggly)} & MAE     & 4.14   & 4.29    & 7.17    & 7.16    & 6.45    & 6.70 \\
                          & CE      & 9.09   & 17.16   & 56.71   & 52.08   & 57.90   & 25.44 \\
                          & Runtime & 2.35 s & 21.3 m  & 1.36 s  & 0.01 s  & 0.11 s  & 0.44 s \\
  \multirow{3}{*}{Flu (smooth)} & MAE     & 2.51   & 2.14    & 4.24    & 3.67    & 2.98    & 5.09 \\
                          & CE      & 2.74   & 5.25& 51.28   & 35.94   & 46.79   & 25.83 \\
                          & Runtime & 2.79 s & 20.5 m  & 1.36 s  & 0.02 s  & 0.12 s  & 0.42 s \\
  \bottomrule
  \end{tabular}
  \end{table}

The retrospective picture carries over to the real-time case.
Once again, ConvRt and EpiNow2 are the most accurate methods.
Among methods that run in seconds, ConvRt wins on both metrics. 
The gap is widest on calibration, where every fast baseline has roughly $5$--$10\times$ its CE.

On the realistic ``Wiggly'' benchmark, ConvRt has slightly lower MAE than EpiNow2.
\cref{fig:realtime-rt-wiggly} shows they share most predictions, but ConvRt better recognizes changes in real time.
For example, it more faithfully tracks the rise and fall in $R_t$ around its November peak. 
While EpiNow2 wins on the smooth dataset, differences are very small, and ConvRt is more accurate on several weeks (\cref{fig:rt-noisy-smooth}).

Beyond accuracy, ConvRt has much better uncertainty bands than EpiNow2.
Their calibration error is nearly 50\% lower than EpiNow2's on both datasets.
ConvRt is again much faster, here by a factor of roughly $500\times$.
In addition to making the conformalized bands practical, ConvRt's runtime aids a further advantage, highlighted in \cref{sec:scenario}: its capacity for real-time scenario analysis.


\subsection{Real-Data Experiments}\label{sec:rt-experiments-real}

Next, we demonstrate ConvRt's utility on real-world influenza data.
To run ConvRt, we tuned $\lambda$ with the 1se rule after noting that the min rule tends to undersmooth on noisy data.

\subsubsection*{Data}

The CDC posts flu $R_t$ in real time, estimated by EpiNow2 \citep{cdccfa2026rt,abbott2020estimating}.
We use public hospitalization counts reported to HHS through the National Healthcare Safety Network (NHSN).
NHSN published inpatient hospitalizations for a variety of pathogens
from the beginning of the COVID-19 pandemic until May 2024.
We focus on the 2022/23 and 2023/24 flu seasons (\cref{fig:retro-flu-both}) because the pandemic flu seasons were considerably smaller.
We conducted a similar analysis on COVID-19 data, discussed further in \cref{apx:covidestim}.

\subsubsection*{Analysis}

In each season, we ran ConvRt and EpiNow2 (the current leading method)  from July 1 through January 31. We parameterized all delay distributions as discrete gamma,
choosing means and standard deviations to align with the literature.
For the infection-to-report delay $\pi$, this was 5.7 and 2.3 days, respectively \citep{rousogianni2025clinical,noh2014viral}.
The generation interval had mean 3.2 and 1.6 \citep{chan2025estimating,cauchemez2009household,cowling2009estimation}.

\subsubsection*{Retrospective Results}

\begin{figure}[!h]
    \centering
    \includegraphics[width=\linewidth]{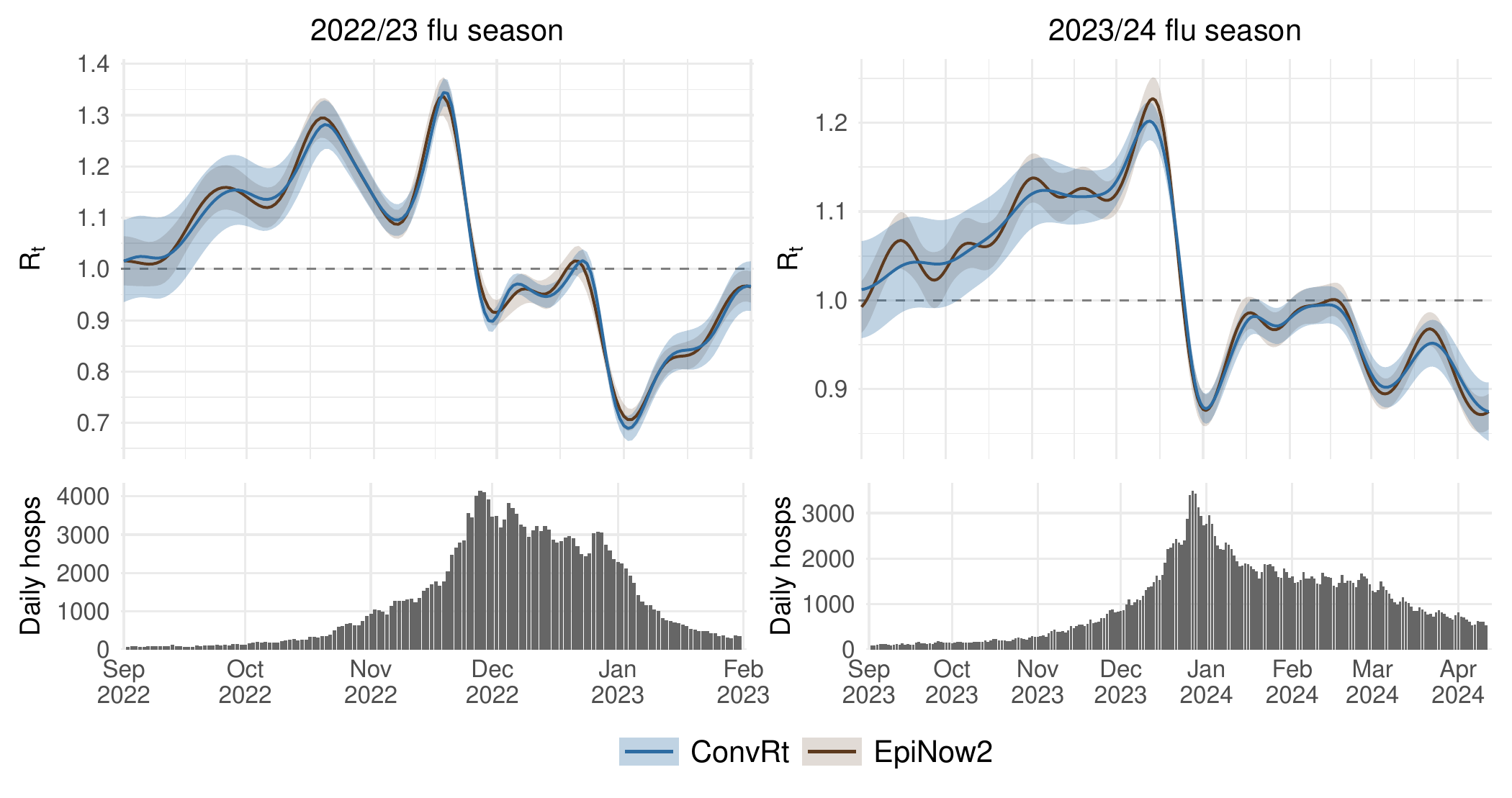}
    \caption{$R_t$ predictions for seasonal influenza. ConvRt and EpiNow2 fit retrospectively on NHSN hospitalization data, with 95\% pointwise uncertainty bands. 2023/24 was a longer flu season than 2022/23, so its plotting window is extended.}
    \label{fig:retro-flu-both}
\end{figure}

Both ConvRt and EpiNow2 generate similar $R_t$ curves when fit retrospectively at the end of each flu season (\cref{fig:retro-flu-both}).
The main difference is that ConvRt predicts a smoother rise in fall 2023---a more intuitive shape, though the true $R_t$ is unknown.
Quantitatively, the two are very close.
Both estimate $R_t$ near $1.3$ in 2022 and $1.2$ in 2023, with discrepancies around $0.02$ at the peaks.
This matches a CDC analysis that found a median peak $R_t$ of $1.28$ for seasonal influenza \citep{biggerstaff2014estimates}.

To fit each season, ConvRt runs in 3 seconds, while EpiNow2 takes 38 minutes.
While evaluating uncertainty bands is complicated by the fact that the true $R_t$ is unknown,
it is worth noting that the two methods' intervals are broadly concordant. 
ConvRt's confidence intervals nearly always cover EpiNow2's point estimates.

\subsubsection*{Real-time Results}

\begin{figure}[!h]
    \centering
    \includegraphics[width=\linewidth]{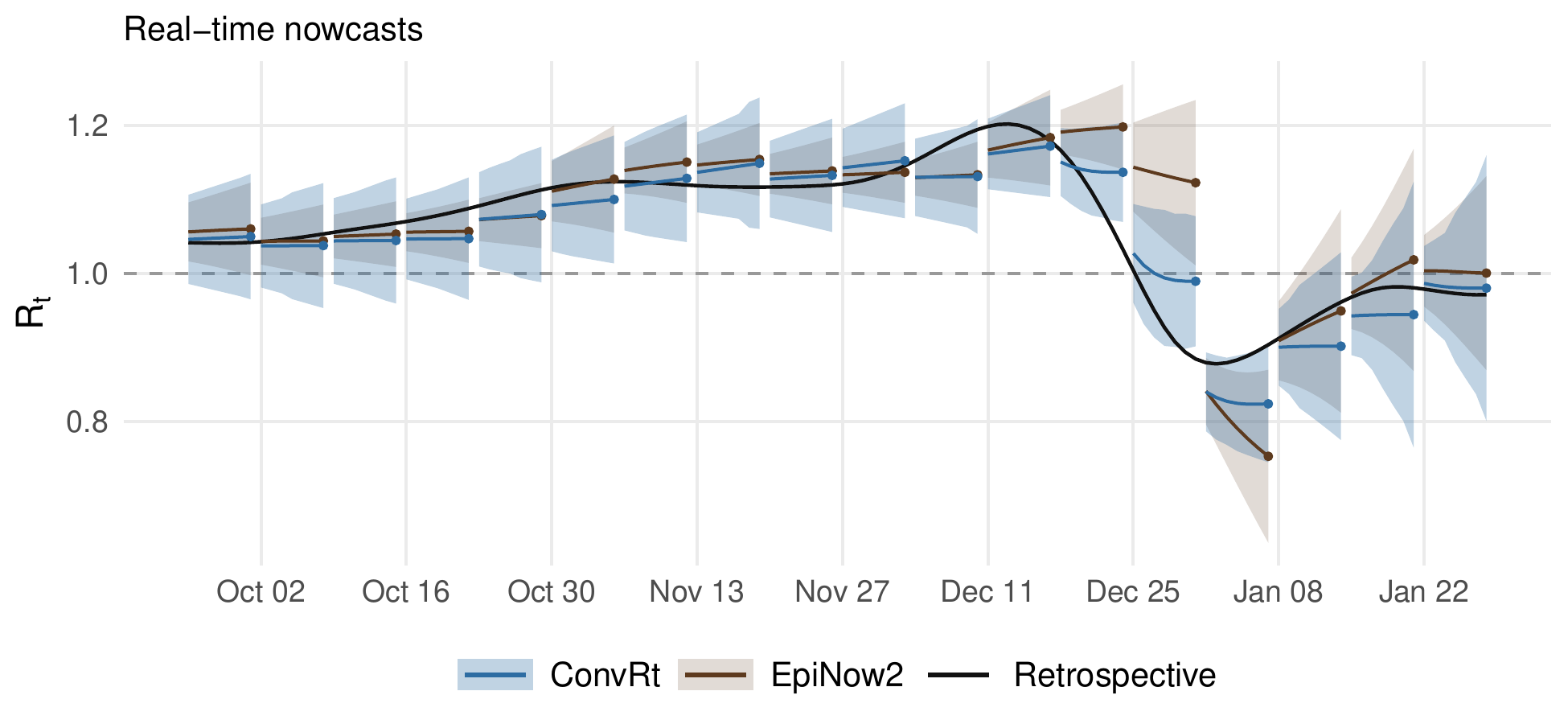}
    \caption{Real-time $R_t$ estimates and uncertainty bands for seasonal influenza in 2023/24.}
    \label{fig:UQ-ConvRt}
\end{figure}

\cref{fig:UQ-ConvRt} shows each method's weekly real-time nowcast against the
finalized end-of-season fit.
Both closely track rising $R_t$ from October through mid-December.
The retrospective fit suggests $R_t$ then declines sharply, before reversing course and leveling out.
In late December, ConvRt predicts the decline more accurately than EpiNow2:
on December 31, when the retrospective $R_t$ is about 0.9, the real-time point estimates are 1.0 and 1.1, respectively.
The following week, as $R_t$ begins to trend upwards, EpiNow2 overcorrects and falls below 0.8;
ConvRt is flat, and has about half the error.


The two methods also differ in uncertainty quantification.
ConvRt's bands
cover the retrospective ConvRt curve 96\% of the time, close to the 95\%
nominal level.
EpiNow2's credible intervals cover its retrospective fit only
76\% of the time.

\section{Policy Analysis}\label{sec:scenario}

\begin{figure}[h]
    \centering
    \includegraphics[width=\linewidth]{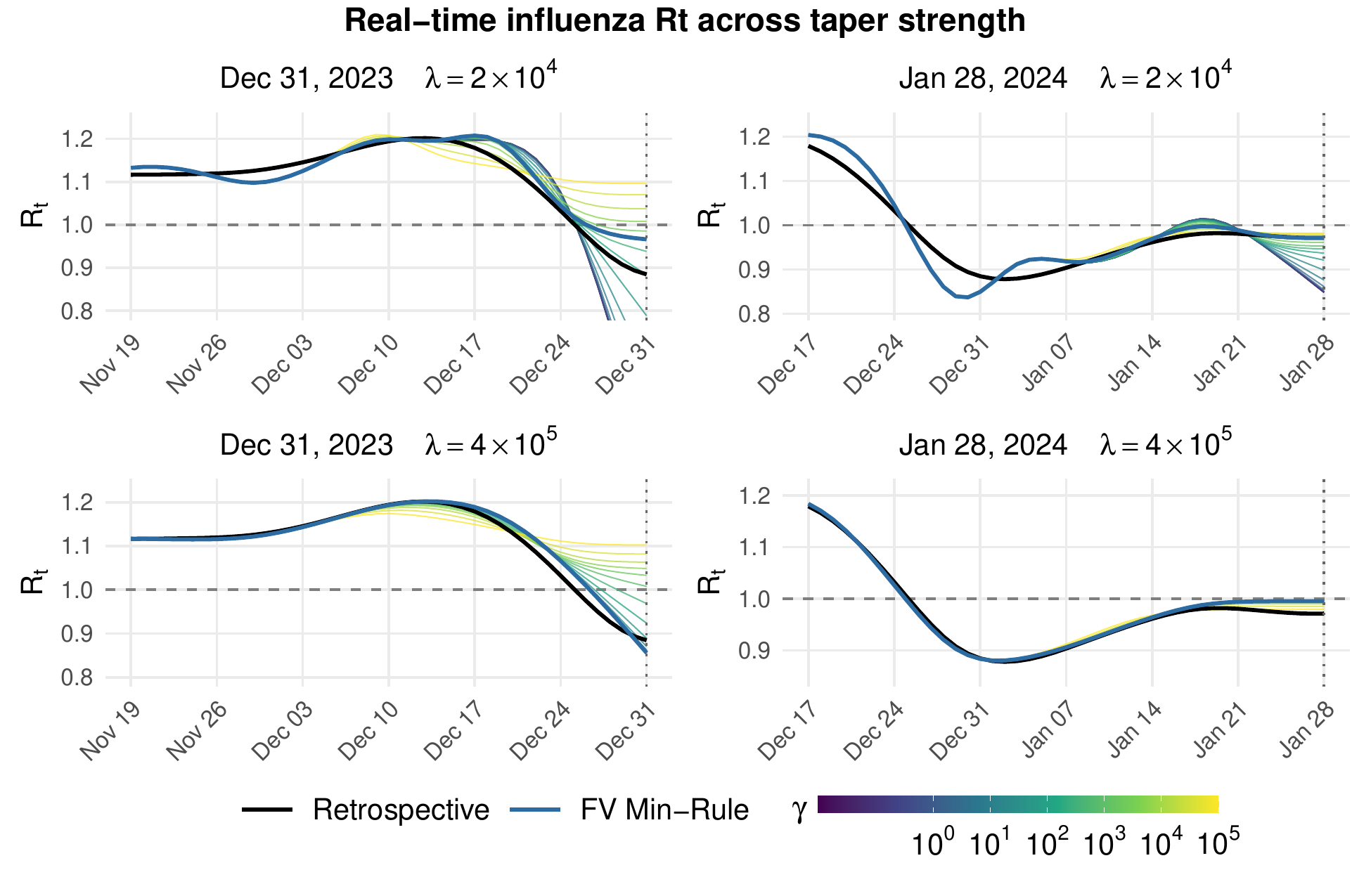}
    \caption{$R_t$ predictions at varying levels of the smoothness hyperparameters $\lambda$ (overall) and $\gamma$ (tail).}
    \label{fig:rt-by-gamma}
\end{figure}

Two knobs govern the boundary predictions in real time.
$\lambda$ controls the overall level of smoothness, while $\gamma$ controls extrapolation at the tail.
Together, these hyperparameters decompose the boundary problem along the axes that matter operationally: how wiggly $R_t$ is allowed to be overall, and what shape the extrapolation takes.

Cross-validation provides a sensible default for $\lambda$ and $\gamma$, but two factors limit its authority in real time. 
First, at the boundary these cannot be well-estimated from data. 
Infections at any interior date $t$ eventually surface in the observed counts, giving enough signal to select a value of $\lambda$ where bias is small.
This is no longer true on recent dates. 
As discussed in \cref{sec:rt-rt}, the right degree of tail regularization $\gamma$ is hard to pin down from the data alone, as it depends on beliefs about future epidemic growth.

Second, estimates corresponding to hyperparameters with the lowest prediction error may not be the most useful for guiding policy decisions. 
The stakes of bad decisions are asymmetric: underestimating a surge is more costly than overestimating one, which may justify biasing toward smaller $\gamma$ when $R_t$ appears to be rising.
A cautious practitioner might make decisions based on the worst case, while another would emphasize the likeliest situation.

As a result, policymakers may analyze $R_t$ predictions across many hyperparameters, especially $\gamma$.
The range of possibilities may be informative in its own right.
Predictions may congregate across reasonable $(\lambda,\gamma)$, instilling confidence in conclusions.
Alternatively, they could diverge significantly, suggesting the data support competing narratives.  In that case, the best policy path depends on the costs and consequences of acting on false positive and false negative signals.
This provides a valuable notion of uncertainty that confidence intervals for a single $(\lambda,\gamma)$ do not capture.

As a simple example, \cref{fig:rt-by-gamma} illustrates these considerations, showing $R_t$ predictions across $\lambda$ and $\gamma$.  (See \cref{fig:real-flu-lambda-realtime} for similar estimates in mid-January, as well as predictions by $\lambda$ across the whole season.) The left subplots depict a sharp decline in $R_t$ near the estimation date of December 31, 2023.
Small $\lambda$ predicts this decline but overshoots dramatically at small $\gamma$, a problem that is remedied with larger $\lambda$.
In both cases, large $\gamma$ misses the decline almost entirely, since it encourages stationarity.
Fortunately, tuning $\gamma$ by forward-validation here happens to identify that $R_t$ dips below 1. 
The right subplots show predictions a month later, at January 28, 2024.
On this date, $R_t$ estimates generally span a tighter range across $\gamma$.
With larger $\lambda$---the cross-validated value, which fits the true curve well---predictions at all $\gamma$ are flat, just shy of 1.

By the end of January, an analyst might conclude that the epidemic is likely to decline slowly in the near future.
The contrast with December 31 is instructive: there, $R_t$ at high $\gamma$ was still elevated, so an analyst wary of underestimating a surge may have maintained low cost, high value measures like masking or intensified air purification in nursing homes.
By January 28, the tighter range removes ambiguity, and the same analyst can transition policies more confidently to reflect a declining epidemic.

\section{Discussion}\label{sec:discussion}

\subsection{Summary}

We developed ConvRt, a frequentist estimator for the effective reproduction number $R_t$.
It deconvolves latent infections from observed counts, then estimates $R_t$ via penalized quasi-Poisson regression with a spline basis.
Tapered penalties and tail structure stabilize real-time predictions.

We compared ConvRt to leading methods to compute $R_t$. In simulations, we showed ConvRt was the most accurate method across benchmarks.
It reduced retrospective MAE by at least 20\% relative to EpiNow2 on most datasets, and the two performed comparably in real time.
Against the remaining estimators, ConvRt's advantage was even larger, often cutting MAE in half.
Furthermore, EpiNow2 and other accurate Bayesian methods \citep{Chitwood2022, bosse2024bayesian} rely on MCMC sampling that can take tens of minutes to hours per fit.
In contrast, ConvRt fits in seconds.
This gap has practical consequences for public health agencies, which refit $R_t$ daily across many jurisdictions. 
Lastly, we show that ConvRt achieves better coverage than every fast baseline.
Its uncertainty bands are comparable to EpiNow2's in retrospect, and are better calibrated in real time.


More broadly, ConvRt decouples two jobs that Bayesian priors conflate: regularizing the historical estimate and expressing assumptions about future transmission.
This distinction matters most in real time, where historical smoothness and tail extrapolation require genuinely different judgments.
The hyperparameter $\lambda$ controls the total smoothness, while $\gamma$ controls how aggressively recent trends are extrapolated.
We provide strategies to tune these by cross-validation, and highlight how practitioners can choose them based on their own assumptions and priorities.
The low cost of refitting lets practitioners sweep over $(\lambda, \gamma)$ to characterize uncertainty for real-time decisions.


\subsection{Extensions}

We implemented two extensions of ConvRt, detailed in \cref{apx:extra-methods}, that are not benchmarked comprehensively in this paper.

The first handles weekly data.
Daily surveillance data were more widely available during the pandemic era, but many systems report only at weekly resolution.
FluSurv-NET has published weekly flu hospitalizations since 2005, and CDC's COVID-NET and RSV-NET follow the same cadence.
NHSN, the source used in our experiments, itself ceased daily reporting in May 2024.
We adjust the likelihood by aggregating daily means within each week, preserving the underlying daily transmission dynamics.
On the simulated flu benchmark, retrospective estimates from weekly data are nearly identical to those from daily data (\cref{apx:weekly}).
Real-time predictions are also encouraging, with forward-validation correctly selecting a tail penalty that stabilizes the final week's estimate.

The second extension replaces the spline basis with trend filtering regularization.
Unlike splines, which place knots at fixed locations, trend filtering selects knot locations adaptively with an $\ell_1$ penalty.
As a result, it may be better at capturing abrupt shifts in $R_t$, such as those around interventions. 
We default to splines because valid inference for trend filtering remains a challenging problem: the $\ell_1$ penalty precludes the closed-form hat matrix needed for our sandwich variance estimator.

In \cref{apx:extra-methods}, we run ConvRt with a fourth-difference trend filtering penalty on the 2022/23 flu season.
The resulting $R_t$ estimates are very similar to the spline's.
A rigorous comparison across datasets is left to future work.
Trend filtering may also be a good choice for the initial step of deconvolving latent infections \citep{Jahja2022}, where we currently use splines but do not rely on their uncertainty bands.

\subsection{Limitations and Future Directions}

We highlight limitations of the current framework and directions for extending it, organized around three themes: data challenges and opportunities, modeling adjustments, as well as open problems in uncertainty quantification.
Developments in $R_t$ estimation could also be applied to other metrics, such as the case-fatality rate \citep{goldwasser2025deconvolution}.


First, real-time counts are typically incomplete at recent dates due to reporting delays and are revised upward as backfill accumulates.
Our evaluations used finalized counts, so real-time performance may degrade when paired with preliminary data subject to backfill.
ConvRt takes observed counts as given; in practice, real-time estimation must be paired with a nowcast \citep{mcgough2020nowcasting, abbott2020estimating}, and
how uncertainty in the nowcast propagates into $R_t$ estimates merits further study.

ConvRt can also generalize to other data sources, such as wastewater viral concentrations.
More generally, multiple data streams---cases, hospitalizations, deaths, wastewater---could be fused by summing their log-likelihoods against the same latent infection curve, similar to
CovidEstim \citep{Chitwood2022}.
The GLM structure also naturally accommodates covariates in the linear predictor for $R_t$, such as mobility indices or vaccination rates.


Second, ConvRt could be adjusted beyond its current specification.
For instance, the tail regularization framework may merit further investigation.
The tapered weights were taken from \citet{Jahja2022}, but other weight functions have not been explored.

More ambitious approaches could learn infections and $R_t$ jointly, rather than in separate stages.
The E-M algorithm may be applicable; it was used by \citet{li2023reconstructing} to estimate latent infections $x$ along with the delay distribution $\pi$.
An alternative approach would model $x_t$ and $R_t$ on the log scale, as well as DoW effects $\omega_{(t \bmod 7)}$.
The unknown parameters then add in the exponent of \cref{eq:mean-cases}, giving
$$
\mu_t = \rho_t \sum_{s<t} \sum_{u<s} \exp\bigl(\underbrace{S(s)^\top \theta_R}_{\log R_s} + \underbrace{S_x(u)^\top \theta_x}_{\log x_u}  + \omega_{(t \bmod 7)}\bigr)\, g_{s-u}\, \pi_{t-s}.
$$
One would then solve jointly for $\theta_R$ and $\theta_x$, regularizing both at once with smoothness penalties.

Third, we may explore other approaches for uncertainty quantification.
While we constructed real-time confidence intervals with a conformal approach (\cref{apx:rt-realtime-ci}),
future work could draw on the online inference literature. 
Quantile tracking methods \citep{angelopoulos2024conformal, ding2026calibrated} maintain adaptive prediction intervals that respond to distribution shift as new data arrive.




In its current formulation, ConvRt does not propagate uncertainty from plug-in quantities into $R_t$ inference.
There are several such quantities, including the deconvolved infections, reporting delay distribution, and generation interval.
We treat these as fixed when fitting $R_t$, so the reported confidence intervals may understate the true uncertainty. 
Our empirical results suggest this omission may not be severe: ConvRt already achieves competitive coverage without accounting for errors in the estimated latent infections $\hat x_t$.
Further work may address this via the parametric bootstrap, resampling from estimated distributions and refitting $R_t$ on each draw.

Despite these limitations, ConvRt fills a practical gap in the $R_t$ estimation toolkit: a method that is fast enough for operational use across many jurisdictions, accurate enough to match or exceed Bayesian alternatives, and transparent enough that practitioners can inspect and tune its assumptions. We release ConvRt as an open-source R package to support reproduction number tracking in future outbreaks.

\section*{Software}
The \texttt{ConvRt} R package is available at \url{https://github.com/jeremy-goldwasser/ConvRt}. Code to reproduce all analyses in this paper is available at \url{https://github.com/jeremy-goldwasser/Convolutional-Rt}.

\bibliographystyle{plainnat}
\bibliography{refs}

\appendix
\section{Optimization}\label[appendix]{apx:optim}

The objective in \eqref{eq:glm-retro} is a penalized Poisson log-likelihood with identity link $\Lambda_t(\theta)=Z_t^\top \theta$ \eqref{eq:linear-mean-cases}. This is concave in $\theta$ wherever the fitted means satisfy $\Lambda_t(\theta) > 0$.                                 
(For now, we ignore the multiplicative terms $\omega$ and $\rho$.)

Iteratively reweighted least squares (IRLS) is the standard way to optimize a GLM: each Newton step on the log-likelihood reduces algebraically to a weighted least squares problem, hence the name.
IRLS involves the score and Fisher information, which we define next.

\subsection{Score, Hessian, and Fisher information.}                                                                                                                

Consider an unregularized, identity-link Poisson GLM. Its log-likelihood contribution at time $t$ is
\[
\ell_t(\theta) \;=\; y_t\,\log \Lambda_t(\theta) \,-\, \Lambda_t(\theta), \qquad \Lambda_t(\theta) \;=\; Z_t^\top\theta,
\]
with $Z_t$ the row of the convolutional design matrix in \eqref{eq:linear-mean-cases}. Differentiating yields the individual score function and its gradient,
\[
s_t(\theta) \;=\; \frac{\partial \ell_t}{\partial \theta} \;=\; \frac{y_t - \Lambda_t}{\Lambda_t}\,Z_t, \qquad
\frac{\partial s_t}{\partial \theta^\top} \;=\; -\,\frac{y_t}{\Lambda_t^2}\,Z_t Z_t^\top.
\]
We stack timesteps into a design matrix $Z \in \mathbb{R}^{T\times p}$ with rows $Z_t^\top$, and note that $\mathbb{E}[y_t] = \Lambda_t$ under Assumption 1. Summing the scores and using $\sum_t a_t Z_t = Z^\top a$ for any vector $a = (a_1,\dots,a_T)^\top$,
\[
s(\theta) \;=\; \sum_t \frac{y_t - \Lambda_t}{\Lambda_t}\,Z_t \;=\; Z^\top W(Y - \Lambda), \qquad W \;=\; \mathrm{diag}(1/\Lambda_t).
\]
Likewise, using $\sum_t w_t Z_t Z_t^\top = Z^\top \mathrm{diag}(w)\, Z$, we define the Fisher information as
\[
\mathcal{I}(\theta) \;=\; -\,\mathbb{E}\!\left[\sum_t \frac{\partial s_t(\theta)}{\partial \theta^\top}\right] \;=\; \sum_t \frac{Z_t Z_t^\top}{\Lambda_t} \;=\; Z^\top W Z.
\]

\subsection{IRLS}

Our method uses a ridge penalty $\lambda\,\theta^\top\Omega\,\theta$ to regularize for smoothness; in the real-time setting there is a second term $\gamma\,\theta^\top\Psi\,\theta$. Differentiating, the gradients are $2\lambda\Omega\theta$ and $2\gamma\Psi\theta$ and the Hessians are $2\lambda\Omega$ and $2\gamma\Psi$. Since $\lambda$ and $\gamma$ are selected by cross-validation, the factor of two is immaterial and we absorb it into the tuning parameters henceforth, writing the gradients as $\lambda\Omega\theta$, $\gamma\Psi\theta$ and the Hessians as $\lambda\Omega$, $\gamma\Psi$. These enter the gradient and Fisher information, respectively. Our notation adheres to the retrospective case in the following derivations (just $\lambda$), though $\gamma$ can simply be added for real-time.

The penalized objective \eqref{eq:glm-retro} adds the ridge term to the negative summed log-likelihood,
\[
f_\lambda(\theta) \;=\; -\sum_t \ell_t(\theta) \,+\, \tfrac{1}{2}\lambda\,\theta^\top\Omega\,\theta.
\]
This has penalized gradient $g_\lambda(\theta) = -Z^\top W(Y - \Lambda) + \lambda\Omega\,\theta$ and Fisher information $\mathcal{I}_\lambda(\theta) = Z^\top W Z + \lambda\Omega$, the penalty contributing its Hessian $\lambda\Omega$ to the latter. 
In general, Fisher scoring updates an iterate by subtracting the information-preconditioned gradient,
\[
\theta^{(k+1)} \;=\; \theta^{(k)} \,-\, \mathcal{I}_\lambda(\theta^{(k)})^{-1}\,g_\lambda(\theta^{(k)}),
\]
which replaces the Hessian in a Newton step with its expectation $\mathcal{I}_\lambda$, guaranteeing a positive-definite, descent-directed system. Substituting our gradient and information gives
\[
\theta^{(k+1)} \;=\; \theta^{(k)} \,-\, \bigl(Z^\top W^{(k)} Z + \lambda\Omega\bigr)^{-1}\!\bigl[-Z^\top W^{(k)}(Y - \Lambda^{(k)}) + \lambda\Omega\,\theta^{(k)}\bigr],
\]
with working weights $W^{(k)} = \mathrm{diag}\bigl(1/\Lambda_t(\theta^{(k)})\bigr)$ and $\Lambda^{(k)} = Z\theta^{(k)}$.
Because $f_\lambda$ is convex (with positive-definite information $\mathcal{I}_\lambda$ once $\lambda\Omega \succ 0$), any stationary point is the global minimizer, so Fisher scoring converges to the unique penalized MLE regardless of initialization.

To simplify, write $\theta^{(k)} = (Z^\top W^{(k)} Z + \lambda\Omega)^{-1}(Z^\top W^{(k)} Z + \lambda\Omega)\theta^{(k)}$ and fold it into the bracket. The $\lambda\Omega\,\theta^{(k)}$ terms cancel, leaving
\[
\theta^{(k+1)} \;=\; \bigl(Z^\top W^{(k)} Z + \lambda\Omega\bigr)^{-1} Z^\top W^{(k)}\bigl[Z\theta^{(k)} + (Y - \Lambda^{(k)})\bigr].
\]
The bracketed term is the working response; under the identity link $\Lambda^{(k)} = Z\theta^{(k)}$, so it collapses to $Y$ and we obtain the penalized WLS normal equations
\[
\bigl(Z^\top W^{(k)} Z + \lambda\Omega\bigr)\,\theta^{(k+1)} \;=\; Z^\top W^{(k)} Y,
\]
where $Y$ appears directly rather than through a canonical working response, because the link is identity. Each iterate is thus a ridge-penalized weighted least squares fit of $Y$ on $Z$ with Poisson variance weights, reweighted as $W^{(k)}$ updates. We iterate until the relative change in $\theta$ falls below a tolerance; convergence is fast in practice, typically within 10--20 iterations from a flat initialization.

One practical wrinkle is that unlike the log link, the identity link does not automatically keep $\Lambda_t$ positive. A full Newton step can overshoot into the infeasible region where the working weights blow up. We guard against this with step-halving: if the proposed update yields any non-positive $\Lambda_t$, we halve the step length and retry, up to a small number of backtracks.

The non-negativity constraint $S\theta \succeq 0$ on $R_t$ itself is not enforced inside the solver; as argued in \cref{apx:rt-uncertainty}, it is asymptotically inactive whenever the true $R_t$ is bounded away from zero, which holds in every setting we consider.

\paragraph{Ascertainment rates}

Of the two multiplicative terms deferred above, the ascertainment rate
$\rho_t \in (0,1]$ is the simpler: it is fixed and known (e.g.\ from external
calibration), not estimated. Entering the mean multiplicatively,
$\mu_t = \rho_t\,\Lambda_t(\theta)$, it acts as a known Poisson offset that
rescales each expected count without adding free parameters. The derivations
above therefore carry through verbatim with $\Lambda_t$ replaced by
$\rho_t \Lambda_t$: the working weights become
$W = \mathrm{diag}(1/(\rho_t\Lambda_t))$ and the design rows are scaled to
$\rho_t Z_t$, leaving the IRLS recursion and its convexity unchanged. The
day-of-week multipliers $\omega$, by contrast, are estimated, and we treat them
next.

\subsection{Day-of-week effects}

When day-of-week (DoW) multipliers $\omega_{(\text{$t$ mod 7})}$ enter the mean via $\mu_t = \omega_{(\text{$t$ mod 7})}\,\Lambda_t(\theta)$ as in \eqref{eq:mean-cases}, the model is invariant to the rescaling $(\omega, \theta) \leftrightarrow (c\omega, \theta/c)$ for any $c > 0$. 
Multiplying every $\omega_j$ by $c$ and dividing $\theta$ by $c$ leaves $\mu_t$ — and hence the likelihood — unchanged.      
This is problematic, as the MLE is non-unique along this one-dimensional ray. To ensure uniqueness,
we impose the geometric-mean constraint $\prod_j \omega_j = 1$, the natural normalization for a multiplicative factor,                                                                                                                                           
Estimation proceeds by block coordinate ascent on the log-likelihood. Holding $\theta$ fixed, partition the Poisson log-likelihood by day-of-week: with $T_j = \{t : \text{$t$ mod 7} = j\}$,                  
  \[                                                                                                                                                                                                             
  \ell(\theta, \omega) \;=\; \sum_j \sum_{t \in T_j}\bigl[y_t\log(\omega_j \Lambda_t) - \omega_j \Lambda_t\bigr].
  \]                                                                                                                                                     
Differentiating reveals that the score for $\omega_j$ is $\sum_{t \in T_j}[y_t/\omega_j - \Lambda_t]$. 
This is maximized in closed form by the day's ratio of observed to expected counts:
  \begin{equation}\label{eq:dow-update}
  \hat\omega_j \;=\; \frac{\sum_{t \in T_j} y_t}{\sum_{t \in T_j} \Lambda_t(\theta)},
  \end{equation}                                                                                                                                                                                                 
To enforce $\prod_j \hat\omega_j = 1$, we divide each $\hat\omega_j$ by $\bigl(\prod_j
  \hat\omega_j\bigr)^{1/7}$. We then take one IRLS step on $\theta$ using the DoW-scaled design $\widetilde Z_t = \hat\omega_{(\text{$t$ mod 7})} Z_t$, and the two blocks alternate until the
  log-likelihood converges. Because each block is convex given the other, the iterates converge to the joint MLE---equivalently, to the maximizer of the profile likelihood with $\omega$ profiled out.

\subsection{Tail constraint}\label{apx:optim-constraints}

Our main experiments parameterize $R_t$ as a natural cubic spline. Linearity past the last knot is structural to the basis: every coefficient vector $\theta$ already yields a linear tail, so no constraint is enforced. This is the default both retrospectively and in real time.

The most useful alternative is a \emph{constant} tail. Higher-order tail behavior (quadratic, cubic) lets the spline keep wiggling past the last knot and tends to overfit, so we focus on the constant case.

  A constant-tail constraint requires a regression-spline basis---specifically,
  a clamped cubic B-spline---because, unlike the natural basis, it imposes no
  boundary behavior of its own. The constant tail is enforced as a linear equality
  $A\theta = 0$, where $A$ collects the first three derivatives of the spline basis
  functions evaluated at a collocation point $t_c$ in the tail segment. Writing
  $S_1, \ldots, S_p$ for the components of the basis row $S(t)$, the rows of $A$ are
  \[
  A \;:=\; \begin{bmatrix}
  S_1'(t_c)   & S_2'(t_c)   & \cdots & S_p'(t_c)   \\
  S_1''(t_c)  & S_2''(t_c)  & \cdots & S_p''(t_c)  \\
  S_1'''(t_c) & S_2'''(t_c) & \cdots & S_p'''(t_c)
  \end{bmatrix}.
  \]
  Because B-splines are local---each basis function is supported on only a few
  knot spans---most columns are zero; only the basis functions whose support
  contains $t_c$ contribute nonzero entries. Since $R_t$ is a cubic polynomial on
  the tail segment, the constraint $A\theta = 0$ sets $R_t' = R_t'' = R_t''' = 0$
  there, which forces every coefficient above the constant term to vanish and
  hence flattens the entire segment.

To enforce both the penalized objective and the constraint $A\theta = 0$ exactly, we solve the constrained optimization problem via its KKT conditions. We introduce Lagrange multipliers $\nu \in \mathbb{R}^3$ (one per derivative constraint) and form the Lagrangian:
\[
L(\theta, \nu) = f_\lambda(\theta) + \nu^\top(A\theta).
\]

At optimality, the gradients with respect to both $\theta$ and $\nu$ must vanish:
\[
\frac{\partial L}{\partial \theta} = \nabla_\theta f_\lambda(\theta) + A^\top\nu = 0, \qquad
\frac{\partial L}{\partial \nu} = A\theta = 0.
\]
The first condition says the gradient of the objective is exactly canceled by the constraint forces $A^\top\nu$. Geometrically, you cannot move in any feasible direction (staying on the constraint surface) without increasing the objective. The second condition simply enforces feasibility.

Recall the gradient of the penalized objective is $\nabla_\theta f_\lambda(\theta) = -Z^\top W(Y - Z\theta) + \lambda\Omega\theta$. Substituting into the first KKT condition and rearranging:
\[
-Z^\top W(Y - Z\theta) + \lambda\Omega\theta + A^\top\nu = 0
\]
gives
\[
(Z^\top W Z + \lambda\Omega)\theta + A^\top\nu = Z^\top W Y.
\]
Combined with the feasibility condition $A\theta = 0$, we obtain the augmented linear system:
\[
\begin{bmatrix} Z^\top W Z + \lambda \, \Omega & A^\top \\ A & 0 \end{bmatrix}
\begin{bmatrix} \theta \\ \nu \end{bmatrix}
=
\begin{bmatrix} Z^\top W Y \\ 0 \end{bmatrix}.
\]
This system holds the constraint at machine precision throughout the IRLS iteration. The multipliers $\nu$ encode the strength and direction of constraint forces pulling $\theta$ into the feasible region; they are discarded after solving. Note that only basis functions near the collocation point $t_c$ feel significant constraint forces, since far-away basis functions have negligible derivatives there anyway.

When a tapered penalty is also present, the augmented system can be ill-conditioned. Instead, we solve in the constraint's null space. Every feasible $\theta$ lies in $\mathrm{null}(A)$, so let the columns of $N \in \mathbb{R}^{p \times (p - \mathrm{rank}(A))}$ form an orthonormal basis for that null space:
\[
\theta = N\alpha, \quad \alpha \in \mathbb{R}^{p - \mathrm{rank}(A)}.
\]
The constraint is now self-enforcing: the problem becomes \emph{unconstrained} in $\alpha$, with effective design $ZN$ and effective penalty $N^\top \Omega N$. To obtain $N$, take the QR decomposition $A^\top = QR$ with $Q = [Q_1 \mid Q_2]$: the first $r = \mathrm{rank}(A)$ columns span the column space of $A^\top$, and the remaining $p - r$ columns span $\mathrm{null}(A)$. Set $N = Q_2$.

We do not recommend combining a constant-tail constraint with a natural-spline basis. The natural basis is already shape-constrained at the boundary, and stacking a hard equality on top was numerically fragile in our experiments. The B-spline route is cleaner.

\section{Inference in Poisson GLM}\label[appendix]{apx:rt-uncertainty}

Inference for $R_t$ rests on a Gaussian approximation to the sampling distribution of $\hat\theta$. Because $R_t(\theta) = S(t)^\top\theta$ is linear in $\theta$, once we have an estimate                    
$\widehat{\mathrm{Cov}}(\hat\theta)$, we can transport it to the $R_t$ scale by a simple quadratic form. We derive $\widehat{\mathrm{Cov}}(\hat\theta)$ from the penalized score equations for the             
retrospective estimator \eqref{eq:glm-retro} under three working assumptions: 
\begin{enumerate}
  \item The observed counts $y_t$ are conditionally independent and Poisson with mean $\Lambda_t = Z_t^\top\theta$; 
  \item The delay distributions are known; and
\item  The smoothing parameter $\lambda$ is held fixed.                                                    
\end{enumerate}

We perform inference with a standard approach: the sandwich variance for a penalized M-estimator (or quasi-likelihood). 
Without loss of generality, our notation assumes no day-of-week effects or under-reporting. 

\subsection{Confidence intervals for \texorpdfstring{$R_t$}{Rt}}
\subsubsection*{Sandwich variance of $\hat\theta$}
Stacking timesteps into a design matrix $Z \in \mathbb{R}^{T\times p}$ with rows $Z_t^\top$ and $\Lambda_t(\theta) = Z_t^\top\theta$, the total score $U(\theta) = \sum_t s_t(\theta)$ and Fisher information are
\[
U(\theta) \;=\; Z^\top W (Y - Z\theta), \qquad
\mathcal{I}(\theta) \;=\; Z^\top W Z, \qquad W \;=\; \mathrm{diag}\bigl(1/\Lambda_t(\theta)\bigr).
\]
For independent Poisson observations the information equality holds, so $\mathrm{Var}(U(\theta)) = Z^\top W Z = \mathcal{I}(\theta)$. Crucially, $U$ is the gradient of the \emph{likelihood} alone, so this variance carries no $\lambda$.

The estimator $\hat\theta$ solves the first-order condition of \eqref{eq:glm-retro},
\[
-U(\hat\theta) \,+\, \lambda\Omega\,\hat\theta \;=\; 0.
\]
Expanding to first order about the truth $\theta_0$, using $\nabla U(\theta_0) \approx -\mathcal{I}(\theta_0)$,
\[
0 \;\approx\; -U(\theta_0) \;+\; \lambda\Omega\theta_0 \;+\; \bigl(\mathcal{I}(\theta_0) + \lambda\Omega\bigr)(\hat\theta - \theta_0),
\]
where $\lambda\Omega\theta_0$ is a deterministic shift and $\lambda\Omega$ is the penalty's contribution to the curvature. Let
\[
H \;:=\; \mathcal{I}(\theta_0) + \lambda\Omega \;=\; Z^\top W Z + \lambda\Omega
\]
denote the Hessian of the penalized objective \eqref{eq:glm-retro} at $\theta_0$. Inverting,
\[
\hat\theta - \theta_0 \;\approx\; H^{-1}\bigl[U(\theta_0) \,-\, \lambda\Omega\theta_0\bigr].
\]
The term $-H^{-1}\lambda\Omega\theta_0$ is a deterministic ridge-type shrinkage bias and does not contribute to $\mathrm{Var}(\hat\theta)$. The variance of $\hat\theta$ is therefore driven entirely by the stochastic part $H^{-1}U(\theta_0)$. This gives the sandwich variance
\begin{equation}\label{eq:rt-cov-theta}
\mathrm{Cov}(\hat\theta) \;\approx\; H^{-1}\,\underbrace{(Z^\top W Z)}_{\mathrm{Var}(U(\theta_0))}\,H^{-1},
\end{equation}
where the approximation is due to the first-order Taylor expansion of the score about $\theta_0$.

\subsubsection*{Plug-in estimation}

Both $\mathcal{I}(\theta_0)$ and $W = \mathrm{diag}(1/\Lambda_t(\theta_0))$ depend on the unknown $\theta_0$. We use the standard plug-in: substitute $\hat\theta$ throughout, giving fitted means $\hat\Lambda_t = Z_t^\top\hat\theta$, working weights $\widehat W = \mathrm{diag}(1/\hat\Lambda_t)$, and
\begin{equation}\label{eq:sandwich-cov}
    \widehat H \;=\; Z^\top \widehat W Z + \lambda\Omega, \qquad \widehat{\mathrm{Cov}}(\hat\theta) \;=\; \widehat H^{-1}\,(Z^\top \widehat W Z)\,\widehat H^{-1}.
\end{equation}
Under regularity $\hat\theta$ is consistent for the penalized target it estimates, and by the continuous mapping theorem so is the plug-in covariance.

Equation \eqref{eq:rt-cov-theta} has the classical bread--meat form: bread $H^{-1}$, meat $Z^\top W Z$. The penalty sits in both slices of bread, through $H = Z^\top W Z + \lambda\Omega$, because it changes how responsive $\hat\theta$ is to a perturbation of the data. It is absent from the meat, since the meat is the variance of the unpenalized likelihood score. Setting $\lambda = 0$ collapses both breads onto the meat, recovering the usual unpenalized Poisson covariance $(Z^\top W Z)^{-1}$. Increasing $\lambda$ narrows the intervals, correctly reflecting sampling variability of $\hat\theta$ about the penalized target rather than the true $R_t$.

The real-time estimator \eqref{eq:glm-real-time} carries an additional penalty $\gamma\theta^\top\Psi\theta$, damping first differences in the sparse tail. It enters the bread identically to $\lambda\Omega$, giving
\[
\widehat H \;=\; Z^\top \widehat W Z \;+\; \lambda\Omega \;+\; \gamma\Psi,
\]
with the sandwich $\widehat{\mathrm{Cov}}(\hat\theta) = \widehat H^{-1}(Z^\top \widehat W Z)\widehat H^{-1}$ otherwise unchanged.

\subsubsection*{Wald intervals for $R_t$}

As $T \to \infty$, $U(\theta_0) = \sum_t s_t(\theta_0)$ is a sum of independent mean-zero terms and is asymptotically Gaussian by the CLT. The linearization $\hat\theta - \theta_0 \approx H^{-1}[U(\theta_0) - \lambda\Omega\theta_0]$ is an affine function of $U(\theta_0)$, with $H$ treated as a fixed constant. So $\hat\theta$ is asymptotically Gaussian with covariance \eqref{eq:rt-cov-theta}. The estimator $\hat R_t = S(t)^\top\hat\theta$ is linear in $\hat\theta$ and inherits normality with variance $S(t)^\top\mathrm{Cov}(\hat\theta)S(t)$, giving the Wald interval
\begin{equation}\label{eq:rt-wald}
\hat R_t \;\pm\; z_{1-\alpha/2}\,\sqrt{\,S(t)^\top\,\widehat{\mathrm{Cov}}(\hat\theta)\,S(t)\,}, \qquad z_{1-\alpha/2} \;=\; \Phi^{-1}(1-\alpha/2),
\end{equation}
where $\Phi$ is the standard normal CDF. We can rewrite this as \mbox{$[\hat R_t - z\cdot\mathrm{SE}_t,\, \hat R_t + z\cdot\mathrm{SE}_t]$}, where \mbox{$\mathrm{SE}_t \;=\; \sqrt{S(t)^\top\,\widehat{\mathrm{Cov}}(\hat\theta)\,S(t)}$} denotes the estimated standard error of $\hat R_t$.

\subsection{Day-of-week effects and ascertainment rates}

Having estimated day-of-week effects $\hat\omega$, we must incorporate them into the design in order to perform inference.
To do so, we apply the same sandwich machinery as the no-DoW case. Specifically, we scale $Z_t$ via $\widetilde Z_t =                    
\hat\omega_{(\text{$t$ mod 7})}\, Z_t$ and evaluate the working weights at the scaled fitted mean, $\widetilde W = \mathrm{diag}(1/\hat\mu_t)$ with $\hat\mu_t = \hat\omega_{(\text{$t$ mod 7})}\hat\Lambda_t$.
Formula \eqref{eq:rt-cov-theta} then becomes                 
  \[                                                              
  \widehat{\mathrm{Cov}}(\hat\theta) \;=\; \widetilde H^{-1}\,(\widetilde Z^\top \widetilde W \widetilde Z)\, \widetilde H^{-1}, \qquad \widetilde H \;=\; \widetilde Z^\top \widetilde W \widetilde Z + \lambda
  \Omega,          
  \]
and $R_t$ intervals are formed exactly as before via $\widehat{\mathrm{Var}}(\hat R_t) = S(t)^\top\, \widehat{\mathrm{Cov}}(\hat\theta)\, S(t)$. This is a \emph{conditional} variance: it treats $\hat\omega$ 
as fixed rather than estimated, and so omits the contribution from the joint $(\theta, \omega)$ Hessian that would reflect sampling variability in $\hat\omega$. Fortunately, this omission is typically small in      
practice, since each $\hat\omega_j$ pools across every observation from day $j$ and is far better identified than any single spline coefficient.

  
  The ascertainment rates $\rho_t$ are fixed and known. They fold into the
  sandwich just as $\hat\omega$ does: scale the design to
  $\widetilde Z_t = \rho_t Z_t$ and evaluate the weights at the fitted mean
  $\hat\mu_t = \rho_t \hat\Lambda_t$. (When both terms are present, the two
  scalings compose.) Unlike $\hat\omega$, however, $\rho_t$ carries no sampling
  variability, so no conditional-variance caveat applies: the resulting
  $\widehat{\mathrm{Cov}}(\hat\theta)$ is exact in this respect. We form $R_t$
  intervals as before, via
  $\widehat{\mathrm{Var}}(\hat R_t) = S(t)^\top \widehat{\mathrm{Cov}}(\hat\theta) S(t)$.
  Note that $\rho_t$ acts only on the observation scale; the target
  $R_t(\theta) = S(t)^\top\theta$ is unaffected.

\subsection{Quasi-Poisson adjustment}\label[appendix]{apx:quasi-poisson}

When Poisson dispersion does not hold --- as is typical for real
surveillance data --- we replace Assumption~1 with the quasi-likelihood
assumption $\mathrm{Var}(y_t) = \phi\,\Lambda_t$ for an unknown
dispersion $\phi \geq 1$. The sandwich form above is unchanged except
for the score variance, which becomes $\phi\,Z^\top W Z$ instead of
$Z^\top W Z$. Substituting back, the asymptotic covariance of $\hat\theta$
scales by $\phi$, and every standard error (and hence Wald-CI half-width)
scales by $\sqrt{\phi}$. 

We estimate $\phi$ from Pearson residuals at
the fitted values,
\[
  \hat\phi \;=\; \frac{1}{T - p_{\mathrm{eff}}}
                \sum_t \frac{(y_t - \hat\Lambda_t)^2}{\hat\Lambda_t},
\]
where $p_{\mathrm{eff}} = \mathrm{tr}\bigl((Z^\top W Z + \lambda\Omega)^{-1}
Z^\top W Z\bigr)$ is the effective degrees of freedom of the penalized
fit. All real-data confidence intervals reported in this paper use this
adjustment; most simulation results do not, where the data-generating
process is Poisson.

\subsection{Simultaneous Coverage}
The Wald intervals are \emph{pointwise}: the collection $\{[\hat R_t - z\cdot\mathrm{SE}_t,\, \hat R_t + z\cdot\mathrm{SE}_t]\}_{t=t_0}^{t_1}$ does
not jointly cover the entire $R$-curve at level $1-\alpha$, since family-wise error accumulates over $T$ timesteps. 
To get a band with \textit{simultaneous coverage}, we inflate the critical value from $z_{1-\alpha/2}$ to some larger $c_\alpha$, chosen so that the probability of \emph{any} $R_t$ escaping the band is at most $\alpha$:  \[                                                              
  \mathbb{P}\!\left(\,\max_t \frac{|\hat R_t - R_t|}{\mathrm{SE}_t} > c_\alpha\,\right) \;\leq\; \alpha,
  \]                                                                                                                                                                                                          
While no closed-form expression exists for $c_\alpha$, we can leverage the asymptotic distribution of $\hat \theta$ to estimate it empirically. 
For \mbox{$m = 1,\ldots,M$}, sample \mbox{$\delta^{(m)} \sim
\mathcal{N}(0,\,\widehat{\mathrm{Cov}}(\hat\theta))$}, as a plausible realization of the random error $\hat\theta - \theta_0$. 
Passing this through the spline basis samples deviations $\delta R_t^{(m)} = 
S(t)^\top \delta^{(m)}$, of which we record the standardized supremum \mbox{$Z_m = \max_t |\delta R_t^{(m)}|/\mathrm{SE}_t$}.
Because the $\{Z_m\}$ are i.i.d., we set $c_\alpha$ to the empirical $(1-\alpha)$ quantile to form the simultaneous band $\hat R_t \pm c_\alpha\cdot\mathrm{SE}_t$. 

A simpler alternative to attain simultaneous coverage is the Bonferroni correction.
This strategy simply adjusts the significance level by a factor of $T$, taking \mbox{$c_\alpha \leftarrow z_{1-\alpha/(2T)}$}.
While valid, the Bonferroni correction ignores the strong correlation between adjacent $R_t$ estimates, 
and thus pays a nontrivial width penalty.
In contrast, the empirical approach takes this correlation into account,
and is thus less conservative. 

\subsection{Constraints}
                                                                               
The estimator carries three constraints that restrict the feasible set. 
We ignore all of them for inference, decisions we justify here. 

First is a non-negativity constraint $S\theta \succeq 0$.
This is an inequality constraint on $R_t$, which we ignore it for inference. The justification is standard: when the true $R_t$ is bounded strictly above zero at every timestep --- as is the case in essentially   
every epidemic setting we study --- $\hat\theta$ lies in the interior of the feasible set with probability approaching one, so the constraint is asymptotically inactive and the sandwich derivation goes      
through unmodified.

Secondly, the DoW identifiability constraint $\prod_j \omega_j = 1$ is also not a concern for inference.                                   This is an \emph{identifiability} constraint: it picks a unique representative from the rescaling ray $(\omega, \theta) \leftrightarrow      
(c\omega, \theta/c)$ on which the likelihood is constant, rather than restricting the feasible set. Every statistical quantity we report --- $\hat\mu_t$, $R_t = S(t)^\top\theta$, the day-of-week ratios $\omega_j/\omega_k$ --- is invariant to that same rescaling, so none depends on which representative we pick.

The third constraint is the constant-tail restriction, which is not our default method.
Here the constraint shapes only the point estimate: $\hat\theta$ is fit subject to
$A\hat\theta = 0$ (enforced within the penalized IRLS solve via its KKT system, as
described in \cref{apx:optim-constraints}). The covariance, however, is formed in the
full coefficient space using the same sandwich as our default, unconstrained-tail fits \eqref{eq:sandwich-cov}. The working weights $\widehat W$ are evaluated at the constrained fit, but $A$ does not otherwise enter.

\subsection{Real-time inference}\label[appendix]{apx:rt-realtime-ci}

Our estimates $\hat R_t$ have error
\[
\lvert\hat R_t - R_t\rvert = \left\lvert \underbrace{\mathbb{E}[\hat R_t]-R_t}_{\text{Bias}_t}+\underbrace{\hat R_t - \mathbb{E}[\hat R_t]}_{\text{Noise}_t\ (\varepsilon_t)}\right\rvert, \qquad \varepsilon_t \overset{\cdot}{\sim} \mathcal{N}(0,\sigma_t^2).
\]
Previously, we demonstrated how to compute the sandwich variance for $\sigma_t^2$, and thus form Wald confidence intervals based on the noise $\varepsilon_t$. 
On retrospective analyses, these bands cover the true values, or nearly do.
This is because cubic splines are flexible enough to learn realistic $R_t$ curves, so bias is minimal.
In real time, however, we impose various forms of tail regularization, which produces bias. 
Consequently, the Wald bands fail to cover the tail predictions, so we must take a different approach.

\paragraph{Split-conformal bands.}
Rather than model the edge error with a variance formula, we estimate it directly from errors ConvRt has made before.
For any past date, ConvRt's estimate eventually settles once many more weeks of data arrive.
We trust that settled value as a stand-in for the truth.
Back when the same date sat near the leading edge, the real-time estimate was worse.
The gap between that early estimate and the settled value is a concrete sample of the edge error we now face.
We collect many such gaps and read the band width off their empirical quantile.

Let $\hat R^{W}_{t}$ denote the ConvRt estimate of $R_t$ at date $t$ using data through vintage $W$, and let $b = W - t$ be its horizon behind the edge.
We index everything by $b$ because the error grows as $b \to 0$.
Write $\hat R^{W}(\tau)$ for the settled anchor of the previous paragraph, the estimate at a date $\tau \le W - \Delta$ with $\Delta = 14$ days.
Each earlier vintage $W' < W$ then contributes one calibration example per horizon: its edge-region estimate $\hat R^{W'}_{\tau}$ at $\tau = W' - b$, compared against the anchor $\hat R^{W}(\tau)$.

We score each example in one of two ways,
\begin{align}
\text{Residual:}\quad & s = \bigl\lvert \hat R^{W'}_{\tau} - \hat R^{W}(\tau)\bigr\rvert,\\
\text{Studentized:}\quad & s = \bigl\lvert \hat R^{W'}_{\tau} - \hat R^{W}(\tau)\bigr\rvert \big/ \sigma^{W'}_{\tau},
\end{align}
collect the scores separately for each horizon $b$, and take the split-conformal quantile $\hat q_b = s_{(k)}$ with $k = \lceil (1-\alpha)(n+1)\rceil$.
The band at the target date is then $\hat R^{W}_{t} \pm \hat q_b$ for the residual score, or $\hat R^{W}_{t} \pm \hat q_b\,\sigma^{W}_{t}$ for the studentized score.
Because the revisions are exchangeable across vintages at a fixed horizon, this construction carries the usual finite-sample split-conformal coverage guarantee.
Crucially, both scores absorb the tail bias directly, since the revision $\hat R^{W'}_{\tau} - \hat R^{W}(\tau)$ measures the total error $\lvert\text{Bias}_t+\text{Noise}_t\rvert$, rather than the
noise alone.

\begin{figure}
\centering
\includegraphics[width=\linewidth]{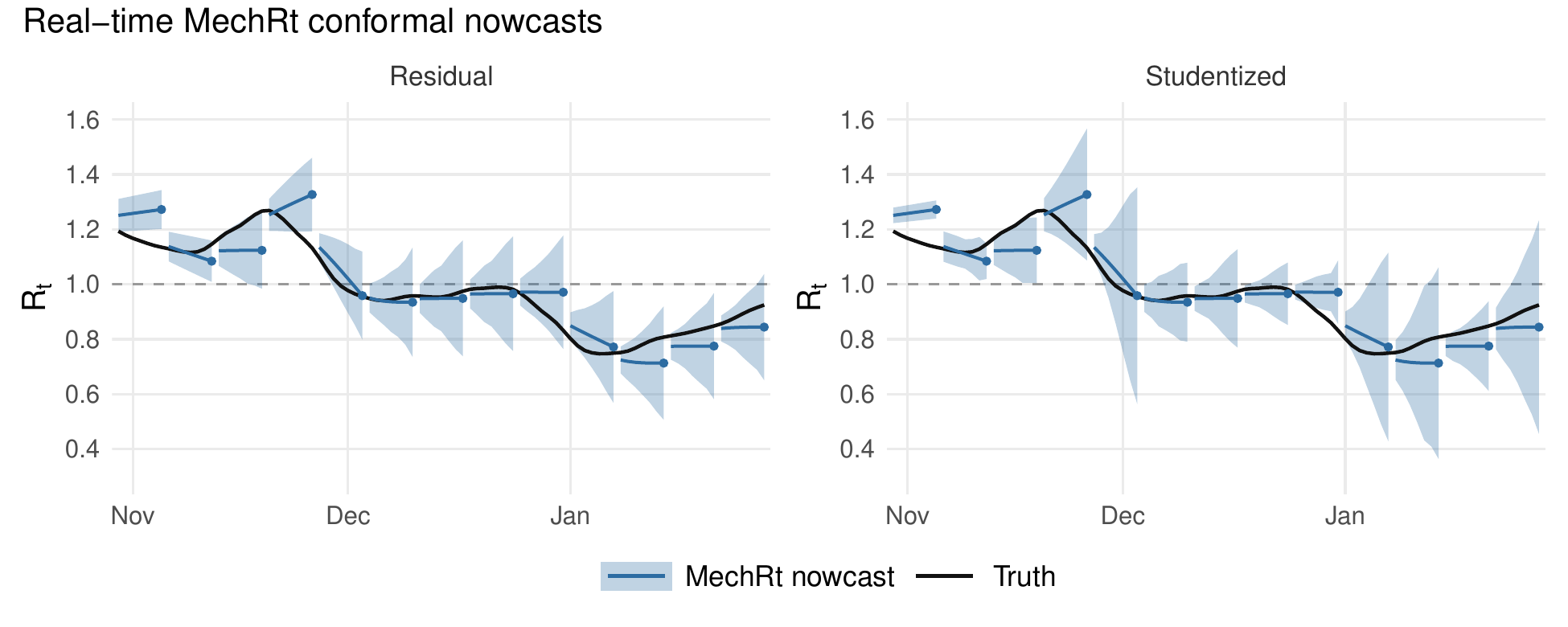}
\caption[Conformal nowcast bands on the wiggly flu simulation.]{Residual (left) and studentized (right) conformal nowcast bands for ConvRt on the wiggly flu simulation, $95\%$ level, last seven days per weekly vintage. Black is the simulation truth; blue is the
real-time mean with its conformal band. The studentized bands swing wide or tight as the edge $\sigma_t$ fluctuates; the residual bands are steadier.}
\label{fig:conf-sim}
\end{figure}

\paragraph{Residual vs. studentized.}
The two scores differ only in whether they rescale each revision by the model's reported precision $\sigma_\tau$.
The residual score adapts its band widths slowly, in response to sustained periods of (mis)coverage. 
The studentized score is designed to make more flexible bands across dates.
Dividing by $\sigma^{W'}_{\tau}$ turns each revision into a standardized residual, so the target band inherits the local scale
$\sigma^{W}_{t}$ and widens exactly where the fit reports itself to be uncertain.
This is appealing in principle, but it presumes that $\sigma$ is a trustworthy gauge of the real-time error.
At the tail, however, the dominant component of the error is the regularization bias, which $\sigma_\tau$ does not reflect. 
The sandwich covariance may also be a less reliable estimator there.

Figure~\ref{fig:conf-sim} shows the consequence on individual vintages.
The studentized bands swing wide or tight as the edge $\sigma_t$ fluctuates.
Meanwhile, the residual bands are more steady over time, yet still generally track the truth.

\begin{figure}[t]
\centering
\includegraphics[width=0.7\linewidth]{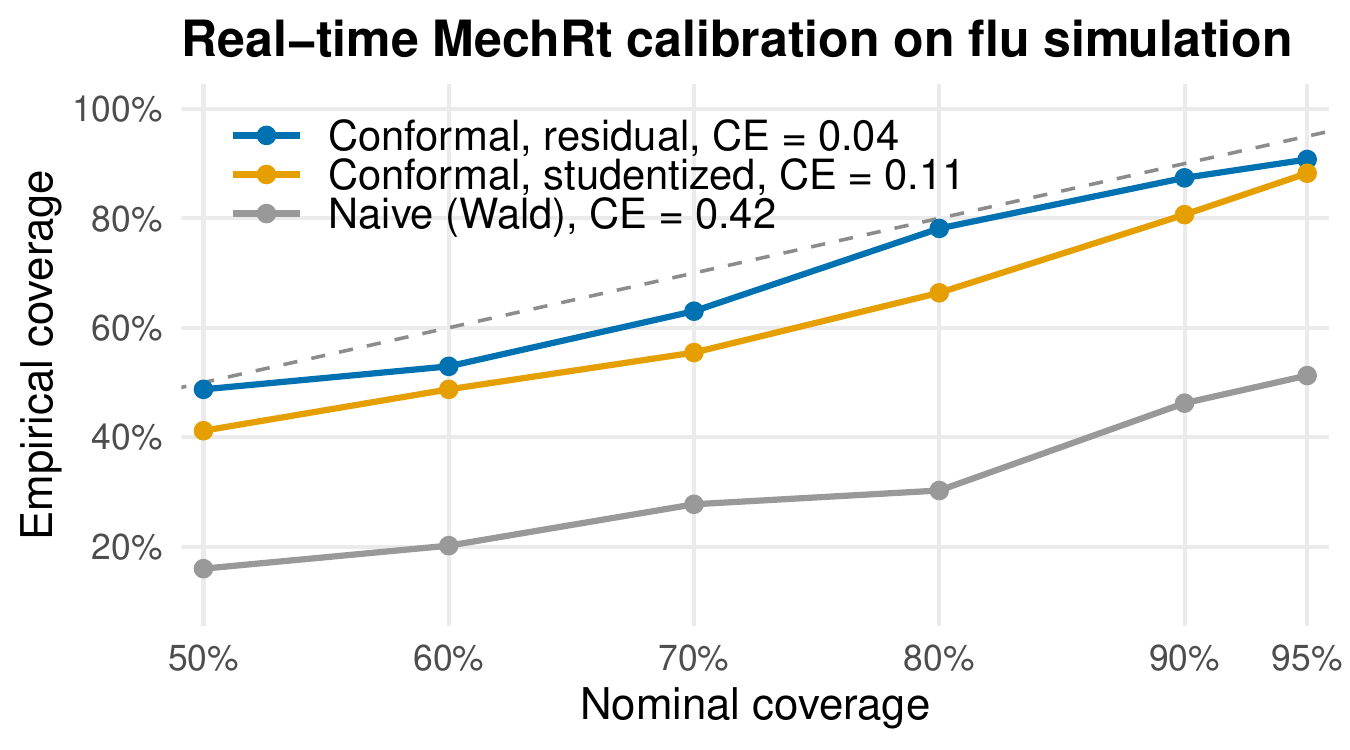}
\caption[Real-time conformal coverage calibration on the flu simulation.]{Real-time coverage calibration on the flu simulation, scored against the true simulated $R_t$.
Empirical versus nominal coverage for the residual and studentized conformal bands and the naive Wald band.
The dashed line marks perfect calibration.}
\label{fig:conf-cal}
\end{figure}

\paragraph{Empirical coverage.}
We use $\ell_1$ coverage error
\sloppy{$\text{CE} = \tfrac{1}{K}\sum_{k} \lvert \widehat{\text{cov}}(\alpha_k) - \alpha_k\rvert$}
to summarize calibration.
This is the mean gap between empirical and nominal coverage over the grid 
{$\alpha_k \in \{0.5, 0.6, 0.7, 0.8, 0.9, 0.95\}$}, 
measured on real-time $(W, t)$ rows.
On the flu simulation, where the true $R_t$ is known, the residual band is nearly perfect at $\text{CE} = 0.04$ ($n = 119$ rows; Figure~\ref{fig:conf-cal}).
The studentized band undercovers, at $\text{CE} = 0.11$.
The naive Wald band collapses to $\text{CE} = 0.42$, reaching only $51\%$ coverage at the nominal $95\%$ level.
On real flu data the ordering is the same, scored against the end-of-season retrospective: residual $\text{CE} = 0.04$, studentized $0.05$, and naive $0.27$ ($n = 175$ rows).

We therefore adopt the residual score as the default ConvRt real-time band.
It is worth noting that nothing in this construction is specific to ConvRt.
The residual band wraps any real-time point predictor, since it needs only past estimates and their revisions, not a variance model.

\section{Deconvolving latent infections}\label[appendix]{apx:deconvolution}

This section provides additional details on the infection deconvolution of \cref{sec:inf-deconv}.
We use the same superscripted notation $\theta^{(x)}$, $\Omega^{(x)}$, $\lambda^{(x)}$ introduced there; for brevity, we drop the superscripts below.

Recall from \eqref{eq:glm-retro} that we model daily infections as a natural cubic spline, $x_t = S(t)^\top \theta$, and jointly solve for the spline coefficients $\theta$ and day-of-week effects $\omega$ by minimizing the penalized Poisson negative log likelihood subject to non-negativity,
\begin{equation}
\hat\theta,\, \hat\omega = \argmin_{\bar\omega_{GM}=1,\; S\theta \succeq 0}\;
\sum_t \bigl(\mu_t(\theta, \omega) - y_t \log \mu_t(\theta, \omega)\bigr)
+ \lambda\, \theta^\top \Omega\, \theta,
\end{equation}
where $\mu_t(\theta,\omega) = \omega_{(t \bmod 7)}\, \rho_t \sum_{s<t} \pi_{t-s}\, S(s)^\top \theta$ is the expected case count.
Here we describe the smoothness penalty $\Omega$ and strategies for tuning $\lambda$.

\paragraph{Hyperparameter tuning.}
$\lambda$ may be selected via K-fold CV or GCV.
For computational convenience, our code uses GCV, though this could be marginally less accurate.
Tuning is less critical here than for $R_t$, since errors in $\hat x_t$ are smoothed out through the convolution to expected cases $\mu_t$ \eqref{eq:mean-cases}.

The smoothness penalty $\Omega$ is a magnitude-weighted integrated squared second derivative, computed by fine-grid quadrature:
\begin{equation}
\Omega = \int v(t)\, S''(t)\, S''(t)^\top\, dt,
\qquad \text{so} \qquad
\theta^\top \Omega\, \theta = \int v(t)\, \bigl(S''(t)^\top \theta\bigr)^2\, dt.
\end{equation}
The natural-spline basis pairs well with this penalty: linear tails are structural to the basis, which stabilizes the under-identified curve ends, and $\int (x{''})^2$ is the canonical smoothing-spline penalty here. (A second-difference $D^{(2)}$ penalty is the analog for B-/P-splines.)

\paragraph{Relative-curvature weights.}
$x$ ranges over several orders of magnitude---a few cases per day on the early limb, thousands at peak.
A uniform penalty ($v_t \equiv 1$) measures curvature in absolute terms, so the peak swamps the low limb in the integral, even when both are equally wiggly in proportional terms.
No single $\lambda$ works: controlling the lower limb oversmooths the peak, while tuning for the peak undersmooths the bottom.

The fix is to penalize \emph{relative} curvature, i.e.\ $\bigl((x)''/x\bigr)^2$.
Dividing by $x$ cancels the scale---$x \to c\, x$ leaves it unchanged---so a proportional wiggle counts the same on the low limb as at the peak.
This corresponds to a weighting $v_t \propto (\tilde x_t)^{-2}$.
While $x$ is unknown, a plug-in estimate $\tilde x$ works.
This makes the integrand $\bigl((x)''/\tilde x\bigr)^2$, matching the relative curvature when $\tilde x \approx x$.

The inverse blows up wherever $\tilde x$ is small, which would over-penalize the low-count tails and prevent the spline from tracking genuine early growth.
We therefore floor $\tilde x$ at a fraction $c$ of its maximum:
\begin{equation}
v_t = \left(\frac{\tilde x_{\max}}{\max(\tilde x_t,\, c\, \tilde x_{\max})}\right)^{\!2},
\qquad \tilde x_{\max} = \max_s \tilde x_s,
\quad c = 0.03.
\end{equation}
The numerator $\tilde x_{\max}$ normalizes the weights so $\lambda$ stays on a fixed, GCV-friendly scale across datasets.
The floor parameter $c$ caps how strongly low-count timesteps are penalized: any $t$ with $\tilde x_t \le c\, \tilde x_{\max}$ inherits the maximum weight $1/c^2$.
We chose $c = 0.03$ by inspection on simulated flu data; fits across the range $c \in [0.01, 0.1]$ are visually indistinguishable, so the estimator is not very sensitive to this choice.

The reference curve $\tilde x$ is deconvolution-free: a centered $7$-day average of the observations, shifted to infection time by the mean delay $\bar L = \mathrm{round}\bigl(\sum_k k\, \pi_k\bigr)$ and held constant past the data edge:
\begin{equation}
\bar Y_t = \frac{1}{|W_t|} \sum_{s \in W_t} Y_s^*,
\quad W_t = \{t-3, \dots, t+3\} \cap [1, n],
\qquad
\tilde x_t = \bar Y_{\min(t + \bar L,\, n)}.
\end{equation}
This avoids any preliminary deconvolution and is valid in real time---the average shrinks at the edge, and the shift never reads beyond $t$.

Splines are not inherently non-negative, so this deconvolution approach risks being inaccurate when counts are low.
Therefore we also implement an alternative that uses a log-link function.
This uses $\log y_t = \mu_t(\theta)$, which is naturally non-negative at the cost of incorrect specification.

\section{Methods configuration}\label[appendix]{apx:methods-config}

\subsection{EpiNow2}\label[appendix]{apx:methods-epinow2}

\begin{figure}
  \centering
  \includegraphics[width=0.8\linewidth]{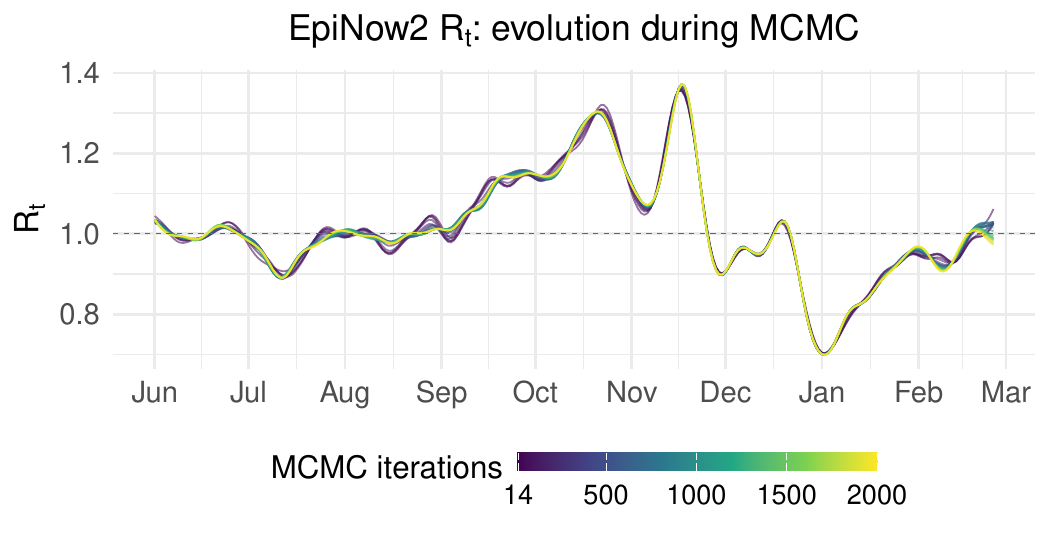}
  \caption[EpiNow2 posterior-mean $R_t$ convergence on the wiggly simulation.]{Posterior-mean $R_t$ from EpiNow2's Gaussian-process model at
  successive MCMC snapshots on the retrospective wiggly-simulation fit (4
  chains; 1500 warmup and 500 post-warmup iterations per chain). Each line is
  one snapshot; color encodes the cumulative number of post-warmup draws
  aggregated across chains, with the corresponding wallclock time in
  parentheses.}
  \label{fig:epinow2-snapshots}
\end{figure}

\begin{table}[ht]
\centering
\caption[EpiNow2 configuration for finalized runs.]{EpiNow2 configuration for the finalized runs. The random-walk (RW)
kernel uses a unit step size on $\log R_t$; the Gaussian-process (GP) kernel is a
Hilbert-space approximation with a Mat\'ern-3/2 lengthscale. `NB'' denotes negative binomial; ``--'' not applicable.}
\label{tab:epinow2-config}
\setlength{\tabcolsep}{5pt}
\begin{tabular}{l cccc cc}
\toprule
& \multicolumn{4}{c}{\textit{rtestim}} & \multicolumn{2}{c}{\textit{Influenza}} \\
\cmidrule(lr){2-5} \cmidrule(lr){6-7}
& \shortstack{Piecewise\\Constant}
& \shortstack{Piecewise\\Exponential}
& \shortstack{Piecewise\\Linear}
& Periodic
& \shortstack{Wiggly\\$R_t$}
& \shortstack{Smooth\\$R_t$} \\
\midrule
Rt prior                & \multicolumn{4}{c}{$\mathrm{LogN}(0, 1.0)$} & \multicolumn{2}{c}{$\mathrm{LogN}(0, 0.5)$} \\
Kernel                  & RW & GP & RW & GP & \multicolumn{2}{c}{GP} \\
\quad Lengthscale       & -- & $\mathrm{LogN}(10,3)$ & -- & $\mathrm{LogN}(10,3)$ & \multicolumn{2}{c}{$\mathcal{N}(21,7)$} \\
\quad Basis prop.       & -- & 0.2 & -- & 0.2 & \multicolumn{2}{c}{0.2} \\
Obs.\ model             & \multicolumn{4}{c}{Poisson} & \multicolumn{2}{c}{NB} \\
Chains                  & \multicolumn{4}{c}{4} & \multicolumn{2}{c}{2} \\
Warmup                  & 2500 & 1500 & 3000 & 1500 & \multicolumn{2}{c}{1000} \\
Samples                 & 2500 & 2000 & 2000 & 2000 & \multicolumn{2}{c}{500} \\
$\delta_{\text{adapt}}$ & 0.97 & 0.95 & 0.97 & 0.95 & \multicolumn{2}{c}{0.95} \\
Max.\ tree              & 13 & 12 & 13 & 12 & \multicolumn{2}{c}{12} \\
\bottomrule
\end{tabular}
\end{table}

We benchmark against EpiNow2 (version 1.8.0), summarizing its configuration for every finalized run in Table~\ref{tab:epinow2-config}. Our aim throughout was to run each fit as cheaply as possible, but this
was in constant tension with EpiNow2's sampler, which failed to converge far more often than not---especially on the rtestim benchmarks. In most experiments $R_t$ follows EpiNow2's default Gaussian-process
prior (Hilbert-space approximation, Mat\'ern-$3/2$ kernel). On the rtestim scenarios with jump discontinuities (piecewise constant and piecewise linear), this prior could not be salvaged at any sampling budget
we tried---fits diverged or returned $\widehat{R}$ in the thousands---so on those datasets we instead modeled $\log R_t$ as a Gaussian random walk. Even where the GP prior was ultimately retained, the
out-of-the-box settings were not enough to reach acceptable diagnostics. 

The observation model matches each dataset's noise---Poisson or negative binomial---and for the real surveillance data we additionally retain EpiNow2's day-of-week reporting effect. MCMC uses 4 chains (2 for
the real data); warmup, \texttt{adapt\_delta}, and \texttt{max\_treedepth} are all raised above EpiNow2's defaults, scenario by scenario, until each fit was free of divergent transitions and well mixed
($\widehat{R} < 1.05$). We fix a single setting per dataset rather than tuning per replicate, accepting some over-provisioning to keep the comparison fair. This makes EpiNow2 by a wide margin the slowest
method evaluated---tens of minutes to over two hours per single fit, against roughly a second for ConvRt---so the inflated sampling budgets compound an already steep runtime cost. The generation-time and
reporting-delay distributions match those used by ConvRt and the other benchmarks.



\cref{fig:epinow2-snapshots} traces EpiNow2's GP posterior mean on the wiggly
retrospective fit as the sampler accumulates draws, computed by polling
cmdstanr's per-chain CSV output every 30 seconds and re-averaging over the
post-warmup draws available at each tick. The oscillatory structure in the final
fit---the short-lived excursions from $R_t = 1$ during the fall and winter---is
present from the earliest snapshot onward and merely tightens as draws
accumulate, confirming it is genuine posterior structure rather than
under-sampled Monte Carlo noise.

The colorbar also exposes a feature of EpiNow2's runtime invisible in a single
end-of-run timing: the first snapshot sits at 1564\,s with only 14 post-warmup
draws, roughly 68\% of the total wallclock. Stan compilation is cached and costs
only seconds; the rest is the 1500 warmup iterations per chain. Because warmup is
the bulk of the iterations and cannot be skipped, EpiNow2's wallclock is bounded
below by warmup and cannot be meaningfully reduced by drawing fewer samples.

\subsection{EpiEstim}\label[appendix]{apx:methods-epiestim}

We call \texttt{EpiEstim::estimate\_R} with \texttt{method = "non\_parametric\_si"},
passing the same generation-interval pmf used by ConvRt and the other
benchmarks (prepended with $\Pr(\mathrm{SI}=0)=0$ to match EpiEstim's indexing
convention). $R_t$ is estimated on a 7-day sliding window
(\texttt{t\_start = 2:(n-6)}, \texttt{t\_end = t\_start + 6}); the gamma prior
on $R_t$ has mean 1 and SD 5, the package defaults. Per the caveat in the main
text, EpiEstim is fit directly to observed reports rather than deconvolved
infections, so its estimates are implicitly shifted by the case-reporting
delay; we do not adjust the time axis to compensate, since doing so would
require a deconvolution step that EpiEstim itself does not perform. 

Credible intervals are taken from EpiEstim's posterior quantiles at the 95\% level. For bands at other levels, we convert each interval into an implied (Gaussian) standard deviation, $\hat\sigma = (q_{0.975} - q_{0.025}) / (2\,z_{0.975})$, where $q_{0.025}$ and $q_{0.975}$ are
the lower and upper posterior quantiles and $z_{0.975} \approx 1.96$. We then form a $100(1-\alpha)\%$ band as $\hat R_t \pm z_{1-\alpha/2}\,\hat\sigma$.

\subsection{EstimateR}\label[appendix]{apx:methods-estimater}

We use \texttt{estimateR} with LOESS pre-smoothing and Richardson--Lucy
deconvolution against the same case-to-report delay distribution $\pi$
($x\!\to\!y$, discrete gamma) used elsewhere. This is followed by an
EpiEstim sliding-window $R_t$ step
(\texttt{estimation\_method = "EpiEstim sliding window"},
\texttt{estimation\_window = 3}, \texttt{mean\_Re\_prior = 1}), using the oracle generation interval distribution $g$. 

For point
estimates and timing comparisons we call
\texttt{estimate\_Re\_from\_noisy\_delayed\_incidence} (a single fit); for
uncertainty intervals we call \texttt{get\_block\_bootstrapped\_estimate} with
50 block-bootstrap replicates, the package's recommended approach for
uncertainty quantification. Because Richardson--Lucy deconvolution truncates
the tail of the incidence series (an unavoidable consequence of the
right-censored convolution), estimateR only estimates $R_t$ up to 5 days
before the vintage date, so in our real-time figures and tables we propagate
its last estimate forward through the remainder of the week.

\subsection{EpiLPS}\label[appendix]{apx:methods-epilps}

We use \texttt{EpiLPS::estimR} with the package defaults: a cubic
B-spline basis with second-order difference penalty, the default
hyperprior on the smoothing parameter, and the LPSMAP backend
(Laplace approximation around the posterior mode), rather than the
slower LPSMALA MCMC alternative. The generation interval $g$ is supplied
as the same pmf used by the other methods. Real-time fits are run on
truncated vintage data without modification.

\subsection{Rtestim}\label[appendix]{apx:methods-rtestim}

We use \texttt{rtestim::cv\_estimate\_rt} with cubic trend filtering
(\texttt{korder = 3}), 3-fold cross-validation, and an explicit
log-spaced $\lambda$ grid spanning $10^{-2}$ to $10^{5}$ with 30
values; we found the package's default auto-grid sometimes terminated
before the 1-SE optimum was located, so we supply the grid explicitly
and raise \texttt{maxiter} to $10^7$ for convergence at the
smaller-$\lambda$ end. $\lambda$ is selected by the 1-SE rule.
Confidence bands come from \texttt{confband()}, which inverts a
quadratic relaxation of the trend-filtering objective; the bands are
pointwise and do not have a frequentist coverage guarantee at the
level suggested by their nominal width. The generation interval is
supplied via \texttt{delay\_distn}, prepended with a 0 so that index
1 corresponds to $\Pr(\mathrm{lag}=0)=0$ (rtestim's convention treats
\texttt{delay\_distn[1]} as the same-day weight, differing from the
Cori convention used by EpiEstim). For series with a long all-zero
leading tail we trim to the first positive count before fitting,
since rtestim's CV objective is undefined when the weighted
past-counts denominator hits 0. Real-time fits are run on truncated
vintage data without modification.

\subsection{CovidEstim}\label[appendix]{apx:methods-covidestim}

We do not refit CovidEstim. The COVID-19 comparisons in
\cref{sec:rt-experiments-real} use $R_t$ trajectories and case
ascertainment rates from the project's public release
\citep{Chitwood2022,covidestim_summer2021}, which were produced by the
authors with their full Bayesian pipeline (random-walk log-$R_t$,
seroprevalence-informed ascertainment, HMC inference). We treat these as
fixed reference outputs; the delay distributions and generation interval
we feed to ConvRt for the same comparison are documented in
\cref{apx:covidestim}.

\section{Rtestim benchmarks}\label[appendix]{apx:rtestim}

Ideally, for both the rtestim and influenza benchmarks, we would resimulate each dataset many times.
We would run all methods on each replicate, and ultimately report average performance. 
However, we did not pursue this because EpiNow2 was slow to fit and frequently failed to converge, making repeated runs across all six datasets impractical.

  \subsection{Data-generating process}
  
   For each of the four ground-truth $R_t$ trajectories from \citet{rtestim}, we
  simulate $n = 300$ days of infections $x_t$ and case reports $y_t$ from a
  discrete-time stochastic SEIR model with explicit, discrete-gamma latent period,
  infectious period, and reporting delay. We set the latent (exposure-to-infectious)
  delay to mean $5.9$ and SD $3.3$ and the infectious period to mean $5.0$ and SD
  $2.2$, chosen so that the implied generation interval matches the mean of $8.4$
  days used by \citet{rtestim}; the implied SD is $3.95$. The exposure-to-report delay has mean $7$ and SD $3$. Each
  scenario is seeded with $10$ exposures per day over a $30$-day warm-up so that
  the infectious pool is near equilibrium at $t = 1$.


\begin{figure}
\centering
\includegraphics[width=0.9\linewidth]{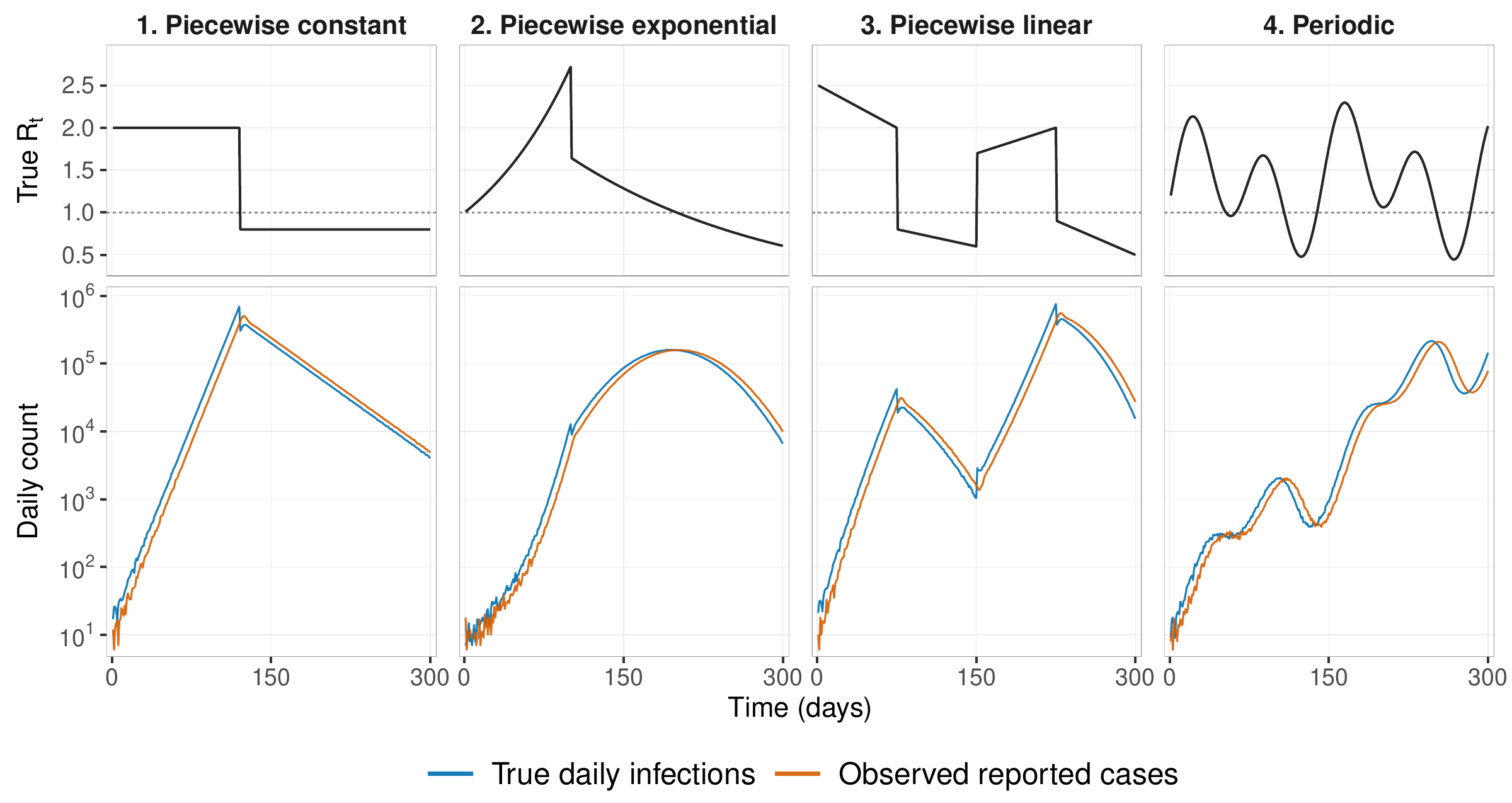}
\caption{rtestim benchmark datasets: $R_t$ curves and the cases they generate.}
\label{fig:rtestim_gt_and_cases}
\end{figure}

The four benchmark datasets from \citet{rtestim}, shown atop \cref{fig:rtestim_gt_and_cases}, are:
\begin{enumerate}
  \item \textbf{Piecewise constant.} A single sharp policy change,
  \[
    R_t =
    \begin{cases}
      2.0, & 1 \leq t \leq 120, \\
      0.8, & 120 < t \leq 300,
    \end{cases}
  \]
  representing four months of unchecked growth before an intervention drops
  transmission well below the critical threshold.

  \item \textbf{Piecewise exponential.} A fast exponential climb followed by a
  slower exponential decay, with a downward jump at the transition:
  \[
    R_t =
    \begin{cases}
      \exp(0.01\, t), & 1 \leq t \leq 100, \\
      \exp\!\bigl(0.5 - 0.005\,(t-100)\bigr), & 100 < t \leq 300.
    \end{cases}
  \]
  An exponential takeoff is partially curbed by a late-stage response that
  bends but does not break the trajectory.

  \item \textbf{Piecewise linear.} Four linear segments separated by jump
  discontinuities,
  \[
    R_t =
    \begin{cases}
      2.5 - \tfrac{0.5}{74}(t - 1),     & 1   \leq t < 76,  \\[2pt]
      0.8 - \tfrac{0.2}{74}(t - 76),    & 76  \leq t < 151, \\[2pt]
      1.7 + \tfrac{0.3}{74}(t - 151),   & 151 \leq t < 226, \\[2pt]
      0.9 - \tfrac{0.4}{74}(t - 226),   & 226 \leq t \leq 300,
    \end{cases}
  \]
  alternating between slowly drifting suppression and growth regimes, with
  abrupt regime switches at each boundary (e.g., the lifting and reimposition
  of restrictions).

  \item \textbf{Periodic.} A sum of three sinusoids on rescaled time
  $\tau = 10(t-1)/(n-1) \in [0,10]$:
  \[
    R_t = 0.2\!\left[\bigl(\sin(\pi \tau/12) + 1\bigr)
                       + \bigl(2\sin(5\pi \tau/12) + 2\bigr)
                       + \bigl(3\sin(5\pi \tau/6) + 3\bigr)\right].
  \]
  The curve oscillates smoothly between $R_t \approx 0.4$ and $\approx 2.4$
  via a slow seasonal envelope modulated by faster multi-week ripples. This is
  the only scenario without discontinuities and serves as a test of smooth,
  non-monotone tracking.
\end{enumerate}

These $R_t$ curves all generate a maximum of just below 1 million daily infections.
For context, COVID-19 crossed this threshold during the Omicron wave \citep{clarke2022seroprevalence}. 
Infections are declining by the end at all but the Periodic benchmark.

\subsection{Additional analysis}
  \begin{table}[ht]
  \centering
  \caption[ConvRt jump-discontinuity performance on the rtestim benchmarks.]{Evaluating ConvRt jump discontinuity on the rtestim benchmarks. Mean absolute error (MAE) and $\ell_1$ calibration error (CE) reported in units of $10^{-2}$.}
  \label{tab:rtestim-jumps}
  \begin{tabular}{l rr rr rr}
  \toprule
  & \multicolumn{2}{c}{S1: const} & \multicolumn{2}{c}{S2: exp}
  & \multicolumn{2}{c}{S3: linear} \\
  \cmidrule(lr){2-3} \cmidrule(lr){4-5} \cmidrule(lr){6-7}
  Method & MAE & CE & MAE & CE & MAE & CE \\
  \midrule
  ConvRt (with jump) & 0.20 &  3.27 & 0.64 & 10.74 & 0.30 &  9.29 \\
  ConvRt (no jump)   & 1.58 &  9.75 & 1.77 &  3.35 & 3.79 & 11.43 \\
  \end{tabular}
  \end{table}

\begin{figure}
    \centering
    \includegraphics[width=\linewidth]{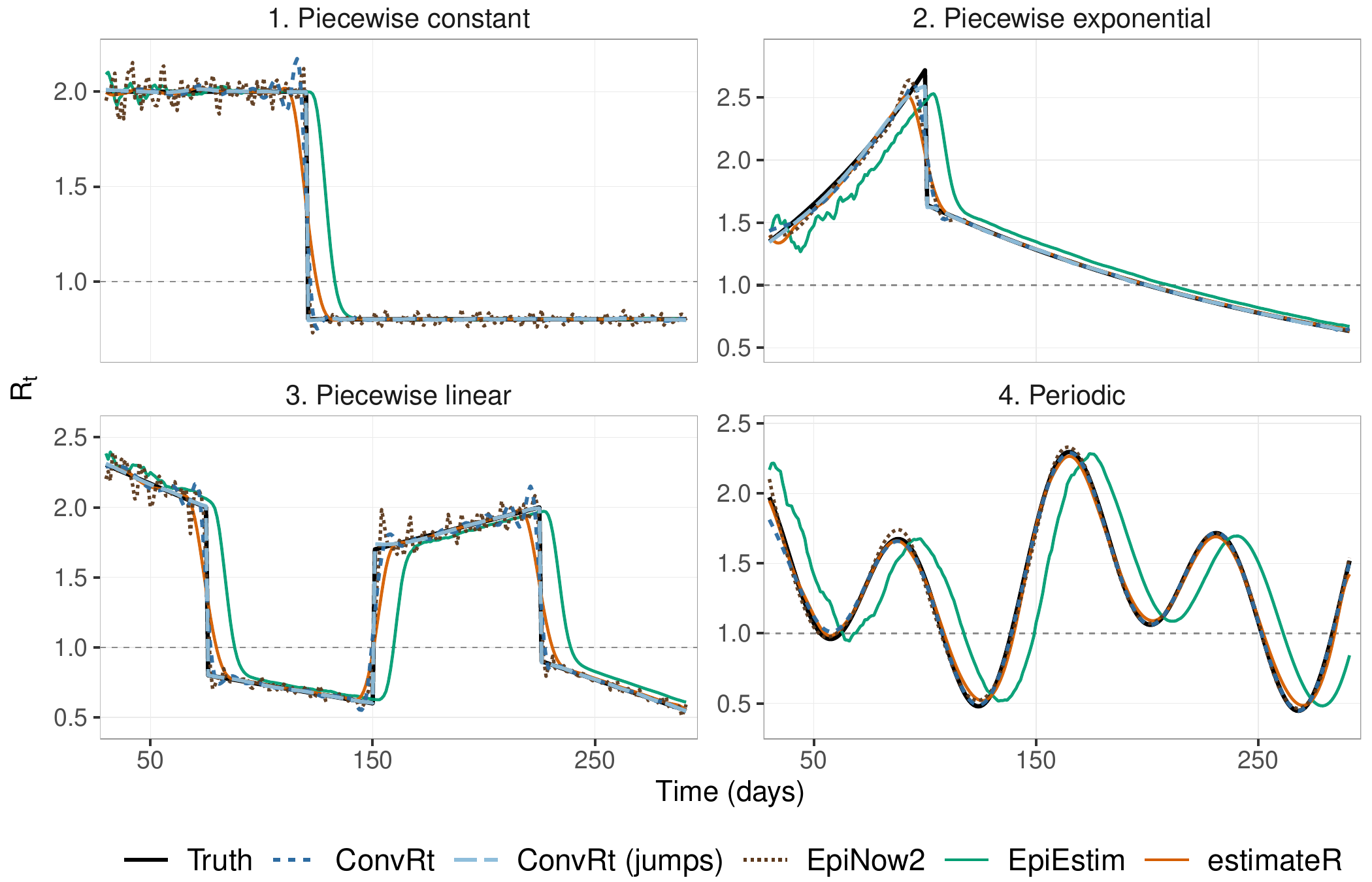}
    \caption[Retrospective $R_t$ estimates on the rtestim benchmarks.]{Retrospective $R_t$ estimates on the rtestim benchmarks, comparing ConvRt with and without the jump augmentation against EpiNow2, EpiEstim, and estimateR. The jump augmentation dramatically sharpens ConvRt's tracking at discontinuities while leaving smooth segments largely unchanged.}
    \label{fig:rtestim-all-results}
\end{figure}

\Cref{tab:rtestim-jumps} evaluates the jump augmentation on the three piecewise scenarios; \cref{fig:rtestim-all-results} provides a qualitative picture of all four benchmarks. The jump augmentation reduces MAE by factors of 8, 3, and 13 on Scenarios 1--3, respectively, with CE falling by 85\%, 76\%, and 72\%. The standard spline rounds off each discontinuity, producing a slow transition where the truth is abrupt; the step-function augmentation eliminates this artifact, yielding near-perfect recovery with tight confidence bands. The improvement on Scenario 2 is more modest in MAE because the jump occurs midway through an exponential decay, where the smooth spline partially compensates by bending aggressively. On the Periodic benchmark, the two ConvRt variants are essentially identical, as expected.

EpiNow2 tracks the broad shape of each curve but fails to recover jumps sharply. EpiEstim lags badly at every transition, substantially underestimating jump magnitudes; on the Periodic benchmark it also exhibits pronounced amplitude attenuation. estimateR oversmooths at the discontinuities but otherwise tracks reasonably well.

\section{Influenza benchmarks}\label[appendix]{apx:flu-sim} 

To prevent any method from having an unfair advantage, we chose to use the \citet{wallinga_teunis} estimator for case $R_t$.
Light smoothing was necessary on the earliest months, when low hospitalization counts led to unsteady estimates.
We analyzed the four months from October through January, when the bulk of hospitalizations occurred.

\subsection{Smooth ground truth, overdispersed observations}\label[appendix]{apx:flu-sim-smooth-od}

The ``smooth'' influenza simulation combines two modifications of its ``wiggly'' counterpart. The first replaces the wiggly Wallinga-Teunis ground truth with a heavily smoothed version, isolating method performance on a simple unimodal $R_t$ curve. The second swaps Poisson observations for a negative-binomial model calibrated to the day-to-day scatter in real flu hospitalizations.

\begin{figure}
    \centering
    \includegraphics[width=\linewidth]{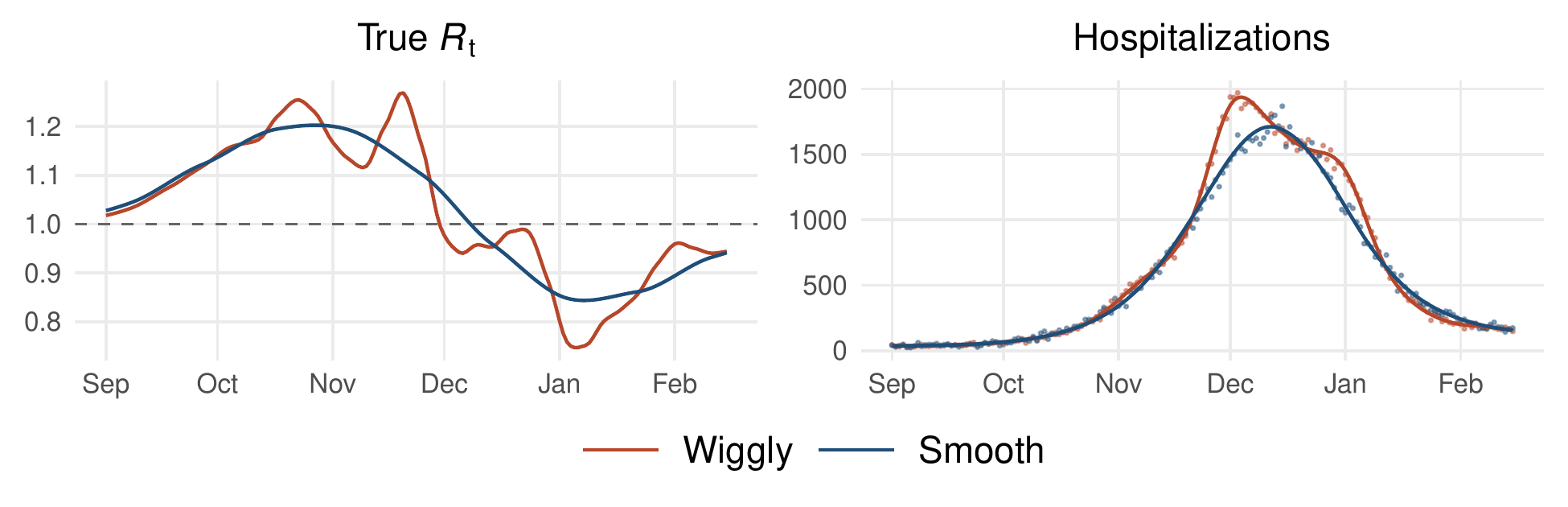}
    \caption{Ground-truth $R_t$ and daily hospitalizations for the flu simulations.}
    \label{fig:flu-sim-both}
\end{figure}

We apply LOESS with span 0.40 to the Wallinga-Teunis $R_t$ from the main text. The resulting curve rises smoothly through October, peaks near 1.2 in late November, and decays back below 1 through January (\cref{fig:flu-sim-both}). The simulated hospitalization wave is correspondingly unimodal, peaking around 1500 daily admissions near the new year.

The observation model is negative-binomial, with $\mathrm{Var}(Y_t) = \varphi\, \mu_t$ and $\varphi = 2.5$. We chose $\varphi$ to be calibrated with real flu hospitalizations. 
We fit a Poisson GAM (mgcv) to each season window, with a smooth time trend and a day-of-week factor. Using a knot count flexible enough to track the epidemic peak, the quasi-Poisson dispersion $\chi^2/(n - \text{edf})$ gives $\hat\varphi \approx 2.4$ for 2022/23 and $2.7$ for 2023/24. 
A Poisson observation model leaves 15--20\% of days outside its 95\% band around the fitted trend; $\varphi = 2.5$ leaves about 5\%, matching nominal.

\subsection{Additional figures}

\begin{figure}
    \centering
    \includegraphics[width=\linewidth]{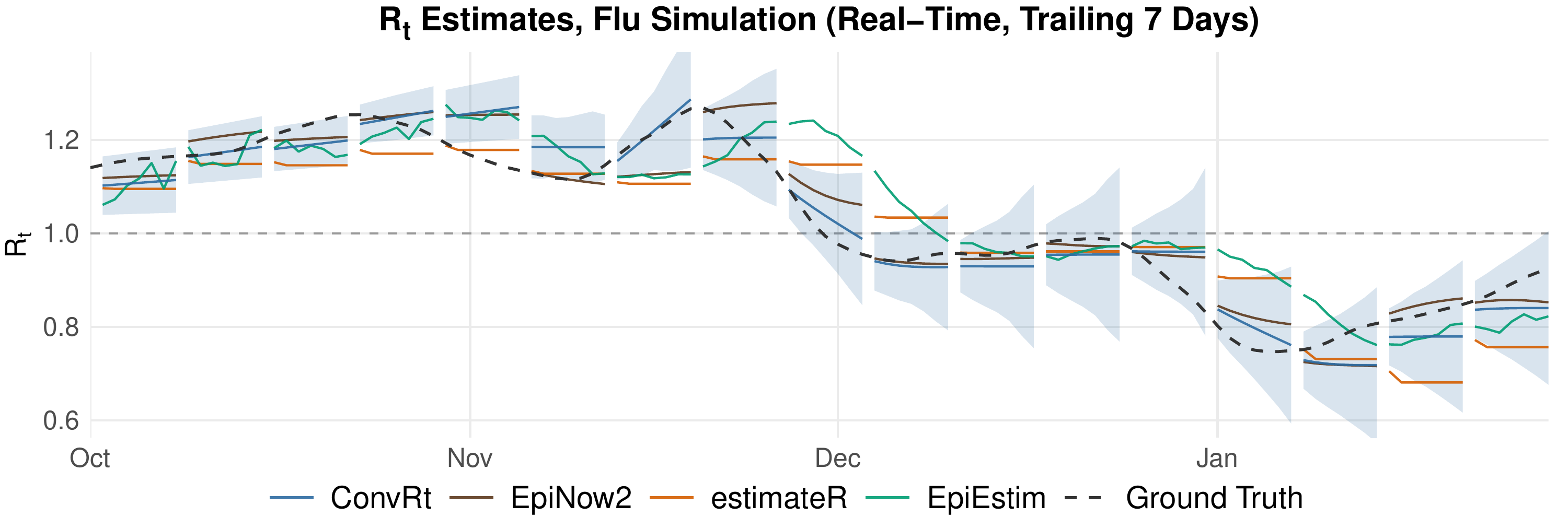}
    \caption[Real-time $R_t$ estimates on the wiggly influenza benchmark.]{Real-time $R_t$ estimates on wiggly influenza benchmark, with ConvRt's 95\% pointwise confidence bands.}
    \label{fig:realtime-rt-wiggly}
\end{figure}

We analyze two figures that compare $R_t$ estimates on the influenza benchmarks.
Matching \cref{fig:retro-comparison}, we omit EpiLPS and rtestim for clarity because they have similar bias as EpiEstim.

For the real-time setting, \cref{fig:realtime-rt-wiggly} shows last-7-day predictions on the ``wiggly'' dataset. 
ConvRt and EpiNow2 remain the most accurate methods.
As before, EpiEstim predicts roughly the true shape, but at a considerable lag. 
EstimateR is relatively unreliable in real time, since it propagates $R_t$ estimates from a noisy deconvolution.

ConvRt gives the best predictions as $R_t$ peaks and falls in November.
It is the sole method that identifies the sharp rise in real time.
Similarly, it follows the descent most faithfully, dropping more quickly than EpiNow2.
Its confidence intervals are also very accurate, covering the truth across most of the season.
Intuitively, they fan out toward the end of each vintage, where the nowcast
extrapolates recent data.
They also widen over the course of the season, as $R_t$ changes direction more frequently. 
Indeed, \cref{tab:combined-realtime} reports that ConvRt has lower MAE than EpiNow2 on this dataset, and considerably lower CE.
The remaining four methods sit well behind.

\begin{figure}
    \centering
    \includegraphics[width=\linewidth]{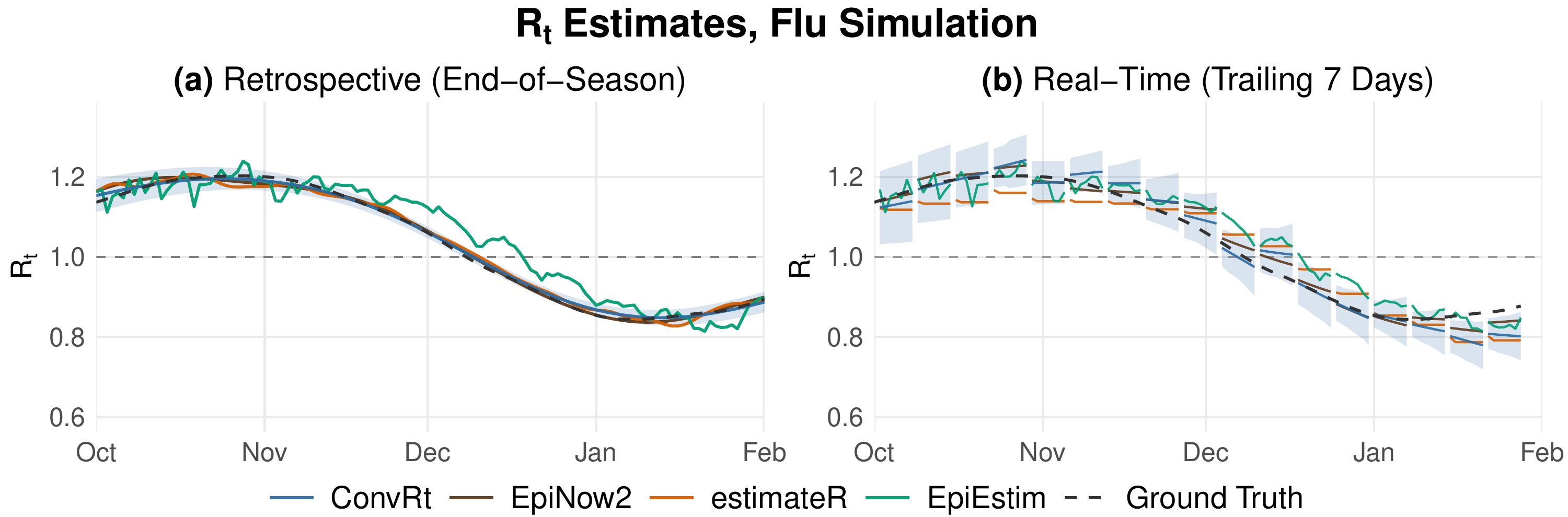}
    \caption[$R_t$ estimates on the smooth flu simulation.]{$R_t$ estimates on the smooth flu simulation. ConvRt's 95\% pointwise interval shaded.}
    \label{fig:rt-noisy-smooth}
\end{figure}

\cref{fig:rt-noisy-smooth} repeats both settings on the ``smooth'' benchmark, whose
true \(R_t\) rises gently to a November peak and bottoms out in January.
Here the problem is easier, and every method except EpiEstim predicts well.
In the retrospective panel (a), ConvRt, EpiNow2, and estimateR are nearly
indistinguishable from the truth. EpiEstim alone stays noisy and lags the
decline.

The real-time panel (b) tells the same story, with the three methods tracking
the truth closely and ConvRt usually getting the directional trend right.
ConvRt edges out EpiNow2 in retrospective MAE and trails it slightly in
real-time, but the gap is small, with both methods performing well (\cref{tab:combined-realtime}).
ConvRt's confidence bands remain well calibrated, with a real-time CE just below 5.
The intervals often cover even when the point estimate misreads direction. 
ConvRt incorrectly predicts a
rise in early November and a fall in early January, and the truth stays inside
the band both times.

\subsection{Results by smoothness parameter \texorpdfstring{$\lambda$}{lambda}}\label[appendix]{apx:flu-sim-lambda}

\begin{figure}
    \centering
    \includegraphics[width=\linewidth]{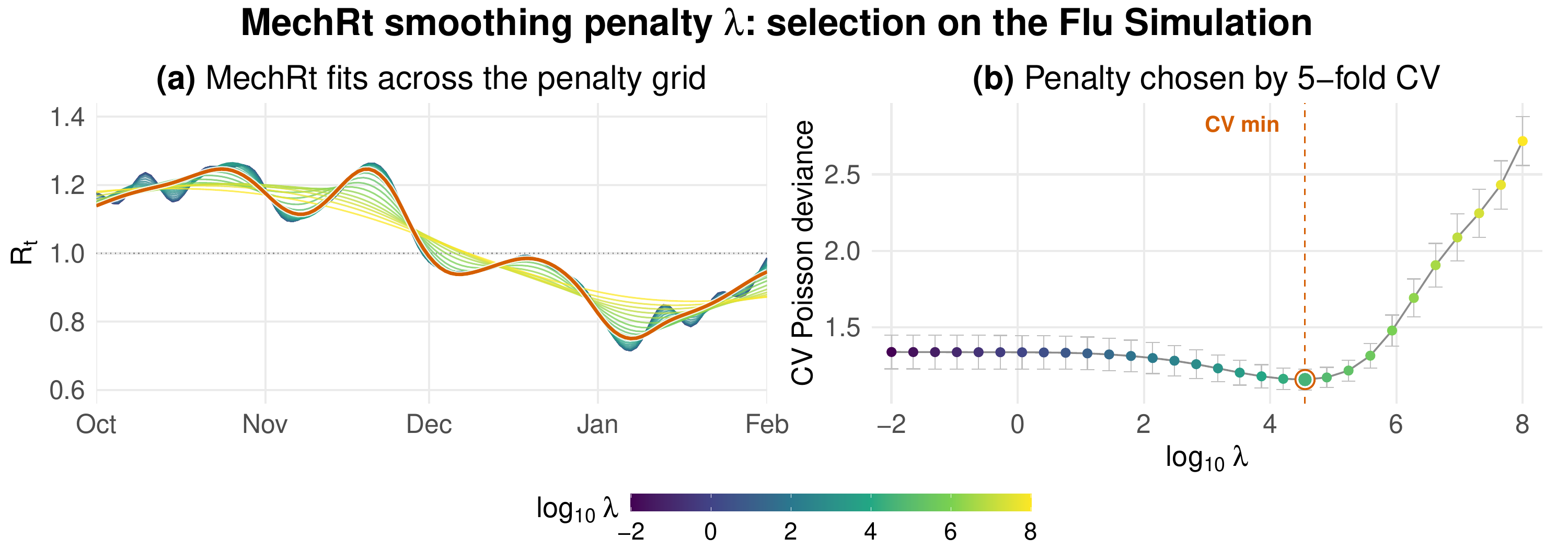}
    \caption[ConvRt predictions by smoothness parameter $\lambda$.]{ConvRt predictions by smoothness parameter $\lambda$. Most $\lambda$'s with lower cross-validation error produce reasonable $R_t$ curves, with minor qualitative differences.}
    \label{fig:results-by-lambda}
\end{figure}

\Cref{fig:results-by-lambda}(a) shows retrospective ConvRt fits on the flu simulation across a grid of curvature penalties $\lambda$, with the min-rule fit highlighted. The min rule slightly oversmooths the peak, leaving $\hat R_t$ about $0.02$ below the truth in early January. A modestly smaller $\lambda$ closes most of this gap, lifting the peak with only minor added fluctuation elsewhere. Pushing $\lambda$ smaller still produces visibly noisy fits that chase short-run fluctuations in the cases, while larger $\lambda$ flattens the curve toward a single hump and erases the brief plateau in November. The min rule trades a small peak underestimate for stability across the rest of the season.

The spread across $\lambda$ is also far from uniform across the season. Near the start and end, where case counts are low, the fits fan out widely and the smaller $\lambda$'s oscillate. Through the high-count middle, the curves collapse onto each other regardless of $\lambda$. This reflects the likelihood: high counts pin $R_t$ down tightly, so the penalty has little room to move the fit, while at low counts the data are weak and $\lambda$ does most of the work.

Panel (b) plots the 5-fold CV Poisson deviance against $\log_{10}\lambda$. It is nearly flat for $\lambda$ below the minimizer near $\log_{10}\lambda \approx 4.5$ and rises sharply above it, so the min rule sits at the foot of a steep wall. Smaller $\lambda$'s incur almost no CV cost, consistent with the muted differences from the min-rule fit on the left.
The flat loss landscape supports the plausibility of $\lambda$ values below the min-rule, and their larger peak $R_t$.

\section{Influenza experiments}\label[appendix]{apx:real-flu}

To run ConvRt, we assumed a constant infection-hospitalization rate.
While a simplification, this assumption is reasonable within a season.
Moreover, estimating a time-varying IHR for a given flu season would likely require non-public data.

We also ran all experiments using finalized hospitalization counts.
These have almost certainly been backfilled, or adjusted as new counts are reported for a given date.
A properly real-time approach would have used appropriately versioned data, but we have not pursued this.

\begin{figure}
    \centering
    \includegraphics[width=\linewidth]{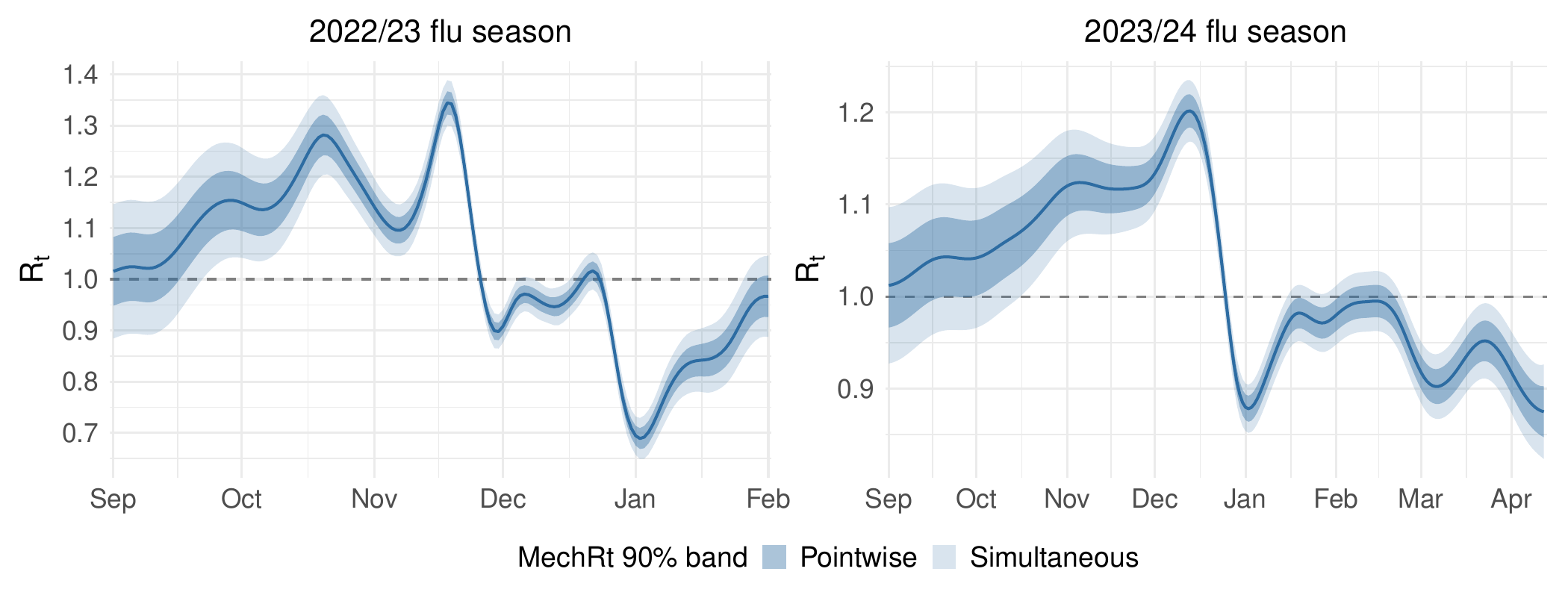}
    \caption[Retrospective ConvRt $R_t$ with pointwise and simultaneous confidence bands.]{Retrospective ConvRt $R_t$ for the 2022/23 and 2023/24 flu seasons, with 90\% pointwise (dark) and simultaneous (light) confidence bands. Pointwise intervals cover each timestep marginally at the nominal level; simultaneous bands cover the entire curve jointly.}
    \label{fig:retro-flu-simband}
\end{figure}
\cref{fig:retro-flu-simband} shows ConvRt's pointwise and simultaneous confidence bands for the two retrospective flu fits. The pointwise intervals are the ones plotted in \cref{fig:retro-flu-both} of the main text; the simultaneous bands are computed from the simulation-based critical value $c_\alpha$ described in \cref{apx:rt-uncertainty}. As expected, the simultaneous bands are uniformly wider --- by a factor of roughly $c_\alpha / z_{1-\alpha/2}$ --- since they must guard against \emph{any} of $\sim$150 timesteps escaping the band rather than a single one. The two bands answer different questions: a pointwise interval supports statements like ``$R_t$ exceeded 1 on Dec 15,'' while a simultaneous band supports statements like ``$R_t$ exceeded 1 throughout December.'' Which is appropriate depends on the inferential target.

\begin{figure}
    \centering
    \includegraphics[width=\linewidth]{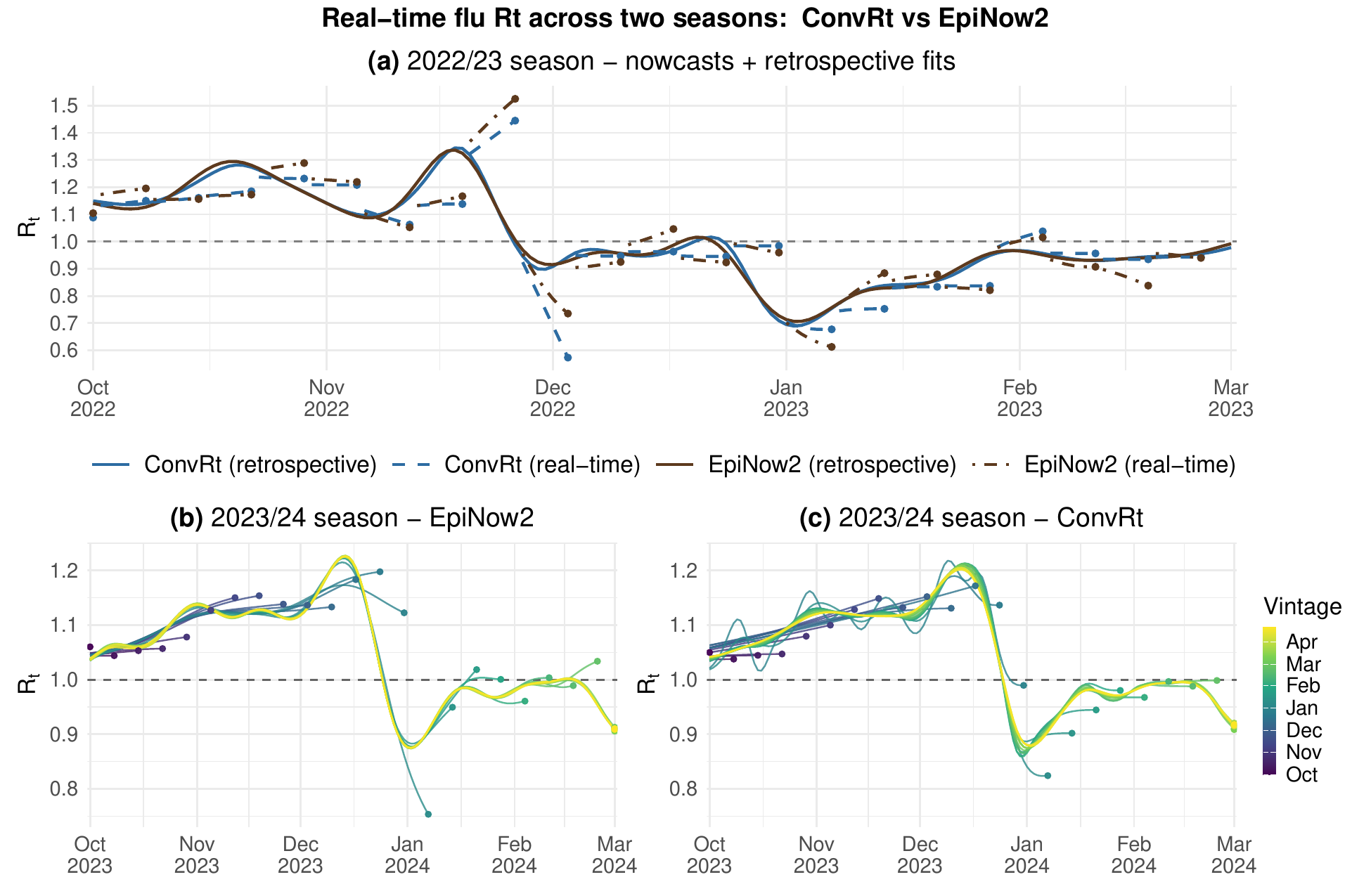}
    \caption[Real-time flu $R_t$, ConvRt vs.\ EpiNow2.]{Real-time flu $R_t$, ConvRt vs EpiNow2.
  (a)~2022/23: last-7-day nowcasts, contextualized by end-of-season fits.
  (b,c)~2023/24: weekly real-time vintages for EpiNow2 and ConvRt.}
    \label{fig:real-time-flu}
\end{figure}

\cref{fig:real-time-flu} shows ConvRt and EpiNow2's real-time predictions across both seasons, tuning $\gamma$ with the min rule.
As in the retrospective case, they are usually close to one another.
Both are quite strong when $R_t$ moves consistently.
In certain instances, ConvRt has superior performance.
For example, it correctly identifies that $R_t$ plunges in late December 2023, while EpiNow2 drops modestly.

\begin{figure}
    \centering
    \includegraphics[width=\linewidth]{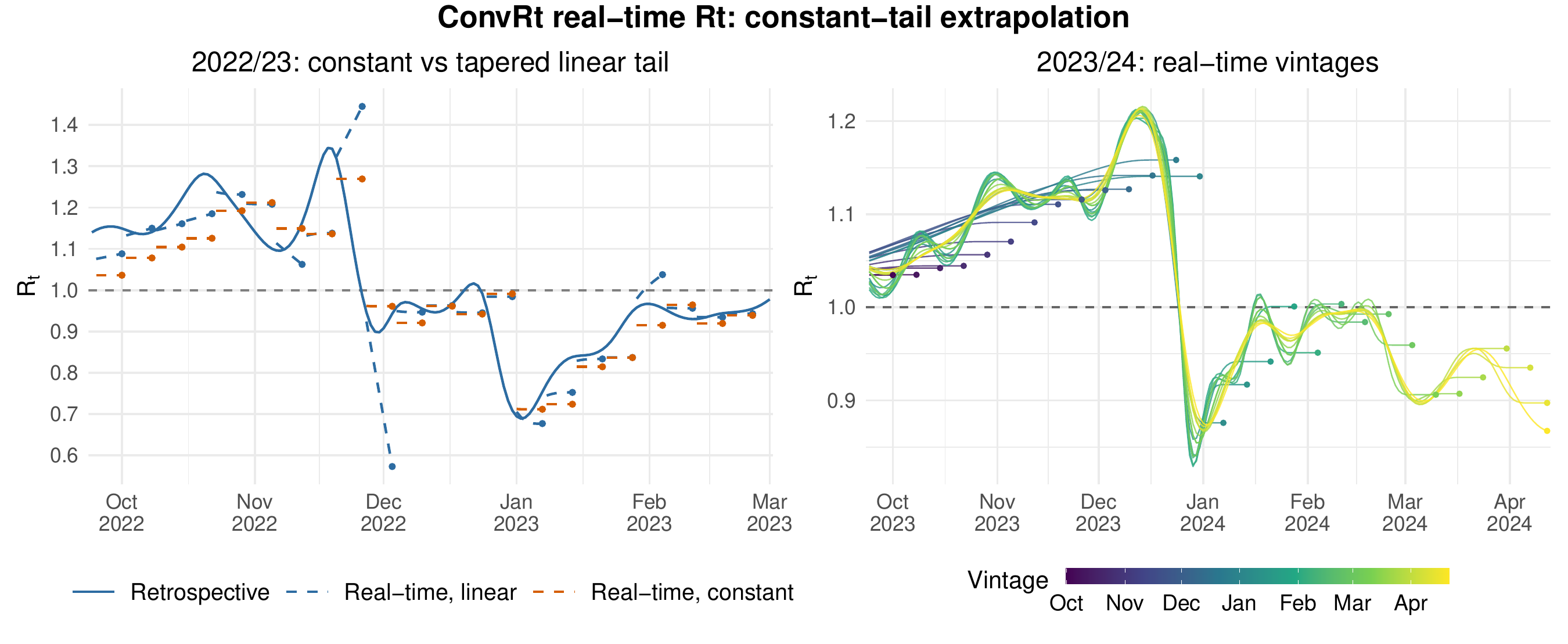}
    \caption[Real-time ConvRt nowcasts with a constant-tail constraint.]{Real-time ConvRt nowcasts with a constant-tail constraint (B-spline basis, $D^{(1)} S_\text{tail}\theta=0$; see \cref{apx:optim-constraints}). Left: 2022/23, comparing the constant-tail variant (orange) to the tapered-linear default (blue dashed) and the retrospective fit (solid blue). Right: full 2023/24 vintages under the constant-tail constraint, colored by vintage month, with dots marking each weekly nowcast.}
    \label{fig:constant-tail}
\end{figure}

ConvRt's nowcasts are roughly as stable as EpiNow2's.
They almost never change sharply, except when the data exhibits a longstanding trend.
This causes errors when $R_t$ changes directionality, which takes over a week to appear in the data.
For example, both methods overshoot the peak and trough in late 2022.
Those swings can be mitigated by tuning ConvRt with the 1se rule.
Alternatively, one can enforce stability by fitting ConvRt with a constant tail constraint.
Doing so produces strong predictions, visualized by \cref{fig:constant-tail} in \cref{apx:real-flu}.
However, the tapered linear tail can better estimate trends, such as rising $R_t$ in October 2023.

\begin{figure}
    \centering
    \includegraphics[width=\linewidth]{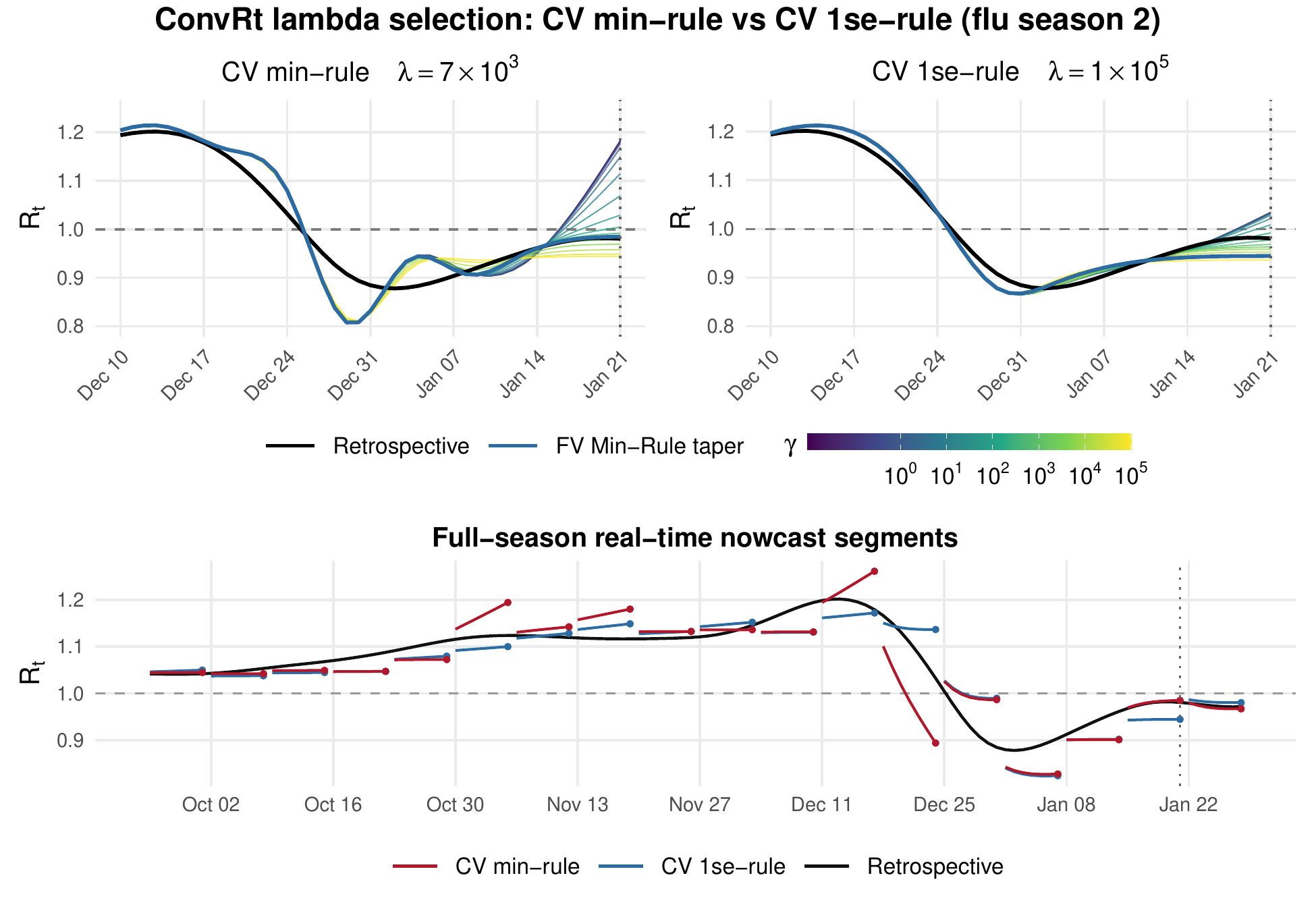}
    \caption{Real-time $R_t$ predictions by hyperparameter values on 2023/24 flu season. Top row shows predictions across $\gamma$, for $\lambda$ tuned by the min-rule and 1se-rule. Bottom row shows all last-7-day nowcasts.}
    \label{fig:real-flu-lambda-realtime}
\end{figure}

The top row of \cref{fig:real-flu-lambda-realtime} illustrates how $\lambda$ and $\gamma$ selection affects real-time predictions. Both rows show $\lambda$ tuned via cross-validation with the min-rule and 1se-rule. On the top row, each colored curve shows a single real-time nowcast at a given $\lambda$ across the range of possible $\gamma$.
Tuning $\lambda$ with the min-rule, the $R_t$ curve is very wiggly, more so than the final retrospective curve. 
Predictions range widely with $\gamma$, though forward-validation correctly tunes it as essentially unchanging.
The 1se-rule produces more sensible $R_t$ estimates, and tail predictions vary less with $\gamma$.
However, forward-validation now selects a high $\gamma$ that predicts $R_t$ slightly too low.

The bottom row compares $\lambda$ selection across the whole season, maintaining the min-rule for tuning $\gamma$.
The min- and 1se-rules usually generate similar $R_t$, but they occasionally diverge.
Intuitively, last-7-day nowcasts are more flexible with the lower $\lambda$'s tuned by the min-rule. 
Sometimes this hurts estimation, as in the three small unnecessary swings before the peak.
But min-rule predictions are more accurate at least twice, after the peak.

\begin{figure}
    \centering
    \includegraphics[width=0.9\linewidth]{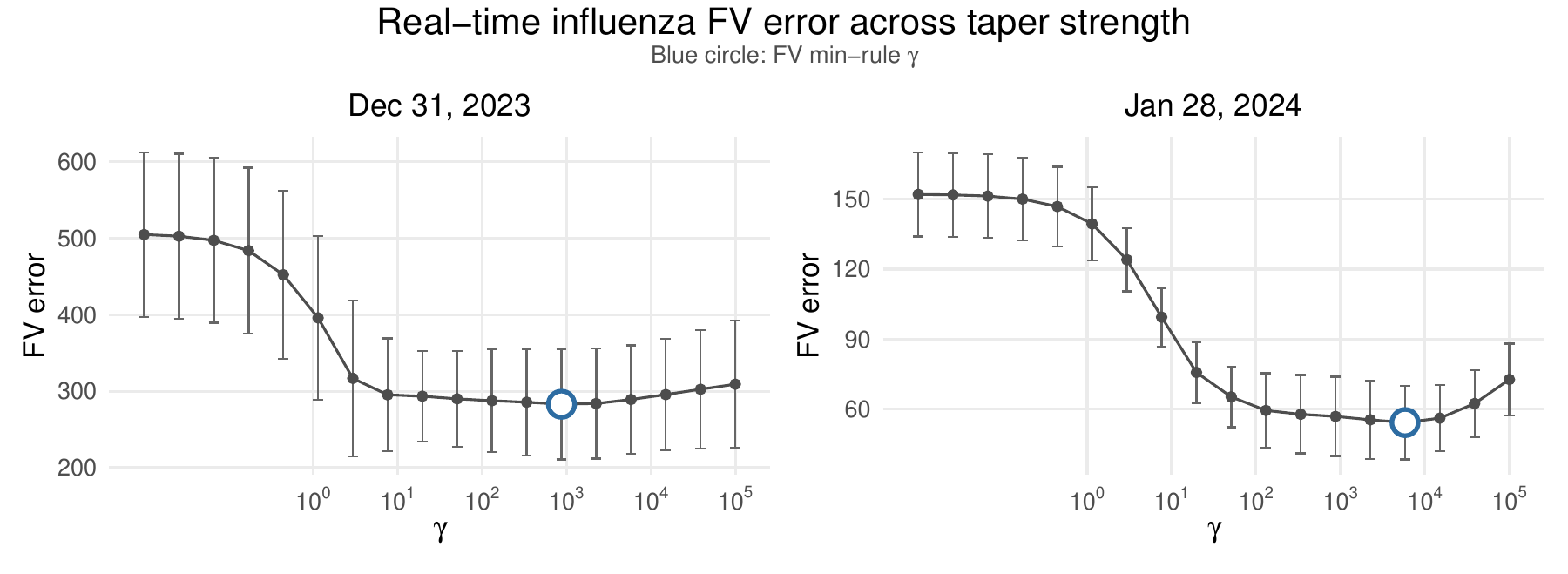}
    \caption{Forward-validation error curves (Poisson deviance) for taper hyperparameter $\gamma$.}
    \label{fig:fv-errors-flu}
\end{figure}

\cref{fig:fv-errors-flu} shows the forward-validation error curves that accompany the $R_t$ curves in \cref{fig:rt-by-gamma}. 
Notably, the error landscape is very flat. 
Barring dramatic undersmoothing (low $\gamma$),
different hyperparameter values have similar FV error.
In our experiments, we use a short 7-day tuning window to emphasize adaptivity to recent trends, 
which may contribute to the large standard errors. 
However, we find this trend persists when using a 14-day window (not shown).

This flat landscape has important consequences.
By these error metrics, hyperparameters producing very different qualitative results have comparable goodness-of-fit.
Moreover, systematic approaches like the min-rule may not be able to discern the most reasonable option.
The 1se-rule will almost always impose near-constant tail smoothing.
As a result of these findings, we find it is most prudent to use the min-rule as a best guess,
while considering predictions at all reasonable options.

The left panel of \cref{fig:constant-tail} highlights both the similarities and the qualitative differences between the two real-time variants. Across most of the season, the constant-tail and tapered-linear nowcasts agree closely. While they occasionally diverge, neither exclusively dominates the other. For example, as $R_t$ climbs in October 2022, the tapered-linear variant correctly extrapolates upward every week.
In contrast, the constant-tail variant stays flat, underestimating by a larger margin. 
But after the peak in mid-November, the true $R_t$ has begun a sharp decline that the trailing-window data does not yet reflect. The tapered-linear estimator extrapolates the still-rising trend and overshoots. The constant-tail estimator is also too high but, by refusing to extrapolate, remains closer. 

The right panel shows that the constant-tail constraint produces coherent results across vintages on the longer 2023/24 season. 
Successive nowcasts track the final retrospective trajectory smoothly.
Comparing to \cref{fig:real-time-flu}, the constant-tail fits are somewhat more stable on this season than their tapered linear counterparts, at the cost of reduced adaptivity. 

The two approaches are closely related. As the tapered penalty weight $\gamma$ grows large, the first-order differences in $R_t$ near the boundary are driven toward zero and the tapered-linear estimator approaches a constant tail. The two are not identical: the tapered penalty's weights extend a few days \emph{before} the last knot rather than acting only past it, so a high-$\gamma$ tapered-linear fit is roughly flat over a slightly wider window than a hard constant-tail constraint imposes. We default to the tapered-linear variant because it spans a richer family of tail behaviors --- approaching flat as $\gamma \to \infty$, reverting to the natural-spline linear tail as $\gamma \to 0$, and interpolating between them --- whereas a constant-tail constraint commits to a single shape. The qualitative differences above suggest this added expressivity is a real, if modest, advantage.

\section{COVID-19 experiments}\label[appendix]{apx:covidestim}

\begin{figure}[h]
    \centering
    \includegraphics[width=\linewidth]{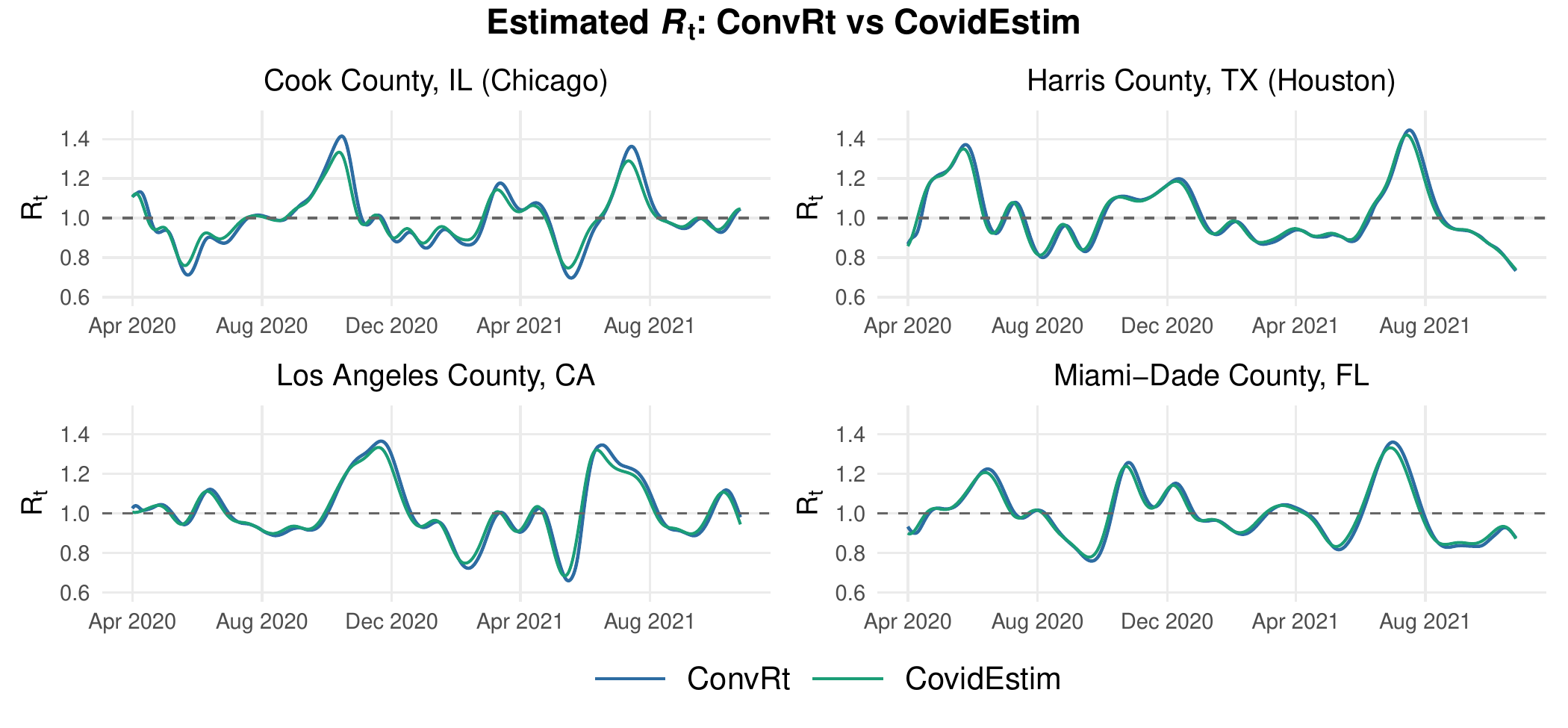}
    \caption{COVID-19 $R_t$ estimates from ConvRt and CovidEstim on four major \US counties.}
    \label{fig:covidestim-counties}
\end{figure}

\begin{figure}
    \centering
    \includegraphics[width=\linewidth]{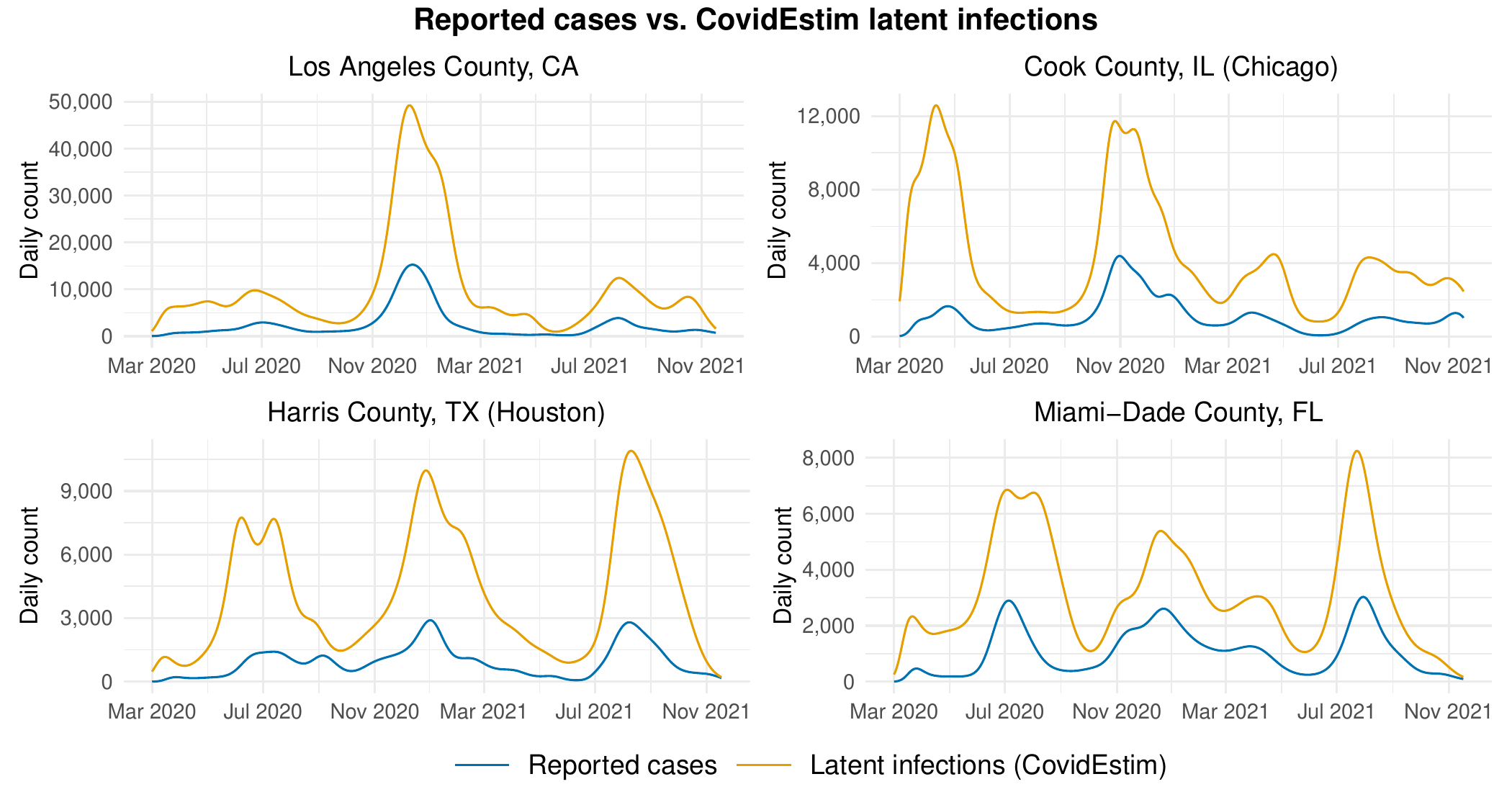}
    \caption[Cases and infection counts for pre-Omicron COVID-19.]{Cases and seroprevalence-based infection counts for pre-Omicron COVID-19 in four \US counties.}
    \label{fig:covid-cases}
\end{figure}

In addition to seasonal influenza, we evaluated our $R_t$ methods on COVID-19.
Estimating $R_t$ for COVID-19 is more challenging, as it is sensitive to changing ascertainment rates over time.
The CovidEstim model addressed this by incorporating seroprevalence data, publishing case ascertainment rates along with $R_t$ \citep{Chitwood2022}.
Both quantities are modeled as splines, with prior distributions placed on their parameters.
This Bayesian method is notoriously slow to run, often taking 10 hours to fit a single state \citep{covidestim_summer2021}.
We plugged in their case ascertainment rates to conduct a retrospective analysis with our method.
The COVID-19 infections these ascertainment rates imply are shown in \cref{fig:covid-cases}, along with positive case reports.

\cref{fig:covidestim-counties} compares ConvRt's $R_t$ estimates on a number of \US counties.
To avoid challenges with data access and model fitting, we used cached results rather than generate our own.
These results only had uncertainty bands at the state level, hence our point estimates.
Overall, predictions are almost identical, always within 0.1 of one another.
Large spikes in $R_t$ clearly correspond to surges in winter 2020, as well as the onset of the Delta variant in summer 2021.
Interestingly, surges happen at slightly different times and magnitudes, mirrored by the infection counts in \cref{fig:covid-cases}.
For example, Houston and Miami had substantial COVID-19 waves in summer 2020, driven by rises in $R_t$ well above 1.
Across the political divide, Chicago and Los Angeles generally did not, with LA's $R_t$ peaking at 1.1.

The values of $R_t$ in \cref{fig:covidestim-counties} are entirely plausible. 
A CDC analysis using EpiEstim found that most state-level peaks sit between 1.2 and 1.5, matching ours of 1.3-1.4 \citep{lopez2023covid}.
\citet{figgins2021sarscov2}, another Bayesian approach, suggests somewhat higher peaks, with a maximum of 1.66.
Gaps in $R_t$ can often be accounted for by the delay distributions,
with longer delays producing higher estimates.
It is thus easier to explore a range of possibilities with our method,
which takes seconds to run rather than hours.

\begin{figure}
    \centering
    \includegraphics[width=0.8\linewidth]{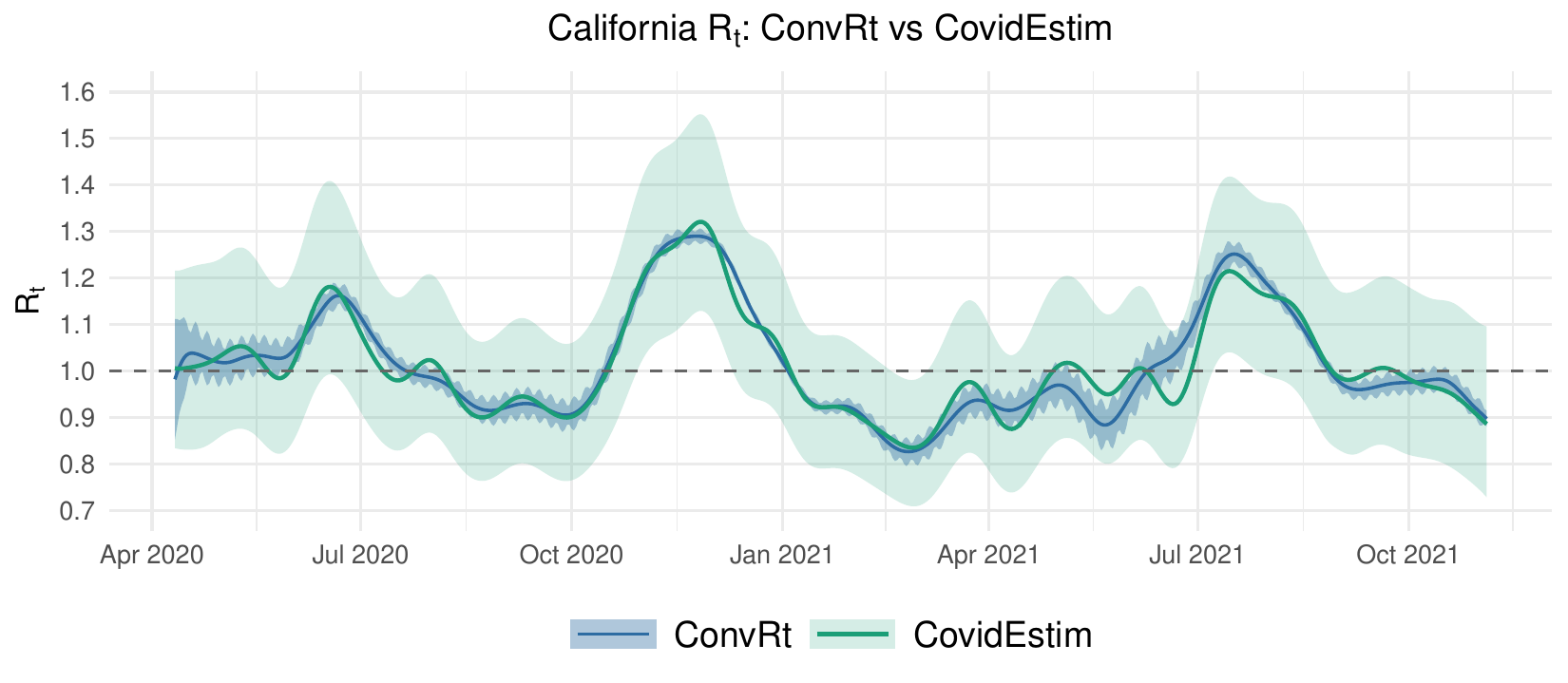}
    \caption{California $R_t$ curve, CovidEstim and ConvRt.}
    \label{fig:california-covidestim}
\end{figure}

We aggregated data to also compare $R_t$ in the state of California. 
\cref{fig:california-covidestim} shows the results, with 80\% pointwise confidence bands. 
Once again, the results are extremely similar throughout the nearly two years tracked.
At the peaks (\textasciitilde1.3 in winter 2020, \textasciitilde1.2 in Delta), the two estimators differ at most 0.05.
Moreover, CovidEstim's credible intervals are far wider than our confidence intervals.
The intervals shown are pointwise, meaning they are valid at any individual timestep but
are not expected to cover jointly. 

To produce Figures \ref{fig:covidestim-counties} and \ref{fig:california-covidestim}, we estimated generation intervals and infection-to-report distributions from the literature.
Parameterizing both as discrete gammas, we set the mean generation time to 4.5 days and an SD of 1.9 \citep{ganyani2020estimating, hart2022generation}.
For the infection-to-report delay, we set its mean to 10.2 days and its SD to 3.6 days \citep{li2020early, he2020temporal,arino2020simple, abbott2020estimating}.

\begin{figure}
    \centering
    \includegraphics[width=0.9\linewidth]{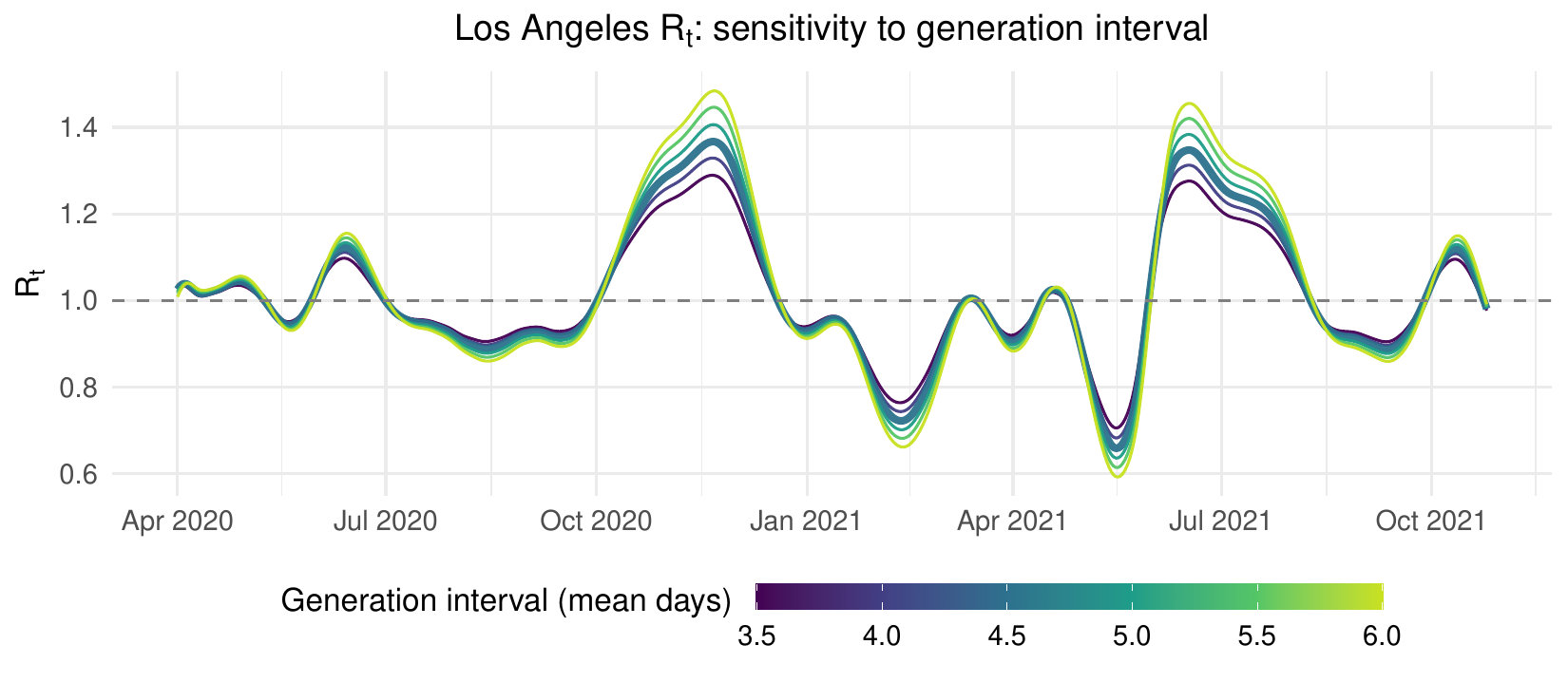}
    \caption[COVID-19 $R_t$ sensitivity to generation interval in Los Angeles.]{ConvRt's predictions of COVID-19 $R_t$ in Los Angeles under various delay distributions.}
    \label{fig:covidestim-delays}
\end{figure}



While the results matched CovidEstim nicely, neither distribution is known with high confidence. For that reason, \cref{fig:covidestim-delays} shows $R_t$ estimates on the Los Angeles CovidEstim data under varying generation intervals, sweeping mean generation times from 3.5 to 6.0 days while holding the SD and infection-to-report delay fixed. The six resulting curves share the same shape, differing only in magnitude: in the winter 2020 peak, estimates range from around 1.3 to 1.5, with longer generation intervals producing higher $R_t$ at peaks and lower at nadirs. This is expected---longer intervals compare current infections to levels further in the past, requiring a more extreme multiplier to explain the larger gap.

\section{Additional methodology}\label[appendix]{apx:extra-methods}

\subsection{Weekly data}\label[appendix]{apx:weekly}

\begin{figure}
    \centering
    \includegraphics[width=\linewidth]{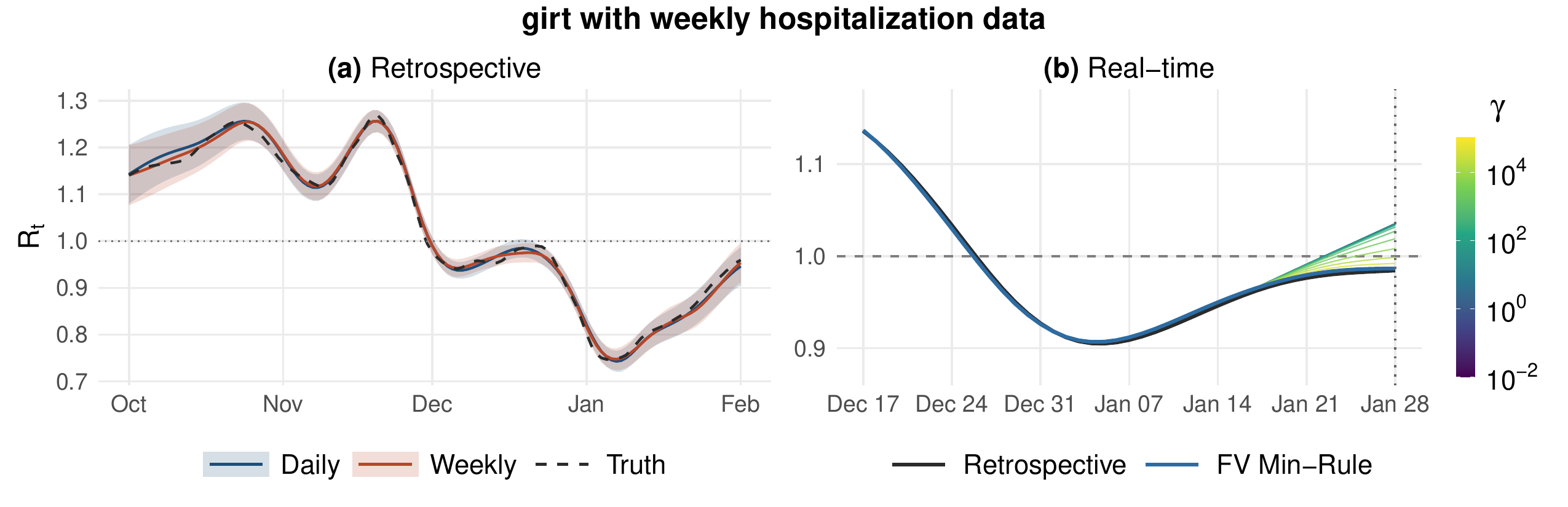}
    \caption{ConvRt predictions given data at a weekly resolution, in retrospect and in real time.}
    \label{fig:weekly}
\end{figure}

Often, data are released at a weekly resolution, not daily. 
For example, the influenza experiments in this paper used hospitalization counts from NHSN.
This pandemic-era program ceased its daily reporting in May 2024.
Alternatively, Flu-Surv NET has reported weekly flu hospitalizations from 2005 to the present, based on a surveillance sample. 
In addition, COVID-NET and RSV-NET publish weekly hospitalizations for COVID-19 and RSV.

Our estimation approach must be adjusted to work with less frequent data.
One option would be to change the discrete-time transmission model to weekly resolution.
However, this risks being significantly too coarse, as most transmissions occur within the first week for many infectious diseases. 
Maintaining daily dynamics, we instead adjust the likelihood to model weekly observations $y_w$.
\cref{eq:mean-cases} related each daily total $y_t$ to its mean $\mu_t$, which had been scaled by the ascertainment rate $\rho_t$ and day-of-week effect $\omega$.
Dropping day-of-week effects for simplicity, we aggregate means over each week $w$:
\begin{equation}\label{eq:poisson-seir-weekly}
    y_w\sim \text{Quasi-Poisson}(\mu_w, \varphi); 
    \qquad \mu_w = \sum_{t\in w} \rho_t \Lambda_t;
    \qquad \Lambda_t = \sum_{s<t} x_{s} \pi_{t-s}.
\end{equation}

Once again, explicitly accounting for overdispersion $\varphi$ is optional. 
The statistical model is justified by the fact that the sum of independent Poissons is still Poisson.
(We assume throughout this work that counts on successive dates are independent.) 

To estimate $R_t$, everything beyond the likelihood is identical to the daily case.
Using the same spline parametrization, ConvRt performs penalized MLE in the retrospective and real-time settings.
It first deconvolves latent infections $\hat x$, as described in \cref{apx:deconvolution}.
Then, it plugs them into $\Lambda_t$ to deconvolve $R_t$ \eqref{eq:exp-cases-rt}, again replacing $x_t$ with the renewal equation \eqref{eq:renewal}.
Hyperparameters are tuned via cross-validation.
A higher degree of smoothing may be necessary, since the number of data points is reduced by a factor of 7.

We ran ConvRt on the simulated flu dataset, fitting on weekly hospitalization totals.
Hyperparameters were tuned via the min-rule.
The left panel of \cref{fig:weekly} shows the retrospective fit’s point estimates and 95\% confidence bands.
Both are almost exactly the same as those from the daily case. 

The right panel displays ConvRt’s real-time predictions, given weekly data through January 28.
The data support a moderate but clear rise in $R_t$ from Jan 7--21. 
However, the infection-to-report delay is such that $R_t$ in the final week is barely observed.
Contrasting predictions across $\gamma$ reflect the range of possible outcomes, and the strength of the evidence supporting them.
With little tail regularization, extrapolation continues linearly and overshoots the truth by \textasciitilde0.05.
Heavier regularization, which forward-validation selects, tamps down the tail, and correctly predicts $R_t$ just under 1.

Together, these results are highly encouraging.
They suggest our method can function perfectly well given only weekly data.
While comprehensive experiments are beyond the scope of this paper, further work should evaluate its quantitative performance, and consider alternate approaches.

\subsection{Trend Filtering}

\begin{figure}
    \centering
    \includegraphics[width=.8\linewidth]{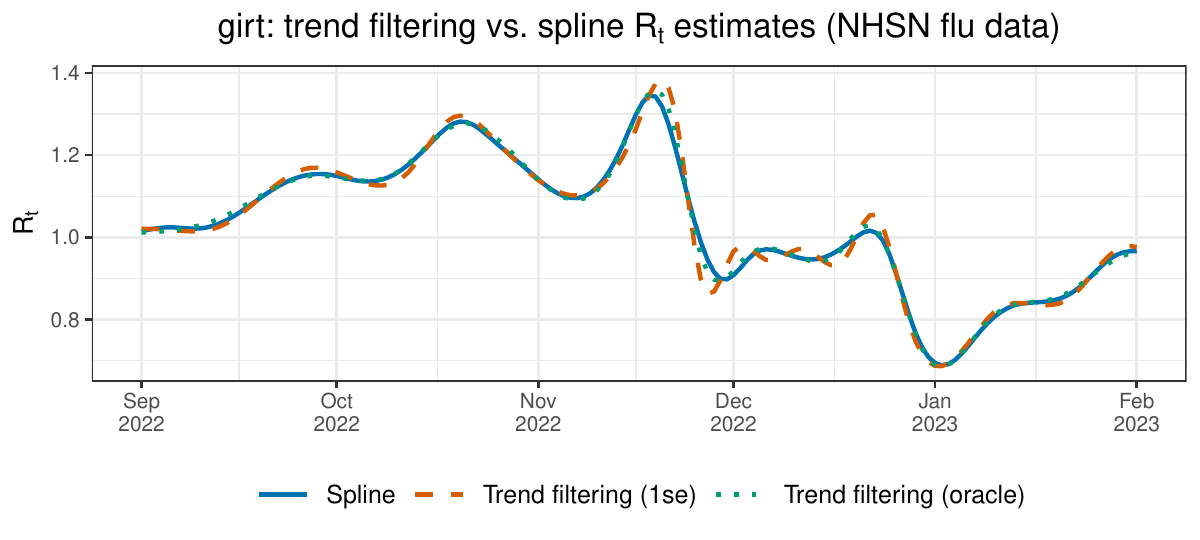}
    \caption{ConvRt predictions on NHSN flu data, modeled as a cubic spline and trend filter.}
    \label{fig:tf-rt}
\end{figure}

We ran ConvRt with trend filtering regularization, instead of as a spline. 
This nonparameteric regression technique, introduced in \cref{sec:rt-retro}, adaptively selects knot locations.
We chose a fourth-difference penalty to produce piecewise cubic functions, matching the splines in our experiments. 
We define  $\mathcal{R}$ as the vector of $R_t$ values, and ignore DoW effects for convenience.
The retrospective fit is
\begin{align}\label{eq:tf-retro}
\hat{\mathcal{R}} &= 
\argmin_{\mathcal{R} \succeq 0}
             \sum_{t=t_0}^{t_1}\!\bigl[ \mu_t(\mathcal{R}) - y_t \log \mu_t(\mathcal{R}) \bigr]
             \;+\; \lambda\lVert D^{(4)} \mathcal{R} \rVert_1 \\
             &\qquad\text{where}\quad \mu_t(\mathcal{R}) = \rho_t\, \sum_{s<t} R_s \left(\sum_{u<s} \hat x_u\, g_{s-u}\right)\pi_{t-s}.
\end{align}
The real-time fit was defined analogously with tail constraints and regularization. Other design decisions also matched the spline experiments. 
The likelihood model for the data stayed the same, as did hyperparameter tuning. 

IRLS is no longer applicable for optimization, since the objective \eqref{eq:tf-retro} is not twice-differentiable due to its non-smooth $\ell_1$ penalty.
Instead, CVXR optimized the objective with the ECOS solver.
For a retrospective fit, this took 30 seconds to tune over a grid of 25 lambdas.
While relatively fast, the spline took only 2 seconds to tune, since it converged in only a few iterations of IRLS.

As with weekly data (\cref{apx:weekly}), we did not conduct thorough experiments to evaluate trend filtering. 
However, we established a meaningful proof-of-concept, this time on real data for contrast.
\cref{fig:tf-rt} displays trend filtering’s predictions on the 2022/23 flu season. 
By and large, they are very similar to the spline’s $R_t$ estimates.
Tuning $\lambda$ with the 1se rule, trend filtering is slightly wigglier, but this can be attributed to the particular value of $\lambda$ that was tuned here; selecting a larger lambda, as shown in the dotted line, yields an $R_t$ curve that is nearly identical to the spline.

In general, differences in prediction are subtle, not systematic.
This supports our choice to focus on splines due to their natural uncertainty quantification. 
In contrast, asymptotic confidence intervals cannot be derived for trend filtering problems, as they are not twice-differentiable.
Nevertheless, trend filtering remains a reasonable alternative if the true signal changes dynamically.

Moreover, several avenues for uncertainty quantification exist, though we have not pursued these here.
Rtestim obtains confidence intervals---albeit wide ones---using a quadratic relaxation of their trend filtering objective \citep{rtestim}.
Data fission \citep{leiner2025data, dharamshi2025generalized} offers a more sophisticated path forward. 
This technique splits the information in each observation to create independent copies for selection and inference.
Applying it to trend filtering under a Poisson inverse problem, as arises in our deconvolution setting, is an open problem.

\end{document}